\documentclass[12pt,a4paper]{article}
\pdfoutput=1
\usepackage{epsfig,axodraw}
\usepackage{array,cite,amsmath,amssymb}
\usepackage[flushleft]{threeparttable}
\usepackage{subcaption}
\usepackage{graphicx,epsfig,dcolumn,bm,epic,eepic,float}% Include figure files
\usepackage{amsfonts}
\usepackage{amssymb}
\usepackage{bbm}
\usepackage{xcolor}
\usepackage{booktabs}
\usepackage{autobreak}
\usepackage{multirow}
\usepackage{longtable}
\usepackage[T1]{fontenc}
\usepackage[utf8]{inputenc}
\usepackage{mathtools} 
\usepackage{pifont}
\usepackage{latexsym}
\usepackage{color}
\usepackage{cancel}
\usepackage{scrextend}
\usepackage{amsmath}
\usepackage{amsthm}
\usepackage{eucal}
\usepackage{amssymb}
\usepackage{mathrsfs}
\usepackage[bottom]{footmisc}
\usepackage{makeidx,shortvrb,latexsym}
\usepackage{epstopdf}
\usepackage{mathtools}
\usepackage{ragged2e}
\usepackage{lipsum}
\usepackage{multirow}
\usepackage{lmodern}
\usepackage{array}
\usepackage{caption}
\usepackage{subcaption}
\newcounter{rowno}
\newcommand{\nc}{\newcommand}
\nc{\beq}{\begin{equation}} \nc{\eeq}{\end{equation}}
\nc{\bea}{\begin{eqnarray}} \nc{\eea}{\end{eqnarray}}
\nc{\baa}{\begin{array}}   \nc{\eaa}{\end{array}}
\nc{\bit}{\begin{itemize}}  \nc{\eit}{\end{itemize}}
\nc{\ben}{\begin{enumerate}} \nc{\een}{\end{enumerate}}
\nc{\bce}{\begin{center}}  \nc{\ece}{\end{center}}
\nc{\bpm}{\begin{pmatrix}}  \nc{\epm}{\end{pmatrix}}
\nc{\bvt}{\begin{verbatim}} \nc{\evt}{\end{verbatim}}

\nc{\non}{\nonumber}
\nc{\hsp}{\hspace{0.5cm}}
\nc{\lsp}{\hspace{1cm}}
\nc{\Lsp}{\hspace{2cm}}
\nc{\LLsp}{\lsp\lsp}
\nc{\lra}{\longrightarrow}
\nc{\vs}{v_s}
\nc{\hc}{\text{H.c.}}

\def\lsim{\mathrel{\raise.3ex\hbox{$<$\kern-.75em\lower1ex\hbox{$\sim$}}}}
\def\gsim{\mathrel{\raise.3ex\hbox{$>$\kern-.75em\lower1ex\hbox{$\sim$}}}}

\usepackage{jheppub} 
\title{\boldmath Pseudo-Goldstone Dark Matter Model \\ with CP violation}

\author[a,b]{Neda Darvishi,}
\author[c]{Bohdan Grzadkowski}

\affiliation[a]{Department of Physics and Astronomy, Michigan State University, East Lansing,
MI 48824,USA}
\affiliation[b]{Institute of Theoretical Physics, Chinese Academy of Sciences, Beijing 100190, China}
\affiliation[c]{Faculty of Physics, University of Warsaw, Pasteura 5, 02-093 Warsaw, Poland}

\emailAdd{neda.darvishi@itp.ac.cn}
\emailAdd{bohdan.grzadkowski@fuw.edu.pl}

\abstract{
We consider an explicitly CP-violating model with two Higgs doublets and one complex singlet scalar. 
The singlet $S$ is charged under a global $\rm U(1)$ symmetry which is softly broken by 
a mass term $\mu^2 S^2+\hc$. Imaginary part of $S$ is a stable dark matter candidate which at the tree level, 
in the limit of zero momentum transfer, decouples from nucleons naturally satisfying all existing direct 
detection limits on dark matter scattering cross-section.
It is explicitly shown that within this framework in the alignment limit CP-violation is still present 
in contrast to a popular version of a 2-Higgs doublet model with softly broken $Z_2$ symmetry.
In this context, we investigate dark matter implications of the model both with and without CP violation in the scalar sector.
In particular, dark matter relic abundance is calculated and the possibility for its indirect detection is discussed.
}

\begin{document}
\maketitle
\flushbottom

%%%%%%%%%%%%%%%%%%%%%%%%%%%%%%%%%%%%%%%%%%%%%%%%%%%%%%%%%%%%%%%%%%%%%%%%%%%%%%%%%%%%%%%%%%%%%%%%%%%%%%
\section{Introduction}
Despite the great success of the Standard Model (SM) in describing fundamental interactions of elementary particles, it fails to explain several cosmological observations, such as the baryon asymmetry of the Universe (BAU) or the origin of dark matter (DM). Many of theories beyond the SM (BSM) that have been put forward to solve these problems utilize an extended Higgs sector. 
However, couplings of Standard Model (SM) like Higgs boson in such theories must be very close to the one measured at the LHC~\cite{ATLAS:2016neq,Palmer:2021gmo,ATLAS:2021upq}. This requirement, referred to as the SM Alignment Limit (AL), severely restricts parameter space of possible scalar-sector extensions of the SM~\cite{Ginzburg:1999fb,Chankowski:2000an,Carena:2013ooa, BhupalDev:2014bir, Bernon:2015qea,Darvishi:2019ltl,Darvishi:2021txa}.

In this work we intend to discuss a model that possesses a natural candidate for DM and, at the same time, contains an extra source of CP violation (CPV). A model of this type would address the two obstacles of the SM, namely the presence of DM and the observation of BAU. One of the simplest extensions of the SM that can provide extra sources of CPV is the well known Two-Higgs-Doublet model (2HDM)~\cite{Lee:1973iz}. However, in the 2HDM with softly broken $Z_2$ symmetry, the observation of the SM-like Higgs boson at the LHC implies vanishing CPV~\cite{Grzadkowski:2014ada}. Thus, to maintain CPV (needed to explain BAU) one must abandon the $Z_2$ and allow for the \textit{generic} 2HDM.
The new CP-violating effects would contribute on electric dipole moments (EDMs), the requirement of keeping EDMs below the experimental bounds~\cite{Andreev:2018ayy} is an important constraint of the parameter space of the model.

Furthermore, there have been many experiments searching for DM particles either through direct or indirect detection~\cite{Bertone:2004pz,Feng:2010gw}. The recent XENON1T experiment~\cite{XENON:2018voc} placed a very stringent upper bound on the DM-nucleon scattering cross-section that challenges many BSM models. However, there exists a simple and natural scenario that avoids the limit, namely the pseudo-Goldstone DM (pGDM). In that case, in the limit of zero momentum transfer, the tree-level DM-nucleon scattering cross-section vanishes, so the limits are easily satisfied. A prototype of the model discussed here has been introduced in \cite{Gross:2017dan} and discussed in detail in \cite{Azevedo:2018oxv} and \cite{Azevedo:2018exj}. 
There a complex scalar singlet has been just added to the SM and the cancellation of the tree-level DM-nucleon scattering cross-section was proven. In this paper, we will generalize this concept by adding the singlet to the generic 2HDM. It will be shown that even in this context the cancellation still holds together with CPV present in the scalar sector. 
In this class of models the leading contribution to DM scattering off nucleons emerges, in a zero-momentum transfer limit, as a finite 1-loop effect~\cite{Azevedo:2018exj}.
From the 2HDM perspective the model discussed here is a generalization of the $Z_2$-symmetric 2HDM supplemented by a real singlet scalar introduced in~\cite{Grzadkowski:2009iz}. 

We will discuss experimental constraints on CPV from the electron EDM and then investigate predictions of the model for indirect detection experiments (as constraints from direct ones are naturally satisfied). 
For these searches, we adopt three benchmark scenarios based on both CP-violating and -conserving models and test our predictions against the current constraint on DM relic density from the Planck data release~\cite{Planck:2018vyg} and the Fermi-LAT experiment~\cite{Fermi-LAT:2015att}. 
Another important requirement that we impose is the possibility of the AL so that there exists a SM-like Higgs boson. We show that indeed the AL could be realized in the model under consideration. 
%We will also calculate the relic abundance of DM and find regions of the model parameter space that are consistent with all the constraints. 
The novel and attractive feature of this model is the simultaneous existence of pGDM and CPV within the AL while fulfilling all experimental limits for the electron's EDM, DM relic abundance observation, and indirect detection.

The layout of this paper is as follows. In Section~\ref{model}, we discuss the basic features of the general 2HDM with a complex singlet scalar (2HDMCS). Thereafter, we re-derive the fermionic couplings pertinent to the general Yukawa interaction. 
In Section~\ref{sec:scat}, we show that the pGDM candidate in this model has a vanishing tree-level scattering amplitude off nucleons at zero momentum transfer. Afterwards, the required constraints for the AL and its consequences for couplings are derived in Section~\ref{sec:al}. 
In Section~\ref{CPI} and Section~\ref{sec:EDM}, we discuss weak-basis-invariant CPV quantities and electron’s EDM, respectively. 
In Section~\ref{sec:Results} we express the parameters of the model in terms of physically observable quantities. Furthermore, we depict constraints on CPV invariants and the electron’s EDM and thus find limits on the allowed CPV. In the same section, we present three benchmark points and provide detailed analyzes for DM relic abundance observation, and indirect detection within the AL. Section~\ref{conc} summarizes our results and discusses the new phenomenological aspects of this model. Finally, technical details are presented in Appendices~\ref{sec:YI},~\ref{app:couplings},~\ref{app:Ji},~\ref{app:eEDM} and~\ref{app:int}.

%%%%%%%%%%%%%%%%%%%%%%%%%%%%%%%%%%%%%%%%%%%%%%%%%%%%%%%%%%%%%%%%%%%%%%%%%%%%%%%%%%%%%%%%%%%%%%%%%%%%%%
\section{General 2HDM with a Singlet Scalar}\label{model}

Here, we consider an extension of the SM by a scalar doublet $\Phi_2$ and a complex singlet scalar $S$, where the singlet under a U(1) symmetry transforms as
\beq
S \hsp \stackrel{{\rm U(1)}}{\longrightarrow} \hsp e^{i\alpha}S\,,
\label{Uone}
\eeq
and other fields are neutral. We will assume that the \text{U(1)} symmetry is softly broken to a $Z_2$ symmetry 
by a mass term $\mu^2$, thereby $S \to -S$ remains as a symmetry.

The scalar potential may be written as follows
\begin{align}
\mathcal{V} &= -\frac12 \left[ m_{11}^2 |\Phi_1|^2 +m_{22}^2 |\Phi_2|^2 + \left(m_{12}^2 \Phi_1^\dagger \Phi_2+ {\rm H.c.} \right) \right] + {\lambda_1 \over 2} |\Phi_1|^4 +{\lambda_2 \over 2} |\Phi_2|^4 \label{pot} \\
	 &+ \lambda_3 |\Phi_1|^2 |\Phi_2|^2+\lambda_4 |\Phi_1^\dagger \Phi_2|^2 + \left[\frac12 \lambda_5(\Phi_1^\dagger \Phi_2)^2+ \lambda_6 (\Phi_1^\dagger \Phi_2)|\Phi_1|^2+ \lambda_7 (\Phi_1^\dagger \Phi_2)|\Phi_2|^2+ {\rm H.c.} \right]
\nonumber \\
&- \mu_s^2 |S|^2 + {\lambda_s \over 2} |S|^4 + (\mu^2 S^2 + {\rm H.c.}) +
|S|^2 \left[\kappa_1 |\Phi_1|^2 + \kappa_2 |\Phi_2|^2 + \left({\kappa_3 \over 2} \Phi_1^\dagger \Phi_2 + {\rm H.c.}\right) \right]. 
\nonumber
\end{align}
The above potential contains the following complex parameters: $m_{12}^2$, $\lambda_5$, $\lambda_6$, $\lambda_7$, $\mu^2$, and $\kappa_3$.

We expand scalar fields around vacuum by 
$$ 
\Phi_1 = \begin{pmatrix} \phi_1^+ \\ {v_{1} + \eta_1 + i\chi_1 \over \sqrt{2}} \end{pmatrix}, \qquad
\Phi_2 = \begin{pmatrix} \phi_2^+ \\ {v_{2} + \eta_2 + i\chi_2 \over \sqrt{2}} \end{pmatrix}, \qquad
S = {v_s + s + iA \over \sqrt{2}}\,,
$$
where $v_{1,2}$ and $\vs$ are in general complex. However, it turns out to be convenient to choose 
a weak basis such that $\kappa_3=0$ (by unitary rotation in the $(\phi_1,\phi_2)$ space), ${\rm Im \,}{\mu^2}=0$ (by rephasing of $S$), 
and ${\rm Im \,}{v_{1,2}}=0$ (by rephasings of $\phi_{1,2}$), we adopt this basis hereafter.
Then $\vs$ can still be complex, however as it is easily seen, the minimization with respect to the phase of $\vs$ forces $\vs$ to be real, therefore 
from now on $\vs$ is assumed to be real. Phases remain in $m_{12}^2$, $\lambda_5$, $\lambda_6$, and $\lambda_7$
therefore, in this basis, it is clear that the $S$-sector does not introduce any extra sources of CPV as compared to the generic 2HDM. 
There is a comment here in order. The AL of 2HDM is defined by the condition that the $125$ GeV Higgs boson couples to the SM fields with exactly SM strength. Needless to say that the AL is a reasonable approximation to experimental constraints on the SM. As it is well known~\cite{Grzadkowski:2014ada} in the AL of the 2HDM with softly broken $Z_2$ symmetry CPV disappears. One could construct a $Z_2$-symmetric (softly broken) 
version of the model with the
$S$ singlet. However then CPV would disappear in the AL. This is why in this work we are discussing the generic 2HDM coupled to the $S$ singlet.

The minimization conditions resulting from the potential give rise to the following relations:
\begin{align}
m_{11}^2&=v_{1}^2 \lambda_1 + v_{2}^2 \lambda_{345} +v_s^2 \kappa_1- {v_2 {\rm Re}m_{12}^2 \over v_1} + 
 {3} v_1 v_2 {\rm Re}\,\lambda_{6} + {v_2^3\over v_1} {\rm Re}\,\lambda_{7}, \\
m_{22}^2&= v_{2}^2 \lambda_2 + v_{1}^2 \lambda_{345} +v_s^2 \kappa_2- {v_1{\rm Re}m_{12}^2 \over v_2 }+ 
 {3} v_1 v_2 {\rm Re}\,\lambda_{7} + {v_1^3\over v_2} {\rm Re}\,\lambda_{6}, \\
\mu_s^2&={1\over 2} \big( v_{1}^2 \kappa_1+ v_{2}^2 \kappa_2+ v_s^2 \lambda_s+4 \mu^2\big),
\label{min-3}
\end{align}
with $\lambda_{345}=\lambda_3+\lambda_4+{\rm Re}\,\lambda_5$
and from imaginary parts 
\begin{align}
{\rm Im}\,m_{12}^2&= v_2 v_1 {\rm Im}\,\lambda_5 + v_1^2 {\rm Im}\,\lambda_6 +v_2^2 {\rm Im}\,\lambda_7\,.
\label{minim}
\end{align}

The 2HDM part of the potential (\ref{pot}) contains 14 real parameters, the $S$-sector has 4 real parameters
and the interaction between $S$ and 2HDM is parametrized by 4 real numbers. Adopting the basis specified above we remove 2 real parameters (i.e. $\kappa_3$) from the interaction part and 1 from the $S$-sector (${\rm Im}\,(\mu^2)$). Therefore, we can conclude that there are 19 real parameters. 
However, yet another element of choosing the weak basis is the condition that $v_1$ and $v_2$ are relatively real, implying the minimization condition \eqref{minim} reduces the number of independent parameters by 1. Therefore, the number of real parameters is 18\footnote{Note that the minimization of $\mathcal{V}$ with respect to the phase of $S$, that implies the reality of $\vs$, does not reduce the number of parameters.}.

Using the minimization conditions one can obtain the following value of the vacuum energy density
\begin{align}
E_{vac}=&V(\langle \Phi_1 \rangle, \langle \Phi_2 \rangle,\langle S \rangle)=- {
 v_1^4 \lambda_1\over 8} - {v_2^4 \lambda_2\over 8} - {v_3^4 \lambda_s\over 8}-{v_1^2 v_s^2 \kappa_1\over 4} - {v_2^2 v_s^2 \kappa_2\over 4}
\nonumber \\
& - {v_1^2 v_2^2 \lambda_{345}\over4} - {v_1^3 v_2 {\rm Re}\,\lambda_{6}\over 2}- 
 {v_1 v_2^3{\rm Re}\,\lambda_{7}\over 2}.
\end{align}

To determine the mass spectrum, we first diagonalize the mass matrix for charged scalars defining physical $H^\pm$ boson and the would be Goldstone 
boson $G^\pm$):
\beq
 G^{\pm} = c_\beta \phi_1^{\pm} + s_\beta \phi_2^{\pm}\,, \qquad H^{\pm} = -s_\beta\phi_1^{\pm} + {c_\beta} \phi_2^{\pm}\;,
\eeq
for $\tan\beta\equiv v_2/v_1$. Employing the same rotation matrix we define $\eta_3$ and a neutral Goldstone boson $G^0$:
\beq   \\ 
 G^0 = c_\beta \chi_1 + s_\beta \chi_2\;, \qquad \eta_3 = -s_\beta \chi_1 + c_\beta\chi_2\;.
\eeq
The charged-Higgs-boson mass is then given as
 \begin{eqnarray}
M_{H^{\pm}}^2 \ = - {v^2 \over 2}\Big[ \big(\lambda_4 + {\rm Re}\,\lambda_5\big ) + {-{\rm Re}\,m_{12}^2 + v_1^2 {\rm Re}\,\lambda_6 + v_2^2 {\rm Re}\,\lambda_7 \over v_1 v_2}\Big].
\nonumber
\end{eqnarray}

The remaining physical neutral states of definite mass ($H_{1},H_2,H_3,H_4$) can be obtained by an orthogonal transformation upon $\eta_{1,2,3}$, $s$, as follows
\begin{align}
\begin{pmatrix}
H_1 \\
H_2 \\
H_3 \\
H_4
\end{pmatrix}
=
R
\begin{pmatrix}
\eta_1 \\
\eta_2 \\
\eta_3\\
s
\end{pmatrix},
\end{align}
where the orthogonal $4\times 4$ matrix $R$ with six degrees of freedom can be given explicitly by
\begin{align}
R= R_6 R_5 R_4 R_3 R_2 R_1, \label{mix}
\end{align}
with
\begin{align}
R_1&= \left(
\begin{array}{ccccc}
\cos\alpha_1& \sin\alpha_1&0& 0\\ 
-\sin\alpha_1&\cos\alpha_1& 0&0\\
0&0& 1&0 \\
0& 0&0&1
 \end{array}
 \right),
 \quad
R_2 = \left(
\begin{array}{ccccc} 
\cos\alpha_2& 0& \sin\alpha_2& 0\\
0&1 & 0 & 0\\
-\sin\alpha_2 & 0&\cos\alpha_2 & 0\\
0 & 0 & 0&1 \end{array}
 \right),
 \nonumber 
 \end{align}
 \begin{align}
R_3 &= \left(
\begin{array}{ccccc} 
 \cos\alpha_3 & 0 & 0& \sin\alpha_3\\
 0&1 & 0 & 0\\
 0 & 0&1&0\\
 -\sin\alpha_3 & 0 & 0& \cos\alpha_3 
\end{array}
 \right),
 \quad
R_4 = \left(
\begin{array}{ccccc} 
1 & 0 & 0 & 0\\
0&\cos\alpha_4& \sin\alpha_4 & 0\\
0&-\sin\alpha_4& \cos\alpha_4 & 0\\
0 & 0 & 0&1 
 \end{array}
 \right),
 \nonumber 
\end{align}
\begin{align}
R_5 &= \left(
\begin{array}{ccccc} 
1 & 0 & 0 & 0\\
0& \cos\alpha_5 & 0& \sin\alpha_5\\
0 & 0& 1& 0\\
0& -\sin\alpha_5 & 0&\cos\alpha_5 
 \end{array}
 \right),
 \quad
R_6 = \left(
\begin{array}{ccccc} 
1 & 0 & 0 & 0\\
0& 1 & 0 & 0\\
0 & 0& \cos\alpha_6& \sin\alpha_6\\
0 & 0& -\sin\alpha_6& \cos\alpha_6 
 \end{array}
 \right),
\nonumber
\end{align}
where the mixing angles run over interval 0 to $\pi$.

The mass matrix ${M}^2$ for neutral states can be written down as follows
\begin{align}
\footnotesize
{M}^2=\left(
\begin{array}{ccccc}
M_{11}^2 & M_{12}^2 & M_{13}^2 & M_{14}^2 & 0 
\\
M_{12}^2 & M_{22}^2 & M_{23}^2 & M_{24}^2 & 0 
\\
M_{13}^2 & M_{23}^2 & M_{33}^2 & 0 & 0 
\\
M_{14}^2 & M_{24}^2 & 0 & M_{44}^2 & 0 
\\
0 & 0 & 0 & 0 & -4 \mu^2 
\end{array}
\right),
\label{mass_mat}
\end{align}
where 
{\allowdisplaybreaks
\begin{subequations} 
 \label{mfc}
\begin{align}
\footnotesize
M_{11}^2 &= \lambda_{1} v_{1}^2+\frac{v_{2}{\rm Re}~m_{12}^2+3 v_{1}^2 v_{2} {\rm Re}\,\lambda_{6}-v_{2}^3 {\rm Re}\lambda_7}{2 v_{1}},
 \\
M_{22}^2 &= \lambda_{2} v_{2}^2+ \frac{v_{1}{\rm Re}~m_{12}^2-v_{1}^3 {\rm Re}\,\lambda_{6}+3 v_{1}
 v_{2}^2 {\rm Re}\,\lambda_{7}}{2
 v_{2}},
\\
M_{33}^2 &=-v^2 {\rm Re}\,\lambda_{5}+\frac{v^2 \left({\rm Re}~m_{12}^2-v_{1}^2 {\rm Re}\,\lambda_{6}-v_{2}^2 {\rm Re}\,\lambda_{7}\right)}{2 v_{1} v_{2}},
\\
M_{44}^2 &= \lambda_{s} v_{s}^2,
\\
M_{12}^2 &= v_{1} v_{2}\lambda_{345}
+\frac{1}{2} \left(-{\rm Re}~m_{12}^2+3 v_{1}^2 {\rm Re}\,\lambda_{6}+3 v_{2}^2 {\rm Re}\,\lambda_{7}\right),
\\
M_{13}^2 &= -\frac{v}{2} (2 v_{1} {\rm Im}\,\lambda_{6}+v_{2} {\rm Im}~{\lambda_5}),
\\ 
M_{14}^2 &= \kappa_{1} v_{1} v_{s},
\\
M_{23}^2 &= -\frac{v}{2} (v_{1} {\rm Im}\,\lambda_{5}+2 v_{2} {\rm Im}\,\lambda_{7}),
\\
M_{24}^2 &=\kappa_{2} v_{2} v_{s}.
\end{align}
\end{subequations} 
}
Thus, by the following diagonalization, the physical masses squared can be obtained
\begin{align}
\mathcal{M}^2 = 
R
{M}^2
R^{\sf T},
\label{mass}
\end{align}
and the mass squared of the DM candidate is $M_A^2 = -4 \mu^2$.

Thus, the mass relations for all neutral Higgs bosons may be expressed in the following form
\begin{align}
M_{H_i}^2=&\lambda_1 R_{i1}^2 v_1^2 + \lambda_2 R_{i2}^2 v_2^2+ 2 (\lambda_3+\lambda_4) R_{i1} R_{i2} v_1 v_2 + \lambda_s R_{i4}^2 v_s^2 
\nonumber \\
 &+ 2 \kappa_1 R_{i1} R_{i4} v_1 v_s + 2 \kappa_2 R_{i2} R_{14} v_2 v_s - {\rm Re}\,\lambda_5 (R_{i3}^2 v^2 -2 R_{i1} R_{i2} v_1 v_2 ) 
 \nonumber \\
 &+ {{\rm Re}\,m_{12}^2 \over 2 v_1 v_2}(R_{i3}^2 v^2 + (R_{i2} v_1 - R_{i1} v_2)^2)
\nonumber \\
& - {{\rm Re}\,\lambda_6 v_1\over 2v_2} (R_{i3}^2 v^2 + R_{i2}^2 v_1^2 - 6 R_{i1} R_{i2} v_1 v_2 - 3 R_{i1}^2 v_2^2)
\nonumber \\
& - {{\rm Re}\,\lambda_7 v_2\over 2v_1}(R_{i3}^2 v^2 + R_{i2}^2 v_2^2 - 6 R_{i1} R_{i2} v_1 v_2 - 3 R_{i1}^2 v_1^2)
\nonumber \\
 & - R_{i3} v \big[ {\rm Im}\,\lambda_5 ( R_{i2} v_1 + R_{i1} v_2 )+ 2{\rm Im}\,\lambda_6 R_{i1} v_1 + 2{\rm Im}\,\lambda_7 R_{i2} v_2\big ].
\end{align}
Note that, in the limit of CP conservation, the elements $R_{i 3}$ and $R_{3j}$ vanish for $i,j\neq3$.

%%%%%%%%%%%%%%%%%%%%%%%%%%%%%%%%%%%%%%%%%%%%%%%%%%%%%%%%%%%%%%%%%%%%%%%%%%%%%%%%
\section{The DM-Nucleon Scattering Amplitude}\label{sec:scat}

Here, we are going to investigate the tree-level amplitude for DM scattering off nucleons at zero momentum transfer for the case of CP-violating generic scalar potential and Yukawa couplings for neutral Higgs bosons.
Since DM-nucleon scattering is induced by DM-quark scattering, we just need to calculate the DM-quark scattering amplitude in the zero momentum transfer limit. 

From the potential, the trilinear couplings involving the DM candidate $A$ are of the following form,
\begin{align}
\mathcal{L}_{{\rm tri},A^2} = {A^2 \over 4} 
	\big( 2\kappa_1 v_1 \eta_1 + 2\kappa_2 v_2\eta_2 +2\lambda_s v_s s \big)
		= {1 \over 2} \sum_{i=1,2,3,4} g_{H_i A^2} H_i A^2,
\end{align}
where the couplings $g_{H_i A^2}$ may be expressed in the following form
\begin{align}
&\begin{pmatrix}
g_{H_1 A^2} \\ g_{H_2 A^2}\\ g_{H_3 A^2}\\ g_{H_4 A^2}
\end{pmatrix}\,= R \,
\begin{pmatrix}
\kappa_1 v_1 \\ \kappa_2 v_2 \\ 0 \\ \lambda_s v_s
\end{pmatrix}.
\label{ghh}
\end{align}

We define the Yukawa couplings $g_{H_j \bar{f_k} f_k}$ as follows:
\begin{equation}
\mathcal{L}^{H_j}_Y = H_j \bar{f}_k g_{H_j \bar{f_k} f_k} f_k,
\label{hqq}
\end{equation}
where $f=u,d,l$, $k=1,2,3$ and we have already assumed no FCNC as explained in Appendix \ref{sec:YI}. 
It will be useful to define $c^{(j)}_{f_k}+i\gamma_5 \tilde{c}^{(j)}_{f_k}$ ($j=1,2,3$) by rewriting $g_{H_j \bar{f_k} f_k}$ in a way analogous to (\ref{ghh}), see also \eqref{eta1} and \eqref{eta2}:
\begin{align}
\begin{pmatrix}
g_{H_1 \bar{f_k} f_k} \\ g_{H_2 \bar{f_k} f_k} \\ g_{H_3 \bar{f_k} f_k} \\ g_{H_4 \bar{f_k} f_k} 
\end{pmatrix} \equiv R
\begin{pmatrix}
c^{(1)}_{f_k}+i\gamma_5 \tilde{c}^{(1)}_{f_k} \\ c^{(2)}_{f_k}+i\gamma_5 \tilde{c}^{(2)}_{f_k}\\ c^{(3)}_{f_k}+i\gamma_5 \tilde{c}^{(3)}_{f_k}\\ 0
\end{pmatrix} = R
\begin{pmatrix}
\big[{-m_{f_k} v_1\over v^2}+{v_2 \over \sqrt{2}v}{\rm Re \,}\rho^f_{kk}\big]+i\gamma_5 \big[{v_2 \over \sqrt{2}v}{\rm Im \,}\rho^f_{kk}\big] \\ \big[{-m_{f_k} v_2\over v^2}-{v_1 \over \sqrt{2}v}{\rm Re \,}\rho^f_{kk}\big]-i\gamma_5 \big[{v_1 \over \sqrt{2}v}{\rm Im \,}\rho^f_{kk}\big]\\ \big[{-1 \over \sqrt{2}}{\rm Im \,}\rho^f_{kk}\big]+i\gamma_5 \big[{1 \over \sqrt{2}}{\rm Re \,}\rho^f_{kk}\big]\\ 0
\end{pmatrix},
\label{ghu}
\end{align}
where $\rho$ is defined in (\ref{eta1}-\ref{eta2}).
Therefore, we can write down the DM-quark scattering amplitude in terms of the above trilinear couplings as
\begin{align}
i\mathcal{A} = {m_{f_k}\over 2 v} \bar{f_k}(p_2) & 
\Big(
g_{H_1 A^2} {i \over p^2-M_{H_1}^2} g_{H_1 f_k \bar{f_k}}
+ g_{H_2 A^2} {i \over p^2-M_{H_2}^2} g_{H_2 f_k \bar{f_k}}
\nonumber \\ &
+ g_{H_3 A^2} {i \over p^2-M_{H_3}^2} g_{H_3 f_k \bar{f_k}}
+ g_{H_4 A^2} {i \over p^2-M_{H_4}^2} g_{H_4 f_k \bar{f_k}}
\Big)f_k(p_1),
\end{align}
where $f_k(p_1)$ and $\bar{f_k}(p_2)$ are the wave functions for the
incoming and outgoing quarks, respectively. 
In the zero momentum transfer limit, $p^2 \to 0$, $\big( \mathcal{M}^2 \big)^{-1} = 
R\big({M}^2_{} \big)^{-1}R^{\sf T}$, therefore the above amplitude can be re-expressed as
\begin{align}
 i\mathcal{A} = & -i {m_{f_k}\over 2 v} \bar{f_k}(p_2) \begin{pmatrix}
\kappa_1 v_1, & \kappa_2 v_2, &0, & \lambda_s v_s 
\end{pmatrix} 
\big( {M}^2\big)^{-1}
\begin{pmatrix}
c^{(1)}_{f_k}+i\gamma_5 \tilde{c}^{(1)}_{f_k} \\ c^{(2)}_{f_k}+i\gamma_5 \tilde{c}^{(2)}_{f_k}\\ c^{(3)}_{f_k}+i\gamma_5 \tilde{c}^{(3)}_{f_k}\\ 0
\end{pmatrix}f_k(p_1).
\end{align}
Thus, adopting Eqs.~\eqref{mass_mat} and~\eqref{ghu} we find 
\begin{align}
\begin{pmatrix}
\kappa_1 v_1, & \kappa_2 v_2, &0, & \lambda_s v_s 
\end{pmatrix} 
\big( {M_{}}^2\big)^{-1}
\begin{pmatrix}
c^{(1)}_{f_k}+i\gamma_5 \tilde{c}^{(1)}_{f_k} \\ c^{(2)}_{f_k}+i\gamma_5 \tilde{c}^{(2)}_{f_k}\\ c^{(3)}_{f_k}+i\gamma_5 \tilde{c}^{(3)}_{f_k}\\ 0
\end{pmatrix}= 0.
\end{align}
Therefore, we have shown that in the general 2HDM with pGDM the tree-level DM-quark amplitude also vanishes at the zero momentum transfer limit. 
%Note that a similar argument stands for the tree-level DM-vector bosons scattering.
%%%%%%%%%%%%%%%%%%%%%%%%%%%%%%%%%%%%%%%%%%%%%%%%%%%%%%%%%%%%%%%%%
\section{The Alignment Limit}\label{sec:al}
The Higgs-vector boson couplings can be expressed as follows
\begin{align}
e_j &= (v_1 R_{j1} + {v_2}\; R_{j2}),
\label{ei}
\end{align}
with $j=1,2,3,4$. By virtue of unitarity of $R$, $e_j$'s respects the following sum rule
\begin{align}
e_1^2+e_2^2+e_3^2+e_4^2=v^2.
\end{align}
 
Now, we specify the AL requiring that the coupling between $H_1$ and gauge vector bosons that can be written explicitly as 
\begin{align}
e_1 =&v \cos\alpha_2 \cos\alpha_3 \cos(\beta-\alpha_1),
\label{e1}
\end{align}
is of the SM strength i.e. $e_1=v$. In order to satisfy (\ref{e1}) the mixing angles 
must be fixed as
\begin{equation}
\alpha_1 = \beta, \quad \alpha_2=\alpha_3=0.
\end{equation}
Thus, within the AL the general unitary matrix $R$ with six parameters reduces to the $R_{\rm AL}$ containing four parameters with an extra condition $\alpha_1 = \beta$. Thereby, $R_{\rm AL}$ takes the form
\begin{align}
R_{\rm AL}=R_6 R_5 R_4 R_1,
\end{align}
with $R_{i}$'s are specified in (\ref{mix}) with $\alpha_1=\beta$.

Note that, the elements $({R}_{\rm AL})_{13}$ and $({R}_{\rm AL})_{14}$ are zero. Therefore, within the AL the Higgs boson couplings to fermions read
{\allowdisplaybreaks
\begin{subequations}
\begin{align}
g_{H_1\bar{f}_k f_k} \; &= \; -{m_{f_k} \over v},
 \\
g_{H_2\bar{f}_k f_k} \; &= \; -{\cos\alpha_5 \over \sqrt{2}} \Big[\big({\rm Re \,}\rho^f_{kk}\cos\alpha_4+{\rm Im \,}\rho^f_{kk}\sin\alpha_4\big)
\nonumber \\
 \; & \,\,\quad -i \gamma_5 \big({\rm Re \,}\rho^f_{kk}\sin\alpha_4-{\rm Im \,}\rho^f_{kk}\cos\alpha_4\big)\Big],
 \\
g_{H_3\bar{f}_k f_k} \; &= \; {1\over \sqrt{2}}\Big[ \big({\rm Re \,}\rho^f_{kk}\sin\alpha_4-{\rm Im \,}\rho^f_{kk}\cos\alpha_4\big) {\cos\alpha_6} 
\nonumber \\ 
 \; & \,\,\quad
 + \big({\rm Re \,}\rho^f_{kk}\cos\alpha_4+{\rm Im \,}\rho^f_{kk}\sin\alpha_4\big){\sin\alpha_5\sin\alpha_6} 
\non \\
 \; & \,\,\quad
+i \gamma_5\Big( \big({\rm Re \,}\rho^f_{kk}\cos\alpha_4+{\rm Im \,}\rho^f_{kk}\sin\alpha_4\big) \cos\alpha_6 
\nonumber \\ 
 \; & \,\,\quad
 -\big({\rm Re \,}\rho^f_{kk}\sin\alpha_4-{\rm Im \,}\rho^f_{kk}\cos\alpha_4\big)\sin\alpha_5\sin\alpha_6 \Big)\Big],
 \\
 g_{H_4\bar{f}_k f_k} \; &= \; {1\over \sqrt{2}}\Big[ \big({\rm Re \,}\rho^f_{kk}\cos\alpha_4+{\rm Im \,}\rho^f_{kk}\sin\alpha_4\big){\sin\alpha_5 \cos\alpha_6}
\nonumber \\
 \; & \,\,\quad
 - \big({\rm Re \,}\rho^f_{kk}\sin\alpha_4-{\rm Im \,}\rho^f_{kk}\cos\alpha_4\big){\sin\alpha_6}
\non \\
 \; & \,\,\quad
 - i \gamma_5 \Big( \big({\rm Re \,}\rho^f_{kk}\cos\alpha_4+{\rm Im \,}\rho^f_{kk}\sin\alpha_4\big) \sin\alpha_6 
\nonumber \\
 \; & \,\,\quad
 + \big({\rm Re \,}\rho^f_{kk}\sin\alpha_4-{\rm Im \,}\rho^f_{kk}\cos\alpha_4\big) \sin\alpha_5 \cos\alpha_6 \Big) \Big].
\end{align}
\label{AL-hqq}
\end{subequations}
}

The couplings $g_{H_j A^2}$ within the AL may be re-written as follows
\begin{subequations} 
\begin{align}
g_{H_1 A^2} \,&=\,
{1\over 2} v \big(\kappa_1 +\kappa_2+ (\kappa_1 -\kappa_2) \cos 2\beta \big),
 \\
g_{H_2 A^2} \,&=\,
 {1\over 2} 
 v (-\kappa_1 + \kappa_2) \cos\alpha_4 \cos\alpha_5 \sin 2\beta+v_s \lambda_s \sin\alpha_5,
 \\
g_{H_3 A^2} \,&=\,
{1\over 2} v (\kappa_1 - \kappa_2) (\cos\alpha_6 \sin\alpha_4 + \cos\alpha_4 \sin\alpha_5 \sin\alpha_6) \sin 2\beta+v_s \lambda_s \cos\alpha_5 \sin\alpha_6,
 \\
g_{H_4A^2} \,&=\,
{1\over 2} v (\kappa_1 - \kappa_2) (\cos\alpha_4 \cos\alpha_6 \sin\alpha_5 - \sin\alpha_4 \sin\alpha_6) \sin 2\beta+v_s \lambda_s \cos\alpha_5 \cos\alpha_6.
\end{align}
\label{ghh-al}
\end{subequations}
%%%%%%%%%%%%%%%%%%%%%%%%%%%%%%%%%%%%%%%%%%%%%%%%%%%%%%%%%%
\section{CP Invariants}\label{CPI}
The scalar Higgs potential (\ref{pot}) for this model can be re-written in the following compact form 
\begin{align}
V \, &= \, -m_{ab}^2 ({\Phi_{\bar{a}}^\dagger} \Phi_b)+ {1 \over 2} \lambda_{a\bar{b}c\bar{d}} ({\Phi_{\bar{a}}^\dagger} \Phi_b)({\Phi_{\bar{c}}^\dagger} \Phi_d)
\nonumber \\ 
&-\mu_s^2 (S^\star S)+{1 \over 2} \lambda_s (S^\star S)^2+ \kappa_{ab} ({\Phi_{\bar{a}}^\dagger} \Phi_b) (S^\star S)+\mu^2 S^2,
\end{align}
with indices running over $a,\bar{a}, \dots =1,2$.
There is a well known freedom of choosing a basis for the Higgs doublets $(\Phi_1,\Phi_2)^T$:
\beq
\Phi_a \to \Phi_a^\prime = U_{a\bar{b}} \Phi_b,
\label{U2}
\eeq
where $U \in U(2)$. Obviously any physical (observable) prediction of the model can not depend on the basis, i.e. on $U$.

After the $U(2)$ transformation (\ref{U2}), the mass terms and quartic couplings in the new basis read
\bea
m_{a\bar{b}}^2& \to & U_{a\bar{c}} m_{c\bar{d}}^2 U_{d\bar{b}} ^{\dagger},
\\
\lambda_{a\bar{b}c\bar{d}}& \to & U_{a\bar{e}} U_{f\bar{b}} ^{\dagger} U_{c\bar{g}} U_{h\bar{d}} ^{\dagger} \lambda_{e\bar{f}g\bar{h}},
\eea
Following \cite{Lavoura:1994fv}, \cite{Botella:1994cs}, \cite{Davidson:2005cw} and \cite{Gunion:2005ja} we introduce CP-sensitive $U$-invariant (i.e. basis independent) quantities
\begin{eqnarray}
{\rm Im}~J_1&=&-{2 \over v^2} {\rm Im} \, [\hat{v}_{\bar{a}} \, m_{a\bar{b}}^2\, \lambda_{b\bar{d}}^{(1)} \, \hat{v}_{d}],
\label{j1}
 \\
{\rm Im}~J_2&=&{4 \over v^4} {\rm Im} \, [\hat{v}_{\bar{b}} \, \hat{v}_{\bar{c}} \, m_{b\bar{e}}^2 \, m_{c\bar{f}}^2\, \lambda_{e\bar{a}f\bar{d}} \, \hat{v}_{a} \, \hat{v}_{d}],
\label{j2}
 \\
{\rm Im}~J_3&=&{\rm Im} \, [\hat{v}_{\bar{b}} \, \hat{v}_{\bar{c}} \, \lambda_{b\bar{e}}^{(1)} \, \lambda_{c\bar{f}}^{(1)} \, \lambda_{e\bar{a}f\bar{d}} \, \hat{v}_{a} \, \hat{v}_{d}].
\label{j3}
\end{eqnarray}
where $\lambda_{a\bar{d}}^{(1)} = \delta_{b\bar{c}} \, \lambda_{a\bar{b}c\bar{d}}$ and $\hat{v}_{a} = v_a/v$.

In this paper it proves to be convenient to choose 
a weak basis such that $\kappa_3=0$, ${\rm Im \,}{\mu^2}=0$, 
and ${\rm Im \,}{v_{1,2}}=0$.
Then $\vs$ can still be complex however the minimization with respect to the phase of $\vs$ forces $\vs$ to be real. The phase remains in $m_{12}^2$, $\lambda_5$, $\lambda_6$, and $\lambda_7$. However, ${\rm Im}\, m_{12}^2$ could be eliminated using the minimization condition (\ref{minim}). Therefore the ${\rm Im\,}J_i$ invariants may be written in the following compact form
\begin{align}
{\rm Im\,} J_i = \sum^{p+q+r<3}_{p,q,r\geq0} a_{i}^{pqr} ({\rm Im\,} \lambda_5)^p ({\rm Im} \lambda_6\,)^q ({\rm Im\,} \lambda_7)^r.
\end{align}

All the non-zero coefficients $a_{i}^{pqr}$ are shown in Appendix \ref{app:Ji}.
In the simplest version of the model, with vanishing $\lambda_6$, $\lambda_7$, one finds 
that only $q,r=0$ and $p=1$ contribute, thus the invariants take the following simple form
\begin{align}
{\rm Im\,} J_i = a_i^{100} \,{\rm Im\,} \lambda_5,
\end{align}
with the non zero coefficients 
\begin{align}
a_1^{100}&=
\frac{v_1^2 v_2^2}{v^4} (\lambda_{2}-\lambda_{1}),
\\
a_2^{100}&=
\frac{v_1^2 v_2^2}{v^{8}} \Big[2 {\rm Re}\lambda_{5} v_1^2 v_2^2 (\lambda_{2}-\lambda_{1})-v_1^4 \big((\lambda_{1}-\lambda_{3}-\lambda_{4})^2
+|\lambda_5|^2)\big)
\nonumber \\ &
+v_2^4 \big((\lambda_{2}-\lambda_{3}-\lambda_{4})^2-|\lambda_5|^2\big)+2 (\kappa_1 - \kappa_2) (\lambda_3 + \lambda_4) v^2 v_s^2 + (\kappa_1^2- \kappa_2^2) v_s^4\Big],
\\
a_3^{100}&= \frac{v_1^2 v_2^2}{v^4} (\lambda_{1}-\lambda_{2}) (\lambda_1+\lambda_2+2\lambda_4).
\end{align}
For numerical analysis in Sec.~\ref{sec:Results}, we will consider the general model with non-vanishing $\lambda_6$, $\lambda_7$.
We will constrain the invariants using the electron EDM and therefore we will limit the allowed strength of CPV. In the next section, we present the theoretical expressions for the electron EDM in this framework.

%%%%%%%%%%%%%%%%%%%%%%%%%%%%%%%%%%%%%%%%%%%%%%%%%%%%%%%%%%%%%%%%%%%%%%%%%%%%%%%
\section{The Electron Electric Dipole Moment}\label{sec:EDM}
\begin{figure}[h]
\begin{center}
\includegraphics[width=0.6\textwidth]{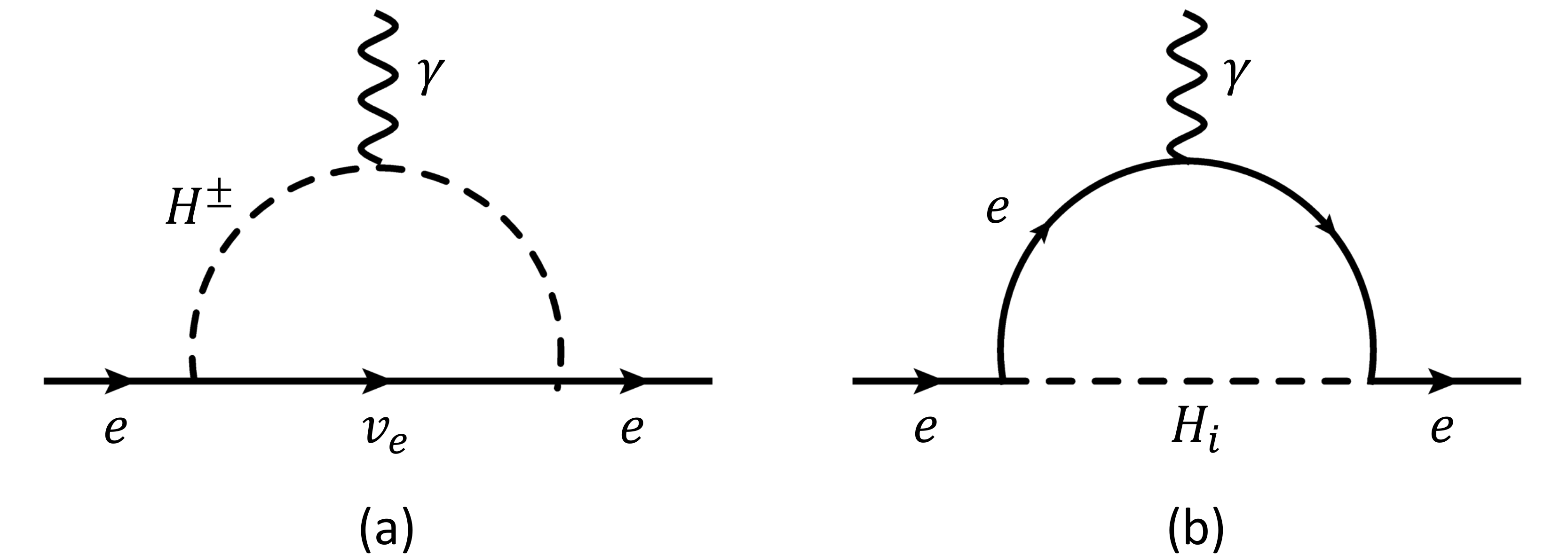}
\end{center}
\caption{The one-loop diagrams that are contributing to the electron EDM are displayed.}
\label{diag-1loopEDM}
\end{figure}

\begin{figure}[h]
\begin{center}
\includegraphics[width=0.75\textwidth]{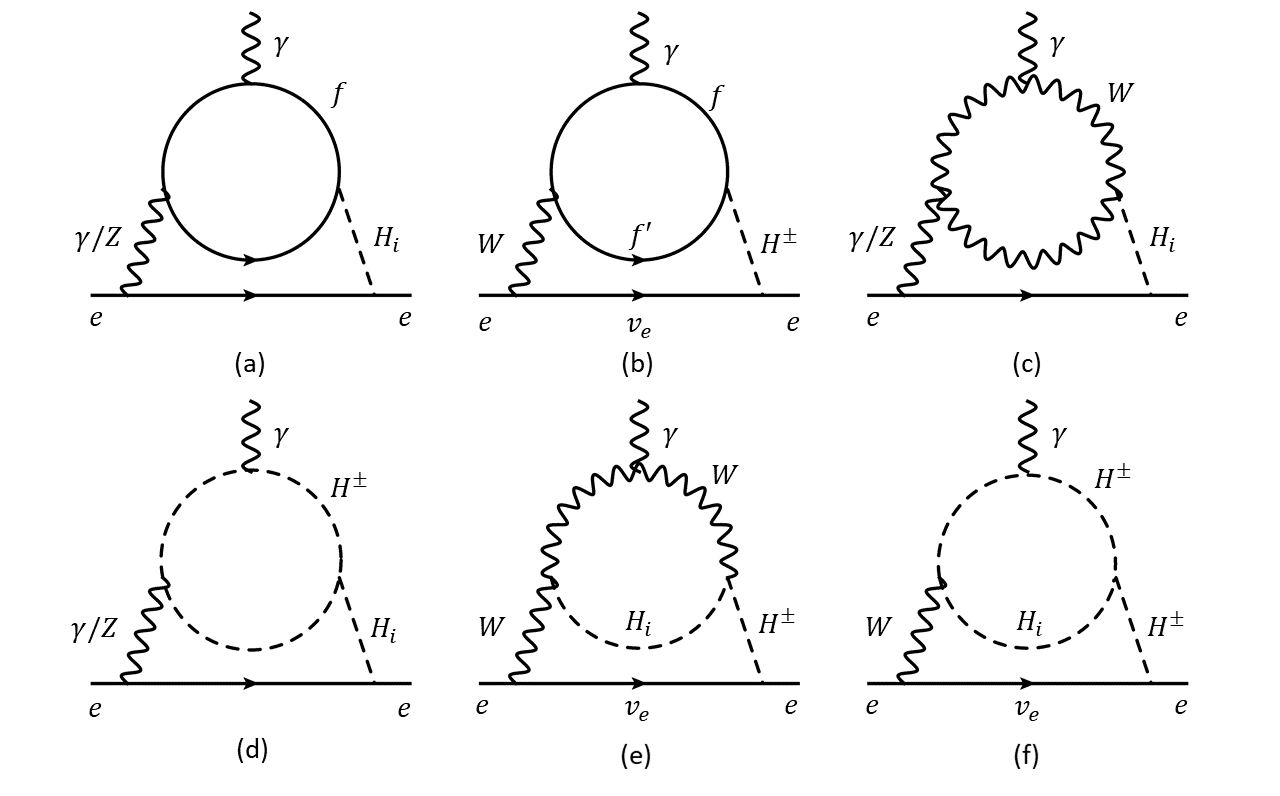}
\end{center}
\caption{The Barr-Zee diagrams that are contributing to the electron EDM are displayed.}
\label{diag-EDM}
\end{figure}

The strength of CPV can be measured by EDM, $d_f$, defined by the following effective interaction
\begin{align}
\mathcal{L}_{EDM}=-{i\over 2} d_f \bar{\psi} \sigma^{\mu \nu} \gamma^5 \psi F_{\mu \nu}.
\end{align}
The current upper bound on the electron EDM at the 90$\%$ confidence level (CL) implies~\cite{Andreev:2018ayy} 
\begin{align}
|d_e|<1.1\times10^{-29}\, e\cdot{\rm cm}.
\label{exp-edm}
\end{align}
It turns out that this measurement places a strong constraint on the parameters of the model. 
As illustrated in Figs.~\ref{diag-1loopEDM} and~\ref{diag-EDM} the electron EDM arises from 1-loop and 2-loop contributions, respectively. Yukawa couplings play an important role in both cases.
Firstly we specify charged Higgs ($H^\pm$) Yukawa couplings
\begin{align}
g_{\bar{u}_m d_k H^+}\; &=\;K_{mk}\Big[c_{q}+i\gamma_5\tilde{c}_{q}\Big]= {1 \over 2} K_{mk}
	\Big[\big[ \big( \rho^u_{mm} \big)^* - \big( \rho^d_{kk} \big)^* \big]- \gamma_5\big[ \big( \rho^u_{mm} \big)^* + \big( \rho^d_{kk} \big)^*\big] \Big],
\non \\
g_{\bar{\nu}_k l_k H^+}\; &=\;\Big[{c_+}+i\gamma_5{\tilde{c}_+}\Big]=- {1 \over 2} \Big[\big( \rho^l_{kk} \big)^* +\gamma_5 \big(\rho^l_{kk} \big)^* \Big],
\non \\
g_{\bar{\nu}_k l_k H^-}\; &=\;\Big[{c_-}-i\gamma_5{\tilde{c}_-}\Big]=- {1 \over 2} \Big[\big( \rho^l_{kk} \big) - \gamma_5 \big(\rho^l_{kk} \big) \Big],
\label{ce}
\end{align}
with $K$ being the CKM-matrix. 

Next we specify neutral-Higgs ($H_i$) Yukawa couplings in a form that will be adopted later. 
We define $c_{i{f}}$ and $\tilde{c}_{i{f}}$ as scalar and pseudoscalar components of the couplings as follows 
\bea
g_{H_i f \bar{f}}\,\equiv\, c_{i{f}}+i\gamma_5\tilde{c}_{i{f}} \,=\, R_{ij}(c^{(j)}_{f_k}+i\gamma_5\tilde{c}^{(j)}_{f_k}),
\label{cie}
\eea 
where $c^{(j)}_{f_k}$ and $\tilde{c}^{(j)}_{f_k}$ are given in \eqref{ghu}.

It is instructive to recall $c^{(j)}_{f_k}$ and $\tilde{c}^{(j)}_{f_k}$ in terms of Yukawa couplings given by \eqref{eta1} and \eqref{eta2}:
\bea
c^{(1)}_{f_k} = &-{m_{f_k} v_1\over v^2}+{v_2 \over \sqrt{2}v}{\rm Re \,}\rho^f_{kk} \lsp & \tilde{c}^{(1)}_{f_k} = {v_2 \over \sqrt{2}v}{\rm Im \,}\rho^f_{kk} \\
c^{(2)}_{f_k} = &-{m_{f_k} v_2\over v^2}-{v_1 \over \sqrt{2}v}{\rm Re \,}\rho^f_{kk} \lsp & \tilde{c}^{(2)}_{f_k} = -{v_1 \over \sqrt{2}v}{\rm Im \,}\rho^f_{kk} \\
c^{(3)}_{f_k} = &-{1 \over \sqrt{2}}{\rm Im \,}\rho^f_{kk} \qquad \quad \lsp & \tilde{c}^{(3)}_{f_k} = {1 \over \sqrt{2}}{\rm Re \,}\rho^f_{kk}
\label{cs}
\eea
and $c^{(4)}_{f_k}=\tilde{c}^{(4)}_{f_k}=0$. Note that Yukawa couplings for light fermions are not constrained experimentally, therefore, since $R_{ij} \sim 1$,
one finds that in the generic 2HDM $c_{if}$ and $\tilde{c}_{if}$ are not necessarily small, in contrast to the case of the SM or versions of 2HDM restricted by a $Z_2$ symmetry.

The 1-loop contributions to the electron EDM, may arise from charged and neutral Higgs boson loops, as have been shown in Fig.~\ref{diag-1loopEDM}. 
However, it turns out that the charged Higgs contribution to $d_e$ vanishes at the 1-loop level while
the remaining neutral Higgs piece is suppressed by the third power of $m_e/M_{H_i}$ (generalizing the 2HDM result of \cite{Bernreuther:1990jx}):
\bea
\left(d_e\right)_{\rm 1-loop}^{H_i} 
&=& 
{m_e \over 16 \pi^2{M^2_{H_i}}} 
\sum_{i=1}^4 D_2({m_e^2 \over {M^2_{H_i}}},{m_e^2 \over {M^2_{H_i}}}) \; c_{ie} \tilde{c}_{ie}
\simeq \non \\
&&-\frac{3e}{32\pi^2} \sum_{i=1}^4 \left(\frac{m_e}{M_{H_i}}\right)^3\left[1+\frac23\ln\left(\frac{m_e}{M_{H_i}}\right)\right] \frac{c_{ie} \tilde{c}_{ie}}{M_{H_i}},
\label{one-loop}
\eea
where $c_{ie}$ and $\tilde{c}_{ie}$ are defined in~\eqref{cie}-\eqref{cs} while the loop function $D_2$ is given in Appendix~\ref{app:int}.
Even though $c_{ie}$ and $\tilde{c}_{ie}$ contain unrestricted experimentally component parametrized by $\rho$ however since perturbativity prohibits $\rho \gsim 4\pi$~\footnote{In fact, here adopted couplings never exceed $2\pi$.} one can conclude that even in the generic 2HDM 1-loop contributions to $d_e$ remain deeply suppressed.

The dominant 2-loop Feynman diagrams contributing to the electron EDM are the Barr-Zee diagrams~\cite{Barr:1990vd,Pilaftsis:1999qt,Abe:2013qla} shown in Fig.~\ref{diag-EDM}. The corresponding analytic results are available in Appendix~\ref{app:eEDM}.

%%%%%%%%%%%%%%%%%%%

In our analysis described in the next section, we perform scanning over the parameter space of the model and show the allowed region that satisfies the EDM constraint~(\ref{exp-edm}). It is worth mentioning already here that an appropriate choice of $\rho$ components might trigger a cancellation (a possibility noticed also in \cite{Kanemura:2020ibp}) between fermionic and bosonic 2-loop contributions to $d_e$ and therefore making the 1-loop contributions dominating (though very small). We are going to explore this option. 

%%%%%%%%%%%%%%%%%%%%%%%%%%%%%%%%%%%%%%%%%%%%%%%%%
\section{Numerical Analysis}\label{sec:Results}
In the model discussed here, choosing the weak basis described in sec.~\ref{model}, there are thirteen quartic couplings in the potential~(\ref{pot}) counting both real and imaginary parts. 
Ten of them could be expressed in terms of physical masses $M_{H_i}$ with $i=1,2,3,4$, $M_{H^\pm}$ and six mixing angles as follows~\footnote{For analogous relations in the generic 2HDM see \cite{Grzadkowski:2014ada}.}.
{\allowdisplaybreaks
\small
\begin{subequations}
\label{phys-couplings}
\begin{align}
\lambda_1 = & {1\over 
 4 v_1^3}\big[4v_1( M_{H_1}^2 \mathcal{R}_{11}^2 + M_{H_2}^2 \mathcal{R}_{21}^2 + M_{H_3}^2 \mathcal{R}_{31}^2 + M_{H_4}^2 \mathcal{R}_{41}^2 )
 - 2 {\rm Re\,}m_{12}^2 v_2
\nonumber \\ &
 - 6 {\rm Re\,}\lambda_6 v_1^2 v_2 + 2{\rm Re\,}\lambda_7 v_2^3 \big], 
\\ 
 \lambda_2 = & {1\over 
 4 v_2^3}\big[4 v_2(M_{H_1}^2 \mathcal{R}_{12}^2 + M_{H_2}^2 \mathcal{R}_{22}^2 + M_{H_3}^2 \mathcal{R}_{32}^2 + M_{H_4}^2 \mathcal{R}_{42}^2) 
 -2 {\rm Re\,}m_{12}^2 v_1 
\nonumber \\ &
 +2 {\rm Re\,}\lambda_6 v_1^3 - 6 {\rm Re\,}\lambda_7 v_1 v_2^2 \big], 
 \\
 \lambda_3 = & {1\over 
 4 v^2 v_1 v_2}\big[4 v^2 (M_{H_1}^2 \mathcal{R}_{11} \mathcal{R}_{12}+ M_{H_2}^2 \mathcal{R}_{21} \mathcal{R}_{22} + M_{H_3}^2 \mathcal{R}_{31} \mathcal{R}_{32} + M_{H_4}^2 \mathcal{R}_{41} \mathcal{R}_{42} )
\nonumber \\ &
+ 8 M_{H^\pm}^2 v_1 v_2 -2 {\rm Re\,}m_{12}^2 v^2 - 2{\rm Re\,}\lambda_6 v^2 v_1^2 - 2{\rm Re\,}\lambda_7 v^2 v_2^2 \big], 
 \\ 
 \lambda_4 = & {1\over 
 4 v^2 v_1 v_2}\big[ 4 v_2 v_1 (-2 M_{H^\pm}^2 + M_{H_1}^2 \mathcal{R}_{13}^2 + M_{H_2}^2 \mathcal{R}_{23}^2 + M_{H_3}^2 \mathcal{R}_{33}^2 + 
 M_{H_4}^2 \mathcal{R}_{43}^2) 
 \nonumber \\ &
 + 2 {\rm Re\,}m_{12}^2 v^2 - 2{\rm Re\,}\lambda_6 v^2 v_1^2 
 - 2{\rm Re\,}\lambda_7 v^2 v_2^2 \big], 
 \\ 
\lambda_s = & {1\over 
 4 v_s^2}\big[M_{H_1}^2 \mathcal{R}_{14}^2 + M_{H_2}^2 \mathcal{R}_{24}^2 + M_{H_3}^2 \mathcal{R}_{34}^2 + M_{H_4}^2 \mathcal{R}_{44}^2\big], \label{lamb_s}
 \\
 \kappa_1 = & {1\over 
 v_1 v_s}\big[M_{H_1}^2 \mathcal{R}_{11} \mathcal{R}_{14} + M_{H_2}^2 \mathcal{R}_{21} \mathcal{R}_{24} + M_{H_3}^2 \mathcal{R}_{31} \mathcal{R}_{34} + M_{H_4}^2 \mathcal{R}_{41} \mathcal{R}_{44}\big], 
 \\ 
 \kappa_2 = & {1\over 
 v_2v_s}\big[M_{H_1}^2 \mathcal{R}_{12} \mathcal{R}_{14} + M_{H_2}^2 \mathcal{R}_{22} \mathcal{R}_{24} + M_{H_3}^2 \mathcal{R}_{32} \mathcal{R}_{34} + M_{H_4}^2 \mathcal{R}_{42} \mathcal{R}_{44} \big], 
 \\
 {\rm Re\,}\lambda_5 = & {1\over 
 2 v^2 v_1 v_2} \big[ -2 v_1 v_2 (M_{H_1}^2 \mathcal{R}_{13}^2 + M_{H_2}^2 \mathcal{R}_{23}^2 + M_{H_3}^2 \mathcal{R}_{33}^2 + M_{H_4}^2 \mathcal{R}_{43}^2) 
\nonumber \\ &
 - {\rm Re\,}\lambda_7 v^2 v_2^2
 + {\rm Re\,}m_{12}^2 v^2 - {\rm Re\,}\lambda_6 v^2 v_1^2 \big], 
\\
 {\rm Im\,}\lambda_6 = & {-1 \over 2 v v_1}\big[
 2 (M_{H_1}^2 \mathcal{R}_{11} \mathcal{R}_{13} + M_{H_2}^2 \mathcal{R}_{21} \mathcal{R}_{23} + M_{H_3}^2 \mathcal{R}_{31} \mathcal{R}_{33} + M_{H_4}^2 \mathcal{R}_{41} \mathcal{R}_{43}) + 
 {\rm Im\,}\lambda_5 v v_2\big], 
 \\
 {\rm Im\,}\lambda_7 = & {-1 \over 2v v_2}\big[
 2 (M_{H_1}^2 \mathcal{R}_{12} \mathcal{R}_{13} + M_{H_2}^2 \mathcal{R}_{22} \mathcal{R}_{23} + M_{H_3}^2 \mathcal{R}_{32} \mathcal{R}_{33} + M_{H_4}^2 \mathcal{R}_{42} \mathcal{R}_{43}) + 
 {\rm Im\,}\lambda_5 v v_1\big].
\end{align}
\end{subequations}
}
Since $M_{34}^2=0$, one mixing angle is redundant, in other words, one can express, e.g. $\alpha_5$ in terms of other parameters. In the AL the appropriate relation reads 
\begin{align}
\alpha_5=\arcsin\big[\frac{(M_{H_3}^2-M_{H_4}^2)\sin2\alpha_6 \cos\alpha_4}{2\sin\alpha_4(-M_{H_2}^2+M_{H_4}^2 \cos\alpha_6^2+M_{H_3}^2 \sin\alpha_6^2)}\big].
\end{align}
Alternatively, $\alpha_5=\pi/2, 3\pi/2$.
%%%%%%%%%%%%%%%%%%%%%%%
\begin{figure}[h]
\begin{center}
\includegraphics[width=0.7\textwidth]{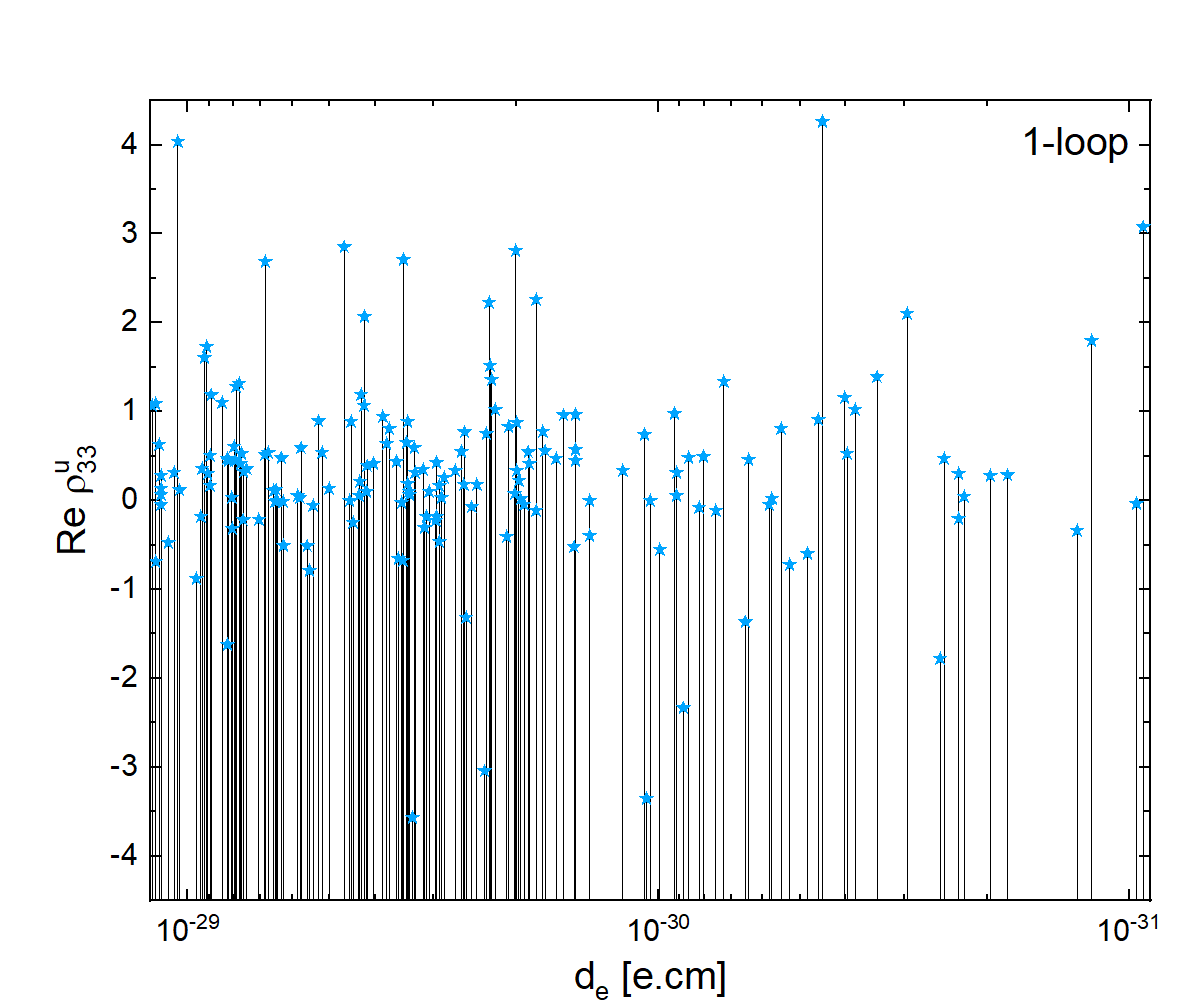}\subcaption{}
\end{center}
\caption{The results of ${\rm Re}\rho^{u}_{33}$ versus the total electron EDM are plotted in the "1-loop dominated" scenario. The vertical lines connect diamonds with corresponding values of $d_e$, this method of illustration will be adopted hereafter.}
\label{ru33}
\end{figure}
%%%%%%%%%%%%%%%%%%%%%
\begin{figure}[h]
\begin{center}
\includegraphics[width=0.7\textwidth]{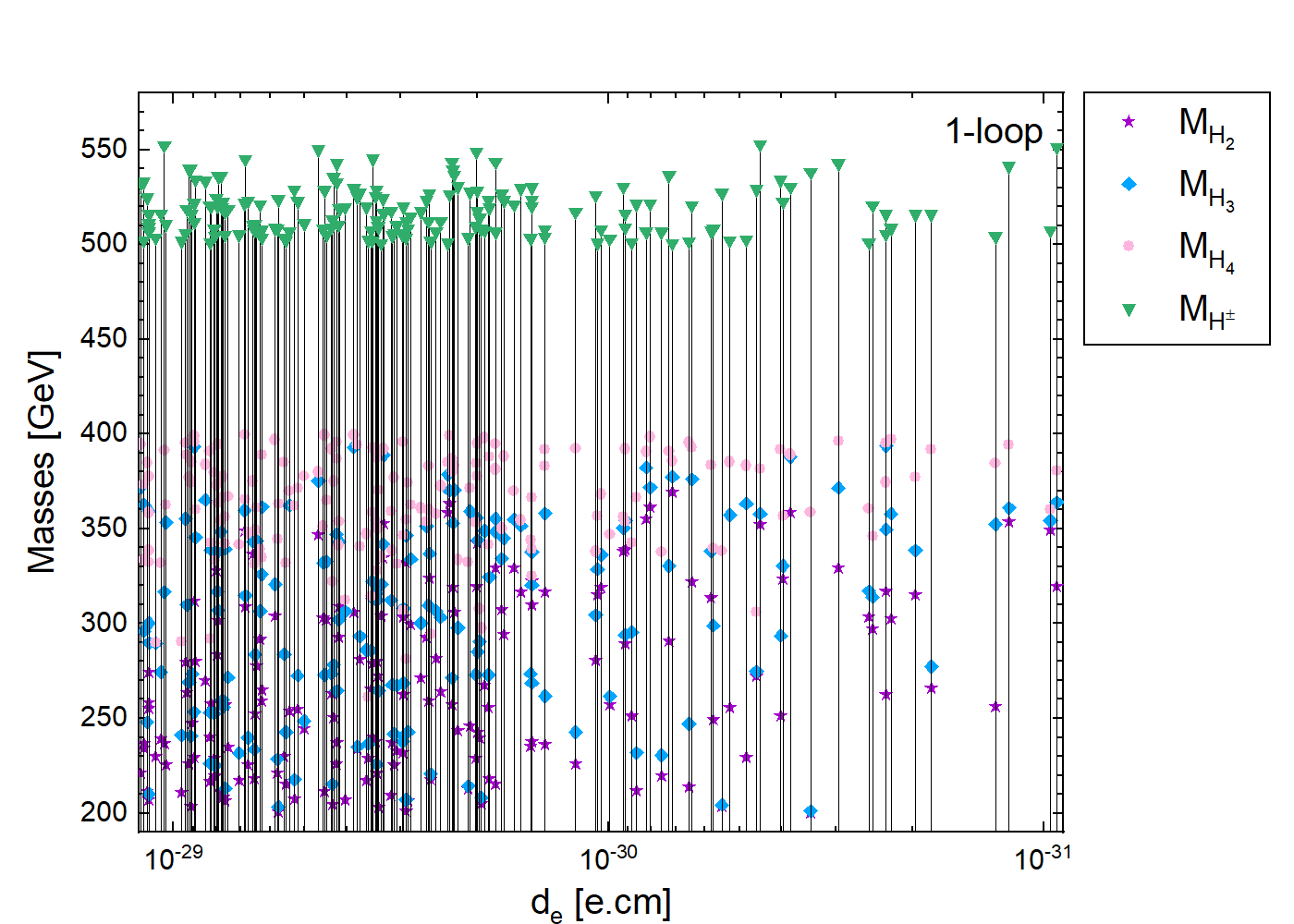}
\end{center}
\caption{Neutral and charged Higgs boson masses versus the total electron EDM are plotted
in the "1-loop dominated" scenario.}
\label{EDM2-lines1}
\end{figure}
%%%%%%%%%%%%%%%%%%%%%%%
\begin{figure}[h]
\begin{center}
\includegraphics[width=0.7\textwidth]{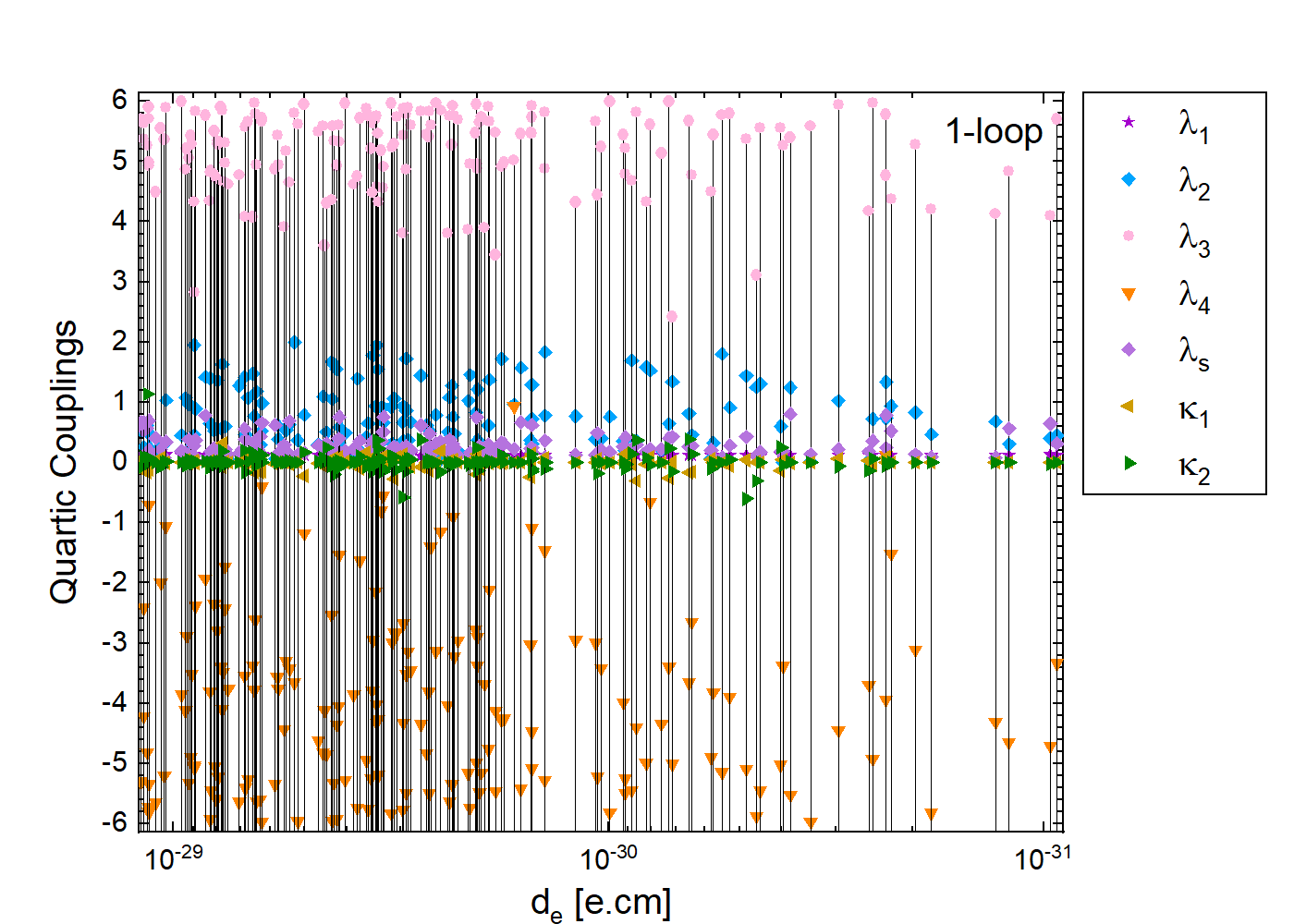}\subcaption{}
\includegraphics[width=0.7\textwidth]{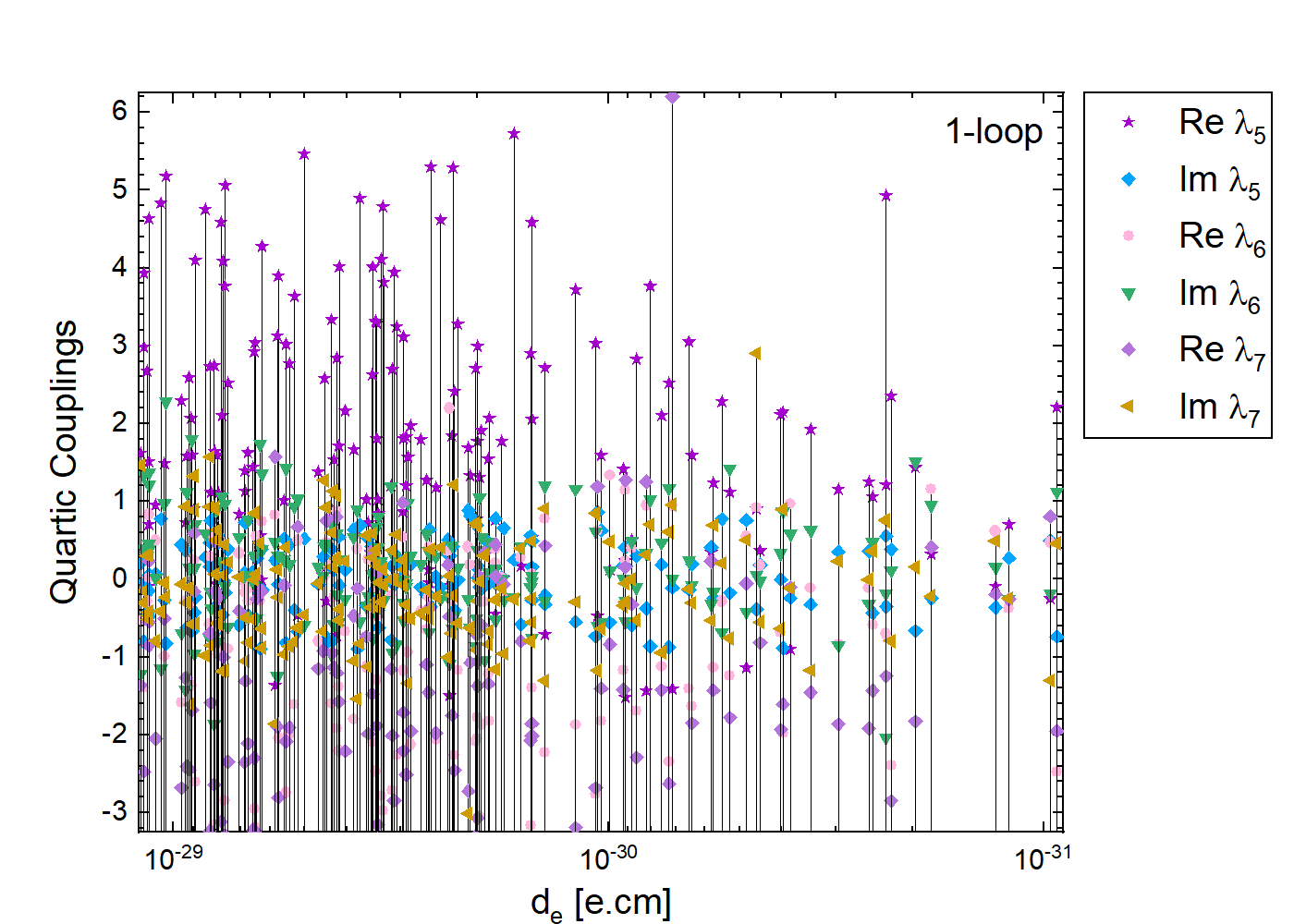}\subcaption{}
\end{center}
\caption{Real and complex quartic couplings versus the total electron EDM are plotted in the "1-loop dominated" scenario.}
\label{EDM2-lines2}
\end{figure}

Eventually, including dimensionfull parameters, in the AL, we adopt for our numerical analyzes the following set of 15 independent quantities:
\beq
\big[\beta,\, v,\, \alpha_{4,6},\, M_{H_j},\, M_{H^\pm},\, {\rm Im\,}\lambda_5, \, {\rm Re\,}\lambda_6, \, {\rm Re\,}\lambda_7,\,{\rm Re\,}m_{12}^2, v_s,\, \mu^2\big]\,,
\label{input}
\eeq
where $j=1,2,3,4$. 
Note that the above counting of the parameters of the potential is consistent with the remark that we have made below \eqref{minim}; here the total number of parameters is additionally reduced to 15 by the three constraints for the AL i.e. $\alpha_1=\beta$ and $\alpha_2=\alpha_3=0$. 

The high-energy scattering matrix of the scalar sector at tree level contains only s-wave amplitudes that are described by the quartic part of the potential. The tree-level unitarity constraints require that the eigenvalues of this scattering matrix be less than the unitarity limit $\sim 4 \pi$~\cite{Ginzburg:2003fe}\footnote{However, here all quartic couplings will be bounded by $2 \pi$.}. We are going to adopt the following regions for scans over model parameters:
\begin{align}
\beta &\in (0,\,{\pi \over 2}),\, &\alpha_{4,6}&\in (0,\,\pi),
\nonumber \\ 
{\rm Im\,}\lambda_{5}&\in (-0.9,\, 0.9), \, &{\rm Re\,}\lambda_{6,7}&\in (-\pi,\,\pi),
\nonumber \\ 
 M_{H_{2,3,4}}&\in (200,\,400)\,{\rm GeV},\, &M_{H^\pm}&\in (500,\,800)\,{\rm GeV},\, 
 \nonumber \\ 
{\rm Re\,}m_{12}^2& \in (\pm 50^2,\,\pm 800^2)\,{\rm GeV}^2, \, &v_s &\in (350,\,900)\,{\rm GeV}.
\label{bound}
\end{align}
Here we assume the mass ordering $M_{H_1}<M_{H_2}<M_{H_3}<M_{H_4}$ with the lightest one being the SM like Higgs particle with mass $M_{H_{1}}= 125.15 \pm 0.24$\,GeV~\cite{Palmer:2021gmo,ATLAS:2021upq}.

There is a comment here in order concerning positivity of the scalar potential.
Often the strong stability condition, i.e. asymptotic positivity of quartic part of the potential, is required, see~\cite{Branco:2011iw} and references therein. The condition severely restricts the parameter space and in fact is usually too strong eliminating interesting models.
Here, as it will be shortly shown by the results of scanning over the parameter space specified above, some quartic couplings are relatively large at the weak scale reaching a value of $5-6$ ($|\lambda_{3,4,5}|$). Even though this is still within the perturbative regime, it may imply the appearance of Landau poles even at an energy scale of the order of a few TeV. Therefore, the scenario considered in this section should be interpreted as an effective theory and its applicability should be limited below a certain UV scale of new physics. This will not only fix the problem of Landau poles but also will influence and (possibly solve) the issue of positivity. Therefore we have decided not to impose conditions for asymptotic positivity of the scalar potential~\footnote{Sufficient and necessary conditions for asymptotic positivity in the model we consider are unknown and finding them lays beyond the scope of this project. Nevertheless it is possible to formulate a set of sufficient conditions. The potential could be rewritten as a sum of a $Z_2$ invariant potential, doublet-singlet pieces and positive-defined parts. Requiring that each of those components is positive the set of positivity conditions can be formulated. We have verified that there exist regions in the parameter space such that the conditions are fulfilled so the positivity is guarantied and all the experimental constraints are satisfied.}

Note also that, there are 18 free parameters, located in $\rho$, in the Yukawa Lagrangian. However, since these parameters are unbounded by experiment and our attention is towards contributions from the bosonic sector we arbitrarily choose the non-zero entries of $\rho$ only for electron, bottom- and top-quark. In addition, in order to avoid the issue of FCNC (irrelevant here) we choose $\rho$ flavour-diagonal.
%%%%%%%%%%%%%%%%%%%%%%%
\begin{figure}[h]
\begin{center}
\includegraphics[width=0.7\textwidth]{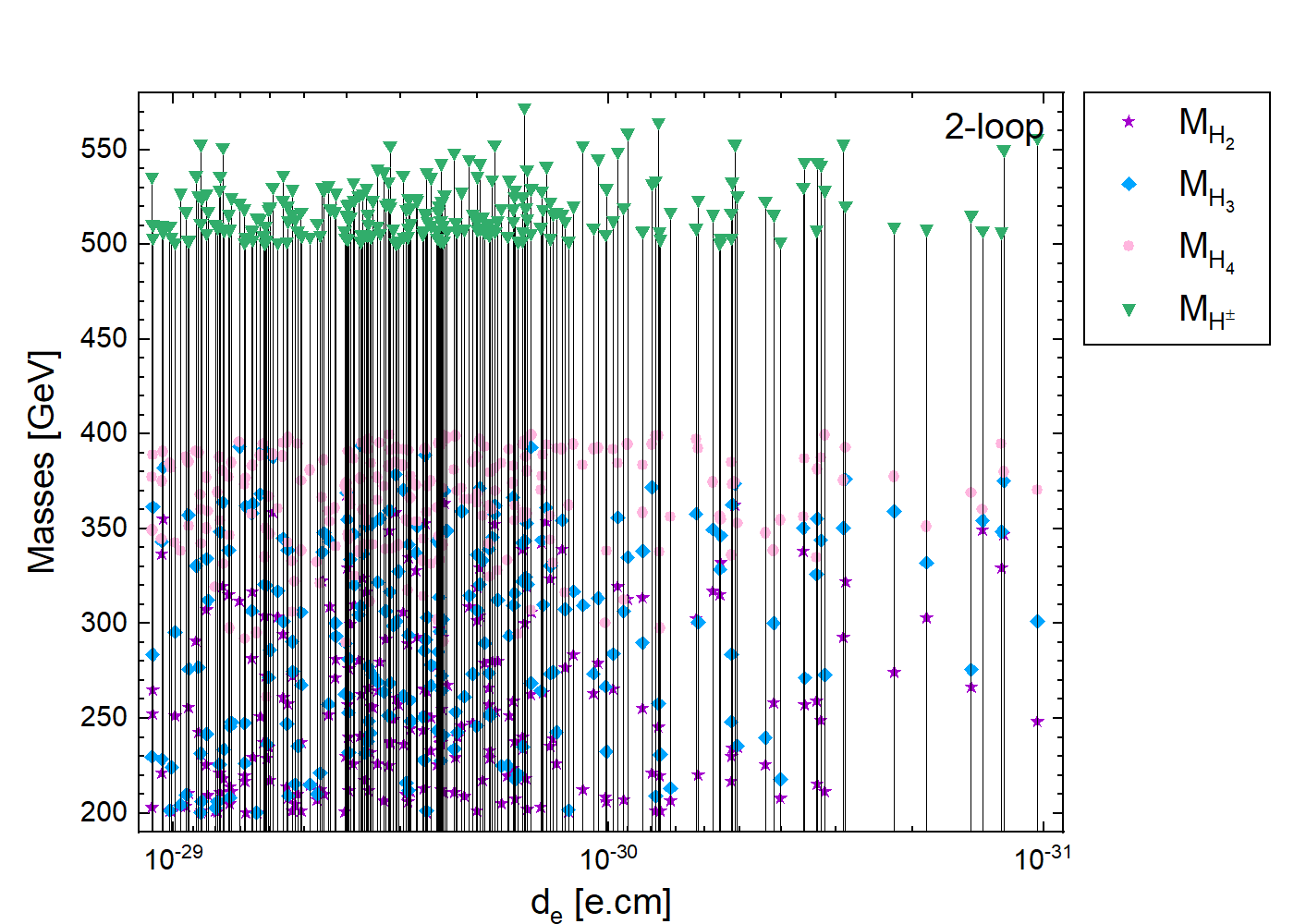}\subcaption{}
\end{center}
\caption{Neutral and charged Higgs bosons masses versus the total electron EDM $d_e$ in the "2-loop dominated" scenario.}
\label{EDM-lines1}
\end{figure}
%%%%%%%%%%%%%%%%%%%%%%%
\begin{figure}[h]
\begin{center}
\includegraphics[width=0.7\textwidth]{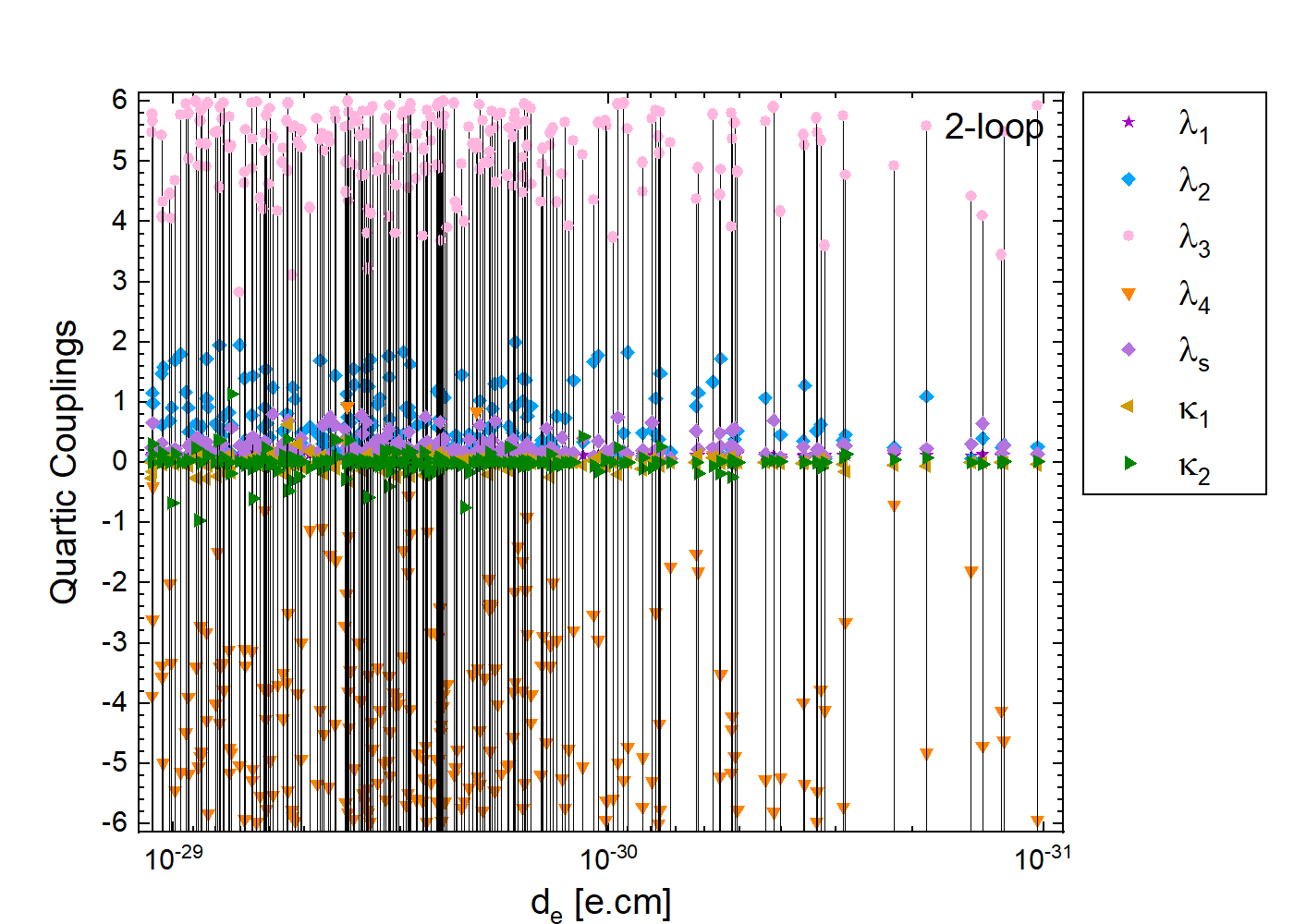}\subcaption{}
\includegraphics[width=0.7\textwidth]{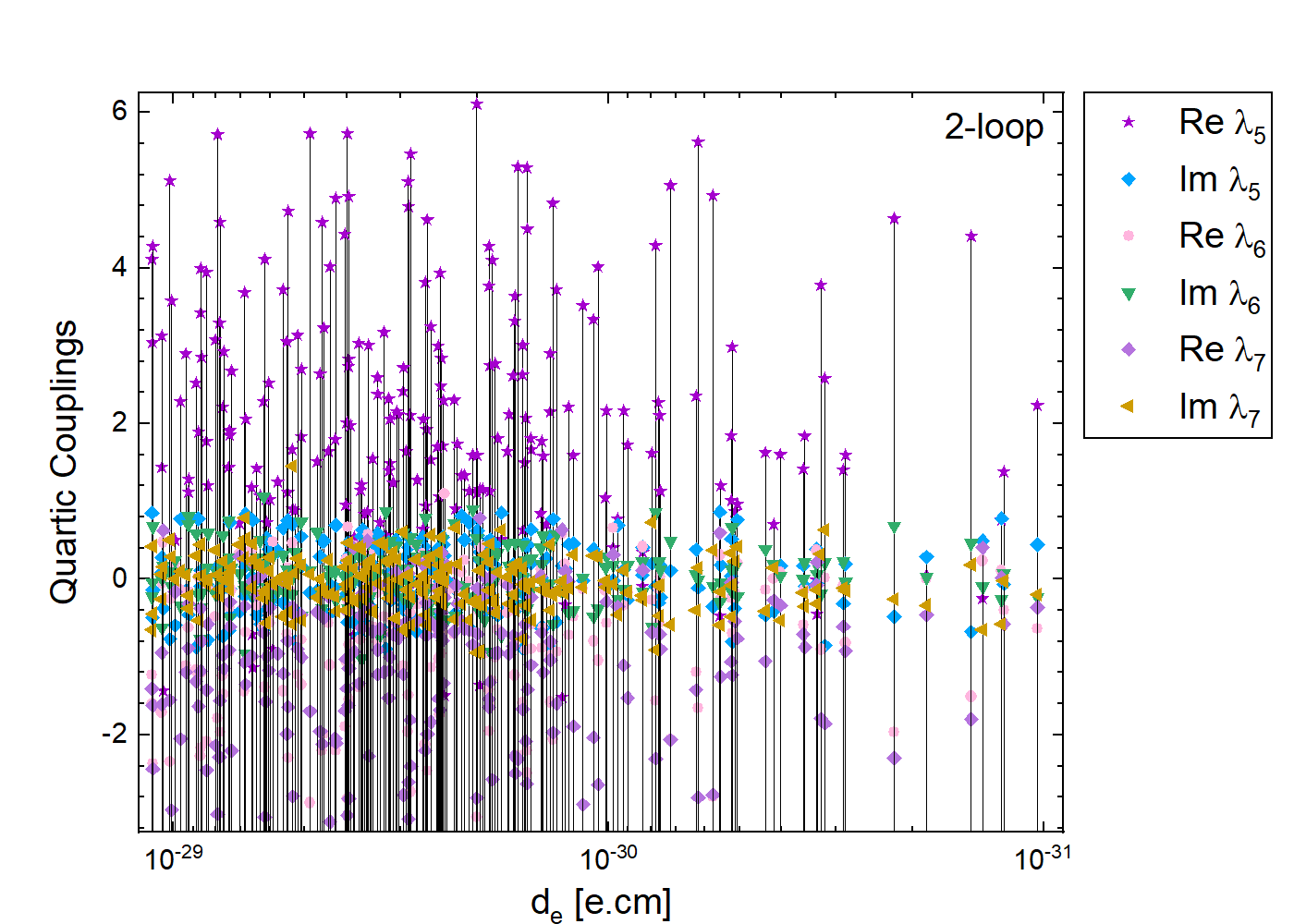}\subcaption{}
\end{center}
\caption{Real and complex quartic couplings versus the total electron EDM are plotted 
in the "2-loop dominated" scenario.}
\label{EDM-lines2}
\end{figure}

We are going to consider the following two classes of scans that differ by our choice of Yukawa couplings. In both cases the scanning regions of non-Yukawa parameters are specified in \eqref{bound}.
\bit
\item {\bf 1-loop dominated}:\\
In this case we scan over the following region of Yukawa couplings
\beq
{\rm Im}\rho^{u}_{33}\sim 1, \quad \rho^{d}_{33}\sim (1+i)10^{-1}, \quad \rho^{l}_{11} \in (1+i)(10^{-4},\, 10^{-2}),
\label{rho-bound2}
\eeq
with ${\rm Re}\rho^{u}_{33}$ tuned to induce a cancellation between bosonic and fermionic contributions in 2-loop diagrams. In Fig.~\ref{ru33} we show values of ${\rm Re}\rho^{u}_{33}$ that are need for the cancellation. It is important to note that $\rho^{u}_{33}$ obtained that way remains in the perturbative regime. 
Results of scans corresponding to this "1-loop dominated" scenario are shown in Figs.~\ref{EDM2-lines1} and \ref{EDM2-lines2}.

\item {\bf 2-loop dominated}:\\
In this case we scan over the following region of Yukawa couplings
\beq
\rho^{u}_{33}\sim (1+i)10^{-3}, \quad \rho^{d}_{33}\sim (1+i)10^{-4}, \quad \rho^{l}_{11} \in (1+i)(10^{-8},\, 10^{-6}).
\label{rho-bound}
\eeq
To set the scale for $\rho$ it is instructive to know e.g. $\rho^{l}_{11} $ in a popular type II version of the 2HDM. Using the appendix E4 of~\cite{Grzadkowski:2018ohf} one finds that $|\rho^{l}_{11}| \sim \tan\beta \cdot 3\cdot 10^{-6}$. 

\begin{figure}[h]
\begin{center}
\includegraphics[width=.7\textwidth]{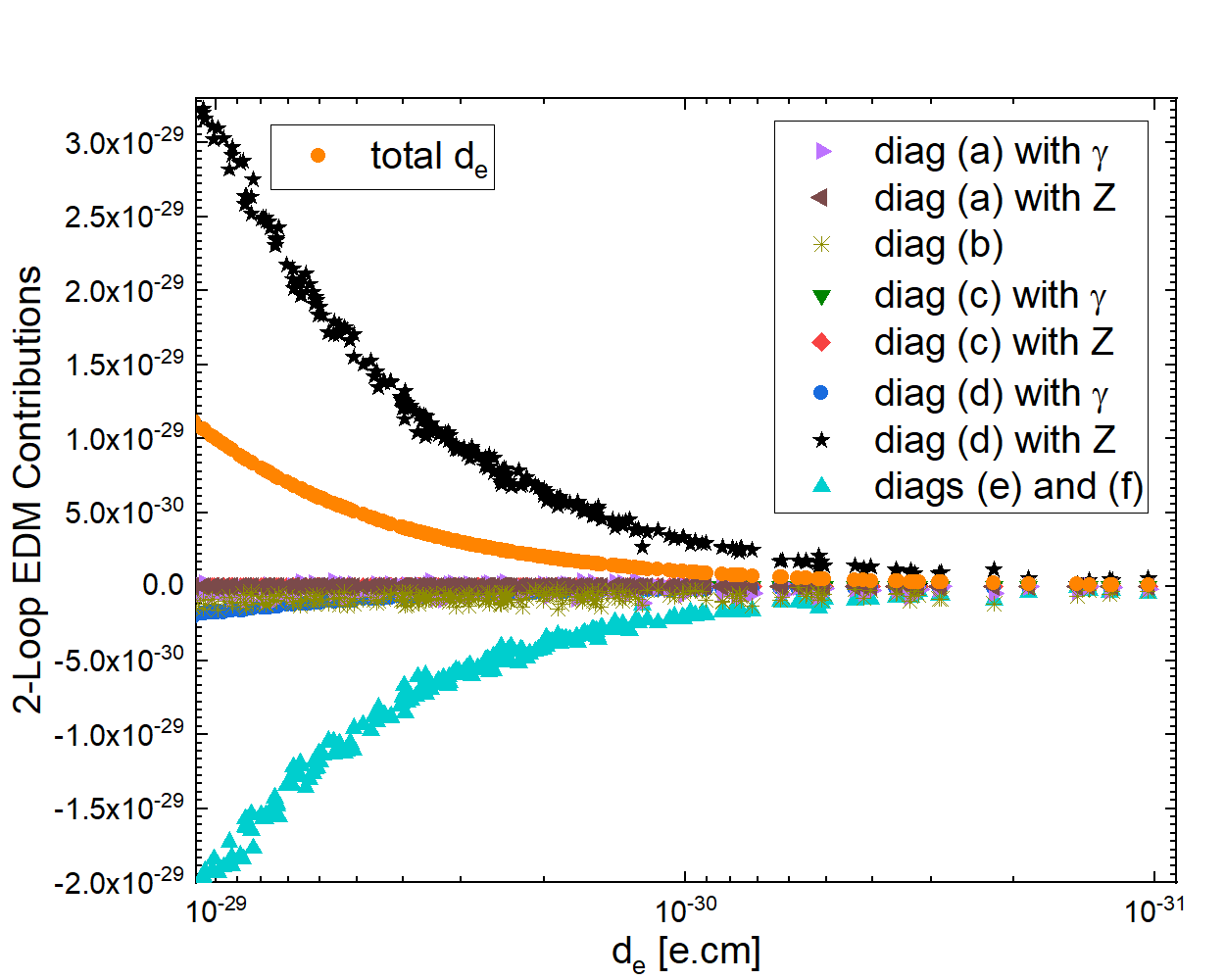}
\end{center}
\caption{Numerical contributions to $d_e$ from Barr-Zee diagrams shown in Fig.~\ref{diag-EDM}
within the "2-loop dominated" scenario.}
\label{EDM-Cont}
\end{figure}

Results of the scan are shown in Figs.~\ref{EDM-lines1} and~\ref{EDM-lines2}. In Fig.~\ref{EDM-Cont} we show contributions of various 2-loop diagrams to the $d_e$.
\eit
%%%%%%%%%%%%%%%%%%%%%%%
%%%%%%%%%%%%%%%%%%%
\begin{figure}[t]
\begin{center}
\includegraphics[width=0.65\textwidth]{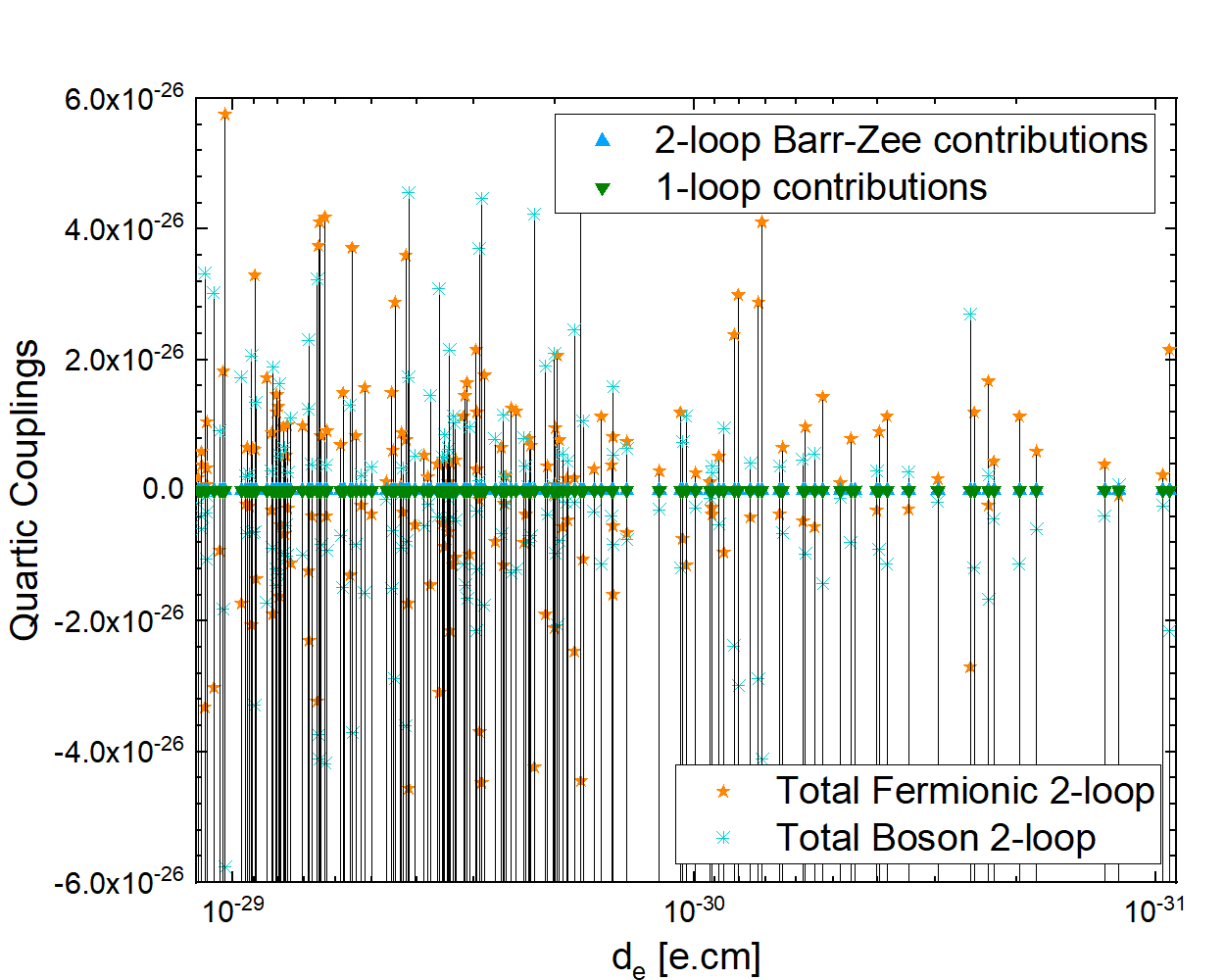}
\subcaption{}
\includegraphics[width=0.65\textwidth]{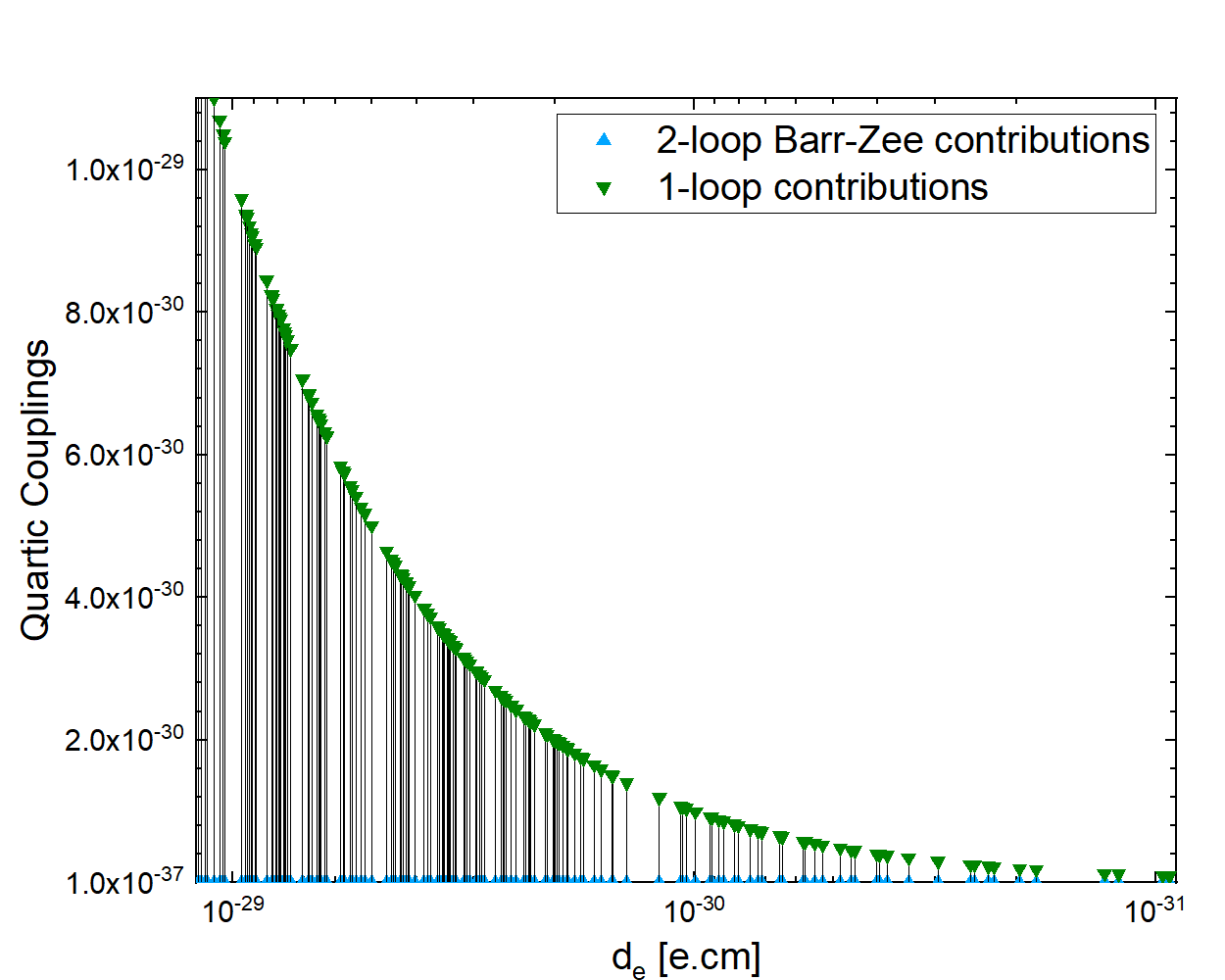}
\subcaption{}
\end{center}
\caption{The upper panel (a) shows fermionic and bosonic 2-loop Barr-Zee contributions to $d_e$ by yellow and cyan stars, respectively. Note that the up-down symmetry of stars is a consequence of the 2-loop cancellation requirement ("1-loop dominated" scenario). Blue and green triangles located around $d_e\sim 0$ stand for total 2- and 1-loop contributions. The lower panel (b) focuses on the band $d_e\sim 0$ of the upper panel and shows the cancellation of the 2-loop contributions. Here $\rho^u_{33}$ is tuned to achieve the cancellation while other parameters are specified by \eqref{bound} and \eqref{rho-bound2}.}
\label{EDM2-Cont}
\end{figure}

Fig.~\ref{EDM2-Cont} illustrates and compares the two scenarios defined above.
%%%%%%%%%%%%%%%%%%%%%%%%%%%%%%%%%%%%%%%%%%%%
\subsection{CP Invariants and Electron EDM}
\label{sec:CP-EDM}
The Barr-Zee type diagrams contributing to the electron EDM are shown in Fig.~\ref{diag-EDM}. 
In Fig.~\ref{EDM-Cont} we separate contributions to EDM that originate from each Barr-Zee diagram. The points that are displayed obey the 
experimental limit \eqref{exp-edm} and are randomly chosen within the region defined by \eqref{bound} and \eqref{rho-bound}. The plot illustrates substantial cancellation between the fermionic diagram (a) (with $\gamma$) and the diagram~(c)~(with~$\gamma$). 

In order to diagnose this case in Fig.~\ref{EDM-lines1} we plot $H_{2,3,4}$ and $H^\pm$ masses corresponding to resulting values of $d_e$.

In addition, in Fig.~\ref{EDM-lines2} we are showing quartic couplings corresponding to the total $d_e$.
It is seen that $|\lambda_{3,4,5}|$ might be substantial, approaching vicinity of the perturbative limit, i.e. $|\lambda_{3,4,5}|\sim 6 \lesssim 2\pi < 4\pi$. 
The regions allowed for quartic couplings are given below:
\begin{align}
 &\quad \,\lambda_2 \in [0.00, 1.99],\quad
 \qquad  \lambda_3 \in [2.42, 5.99], \quad 
 \nonumber \\ 
 & \quad\, \lambda_4 \in [-5.99, 0.93], \quad 
\quad\, \lambda_s \in [0.00, 0.80], \quad
 \nonumber \\
& {\rm Re}\lambda_5 \in [-1.51, 6.10], \quad 
 {\rm Im}\lambda_5 \in [-0.89, 0.88], 
 \nonumber \\
&  {\rm Re}\lambda_6 \in [-3.08, 1.09], \quad 
 {\rm Im}\lambda_6 \in [-1.01, 1.14], \nonumber \\ &
 {\rm Re}\lambda_7 \in [-3.13, 3.09], \quad 
 {\rm Im}\lambda_7 \in [-3.01, 2.90].
\end{align}
and $\lambda_1\sim 0.12$. Additionally, the following ranges for Higgs boson masses has been obtained:
\begin{align}
 &M_{H_2} \in [200.13, 369.37], \quad 
 M_{H_3} \in [200.23, 393.88], \nonumber \\ & 
 M_{H_4} \in [261.22, 399.65], \quad
 M_{H^{\pm}} \in [500.63, 572.28]. 
\end{align}
%%%%%%%%%%%%%%%%%%%%%%%%%%%%%%%
\begin{figure}[t]
\begin{center}
\includegraphics[width=0.7\textwidth]{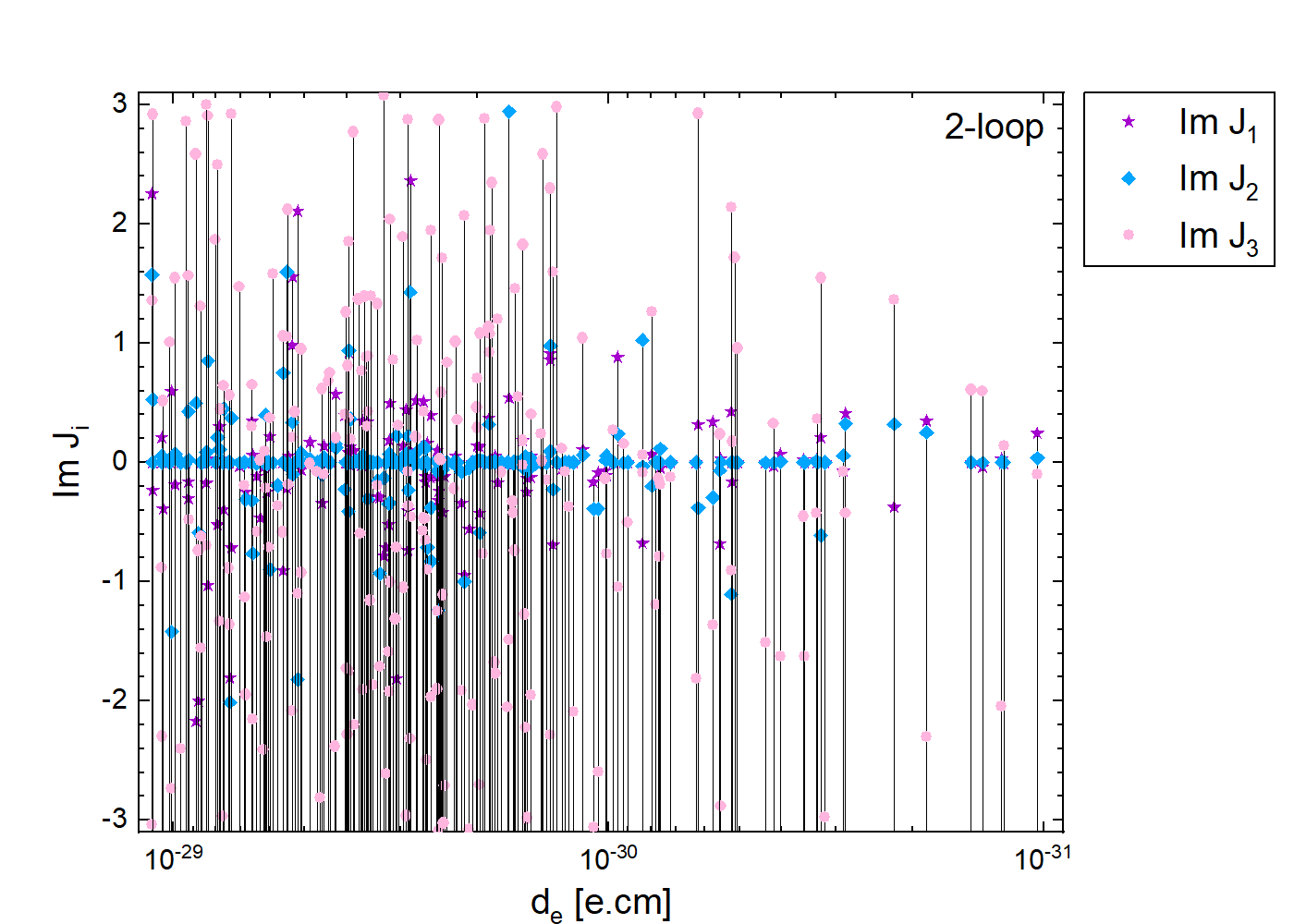}
\end{center}
\caption{The CP invariants ${\rm Im}J_{1,2,3}$ versus the electron EDM are plotted after imposing the current upper bound on electron EDMs (\ref{exp-edm}) in the "2-loop dominated" scenario.}
\label{EDM}
\end{figure}
%%%%%%%%%%%%%%%%%%%%%%%%%%%
Now our intention is to show restrictions imposed by the electron EDM (\ref{exp-edm}) on CP-invariants ${\rm Im}J_{1,2,3}$. In other words we wish to see how much CPV is still allowed in the model after imposing constraints from $d_e$ measurements. 
In Fig.~\ref{EDM}, we present correlations between the EDM, $d_e$, and values of the CP invariants. As expected in order to decrease $d_e$ the invariants must be small enough. This figure is a result of a scan within the "2-loop dominated" scenario over the parameters of the model in the region allowed by the EDM constraint. Note that even for $d_e \sim 10^{-31} \rm{e\; cm}$, ${\rm Im}J_{3}\sim -2$ is still allowed. Using the limit (\ref{exp-edm}) one finds the following (correlated) limits on the invariants 
$$
 -2.17<{\rm Im}J_{1}<2.36,
\qquad
 -2.01<{\rm Im}J_{2}<2.94,
\qquad
 -3.09<{\rm Im}J_{3}<3.07.
$$
Concluding, the EDM constraint still allows for relatively strong CPV, however it should be noted that in order to achieve large invariants, sufficiently strong, although perturbative, quartic couplings are needed.

Since the number of model parameters is substantial it will be convenient in further analysis to introduce three benchmark points BM1, BM2 and BM3 defined in Table \ref{tab1}. In these benchmark scenarios, BM1 and BM2 are CP-violating while BM3 is CP conserving. They have been validated against LHC data using the \textsf{HiggsBounds v5.10.2} package~\cite{Bechtle:2008jh}.

\begin{table}[t]
\begin{center}
\begin{tabular}{|c|c|c|c|c|c|c|c|c|c|} 
 \hline
BM & $\tan \beta$ & $\alpha_1=\beta$ & $\alpha_2$ & $\alpha_3$ & $\alpha_4$ & $\alpha_5$ & $\alpha_6$ & $v_s$ [GeV] \\ \hline \hline
1,2 & 2.565 & 1.199 & 0 & 0 & 1.685& 0.011 & 1.285& 443.173 \\
3 & 1.777 & 1.058 & 0 & 0 & 0 & 0 & 0& 490.929\\
 \hline
\end{tabular}
\end{center}
\begin{center}
\begin{tabular}{|c|c|c|c|c|c|c|c|c|c|c|c|c|} 
 \hline
BM & $\lambda_1$ & $\lambda_2$ & $\lambda_s$& $\kappa_1$ & $\kappa_2$& $\lambda_3$ & $\lambda_4$ 
\\ \hline \hline
1,2 & 0.12 & 0.40 & 0.62 & $-0.18$ & $0.02$ & 5.17 & $-$4.51 \\
3 & 0.12 & 0.19 & 0.53 & 0.00 & 0.00 & 5.83 & $-$5.73
\\
 \hline
\end{tabular}
\end{center}
\begin{center}
\begin{tabular}{|c|c|c|c|c|c|c|c|c|c|c|c|c|} 
 \hline
BM & $\lambda_5$ & $\lambda_6$ & $\lambda_7$ & {\rm Re} $m_{12}^2$ [GeV$^2$] 
\\ \hline \hline
1,2 & $ 3.22+ i 0.49 $ &$-1.99 - i 0.13 $&$- 0.49 - i 0.17 $&347.797$^2$\\
3 & $-$ 0.32 & 0.20 & 0.10 &294.11$^2$\\
 \hline
\end{tabular}
\end{center}
\begin{center}
\begin{tabular}{|c|c|c|c|c|c|c|c|c|c|c|c|c|} 
 \hline
BM & $M_{H_1}$ [GeV]& $M_{H_2}$ [GeV]& $M_{H_3}$ [GeV]& $M_{H_4}$ [GeV]& $M_{H^\pm}$ [GeV]\\ \hline \hline
1,2 & 125.25 & 210.08 & 347.76 & 386.18 & 529.16 \\
3 & 125.25 & 323.98 & 337.01 & 359.91 & 490.92 \\
 \hline
\end{tabular}
\end{center}
\begin{center}
\begin{tabular}{|c|c|c|c|c|c|c|c|c|c|c|c|c|} 
 \hline
BM & $\rho^{l}_{11}$ &${\rm Re}\rho^{u}_{33}$ & ${\rm Im}\,J_1$ & ${\rm Im}\,J_2$ & ${\rm Im}\,J_3$ & $|d_e| \,[{\rm e.\, cm}] $ 
\\ \hline \hline
1\,(1-loop) & $(1+i)10^{-3}$& $-$0.65 &0.14 & 0.04&$-0.09$&$1.11\times 10^{-29}$ \\
2\,(2-loop) & $(1+i)10^{-6}$&$10^{-3}$& 0.14 & 0.04&$-0.09$&$1.02\times 10^{-29}$ \\
 \hline
\end{tabular}
\end{center}
\caption{The three benchmark points within the AL with $M_A=500$ GeV and $\mu^2=-250^2$ GeV$^2$. BM1 and BM2 are CP-violating and correspond to "1-loop dominated" and "2-loop dominated" scenarios, respectively. BM3 is CP-conserving.}
\label{tab1}
\end{table}
%Additionally, the couplings relevant to DM analyses are given in Table~\ref{tab:hiA}.
Note that, the parameter $\mu^2$ is varying in our analyses according to the DM mass, however that does not affect the other parameters in the benchmark points.
%%%%%%%%%%%%%%%%%%%%%%%%%%%%
\subsection{DM Relic Density and Indirect Detection}
As we have shown in Section~\ref{model}, the singlet $S$ is charged under a global $\rm U(1)$ symmetry which is softly broken by a mass term $\mu^2$. Therefore, the imaginary part of $S$ becomes a stable DM candidate which, at the tree level and in the limit of zero momentum transfer, decouples from nucleons naturally satisfying all existing direct detection limits on DM scattering cross-section. However, there is a strong constraint on DM relic density from the Planck data release~\cite{Planck:2018vyg} that restricts the present DM abundance at
\begin{align}
\Omega_{\rm DM} h^2 = 0.120 \pm 0.001. 
\end{align}
In the model discussed here relevant channels for DM annihilation are displayed in Fig.~\ref{relic-diag}. The couplings pertinent to these channels are given in Appendix \ref{app:couplings}.
\begin{figure}[t]
\begin{center}
\includegraphics[width=1\textwidth]{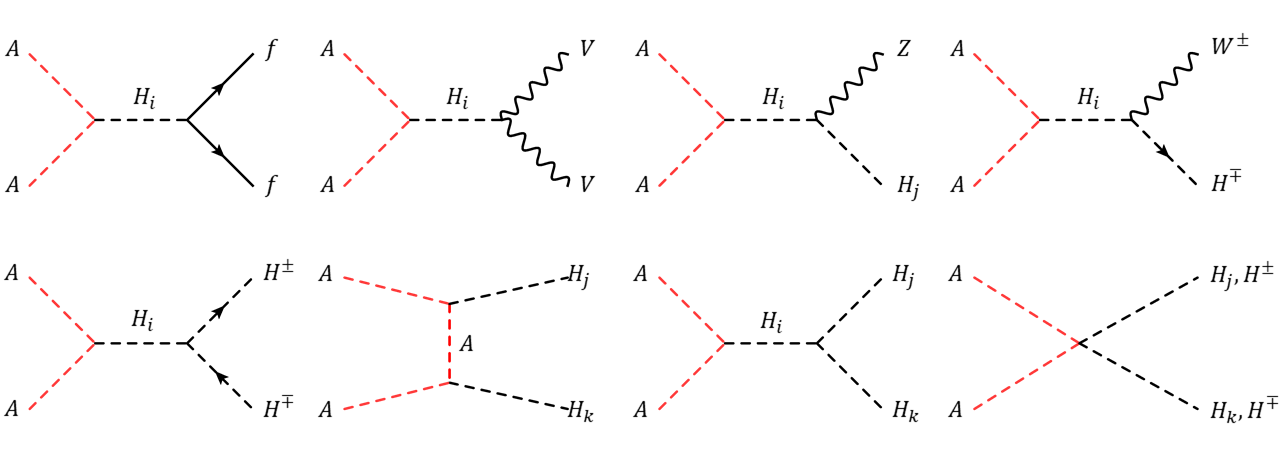}
\end{center}
\caption{Relevant annihilation channels for the calculation of the relic density of DM.}
\label{relic-diag}
\end{figure}

Other important experimental constraints on the DM sector are provided by indirect detection experiments. In this approach, a pair of DM particles annihilate into the SM particles which subsequently decay to produce $\gamma$-rays which could be searched by experiments such as Fermi-LAT~\cite{Fermi-LAT:2015att}. Numerical results shown hereafter have been obtained using \textsf{MadDM v3.1.1}, a code for the calculation of DM properties~\cite{Backovic:2013dpa,Ambrogi:2018jqj}.

\begin{figure}[t]
\begin{center}
\includegraphics[width=0.7\textwidth]{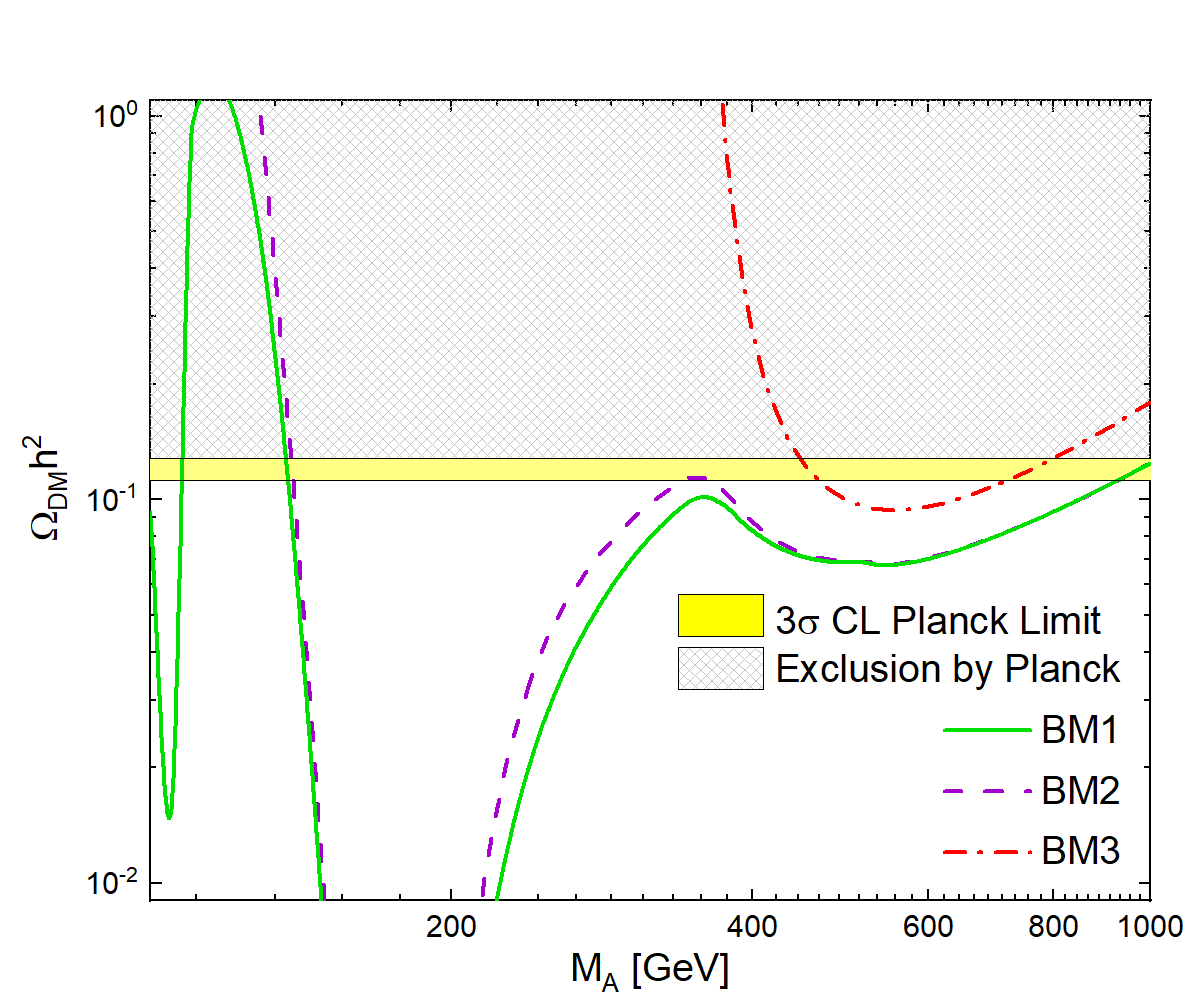}
\end{center}
\caption{
Profiles of relic density of pGDM in 2HDMCS for BM1, BM2 and BM3. The parallel horizontal lines with the yellow region in between indicate the constraint on DM relic density from the Planck data within the 3$\sigma$ confidence level. The hachured region is excluded by over-abundance while the non-hachured one requires some other DM component. These results are produced with the resolution of $1$~GeV for DM mass.}
\label{relic}
\end{figure}

In Fig.~\ref{relic}, we exhibit the behaviour of DM relic density as a function of pGDM mass in the model. In this plot, the horizontal lines limit the experimentally allowed relic density value. As it is seen from the plot, BM1 provides
the correct relic density for DM mass $\sim 107$~GeV, $\sim 135$~GeV and in the range $925-1000$~GeV. DM is under-abundant in ranges $100 - 107$~GeV, and 
$350 - 925$~GeV and over-abundant the range $107 - 135$~GeV. 
BM2 has similar properties, except it predicts an over-abundant relic density for DM masses everywhere below $\sim 135$~GeV and hits the correct values for masses in the short range $335-365$ GeV.
Finally, for BM3 the correct relic density can be obtained for $M_A\sim 440-470$~GeV and $710-810$~GeV with the region between (i.e. $470-710$~GeV) being slightly under-abundant, whereas for masses below $440$~GeV and above $810$~GeV the relic density is over-abundant.

\begin{figure}[t]
\centering
\begin{subfigure}[t]{2.86in}
		\centering
\includegraphics[width=1\textwidth]{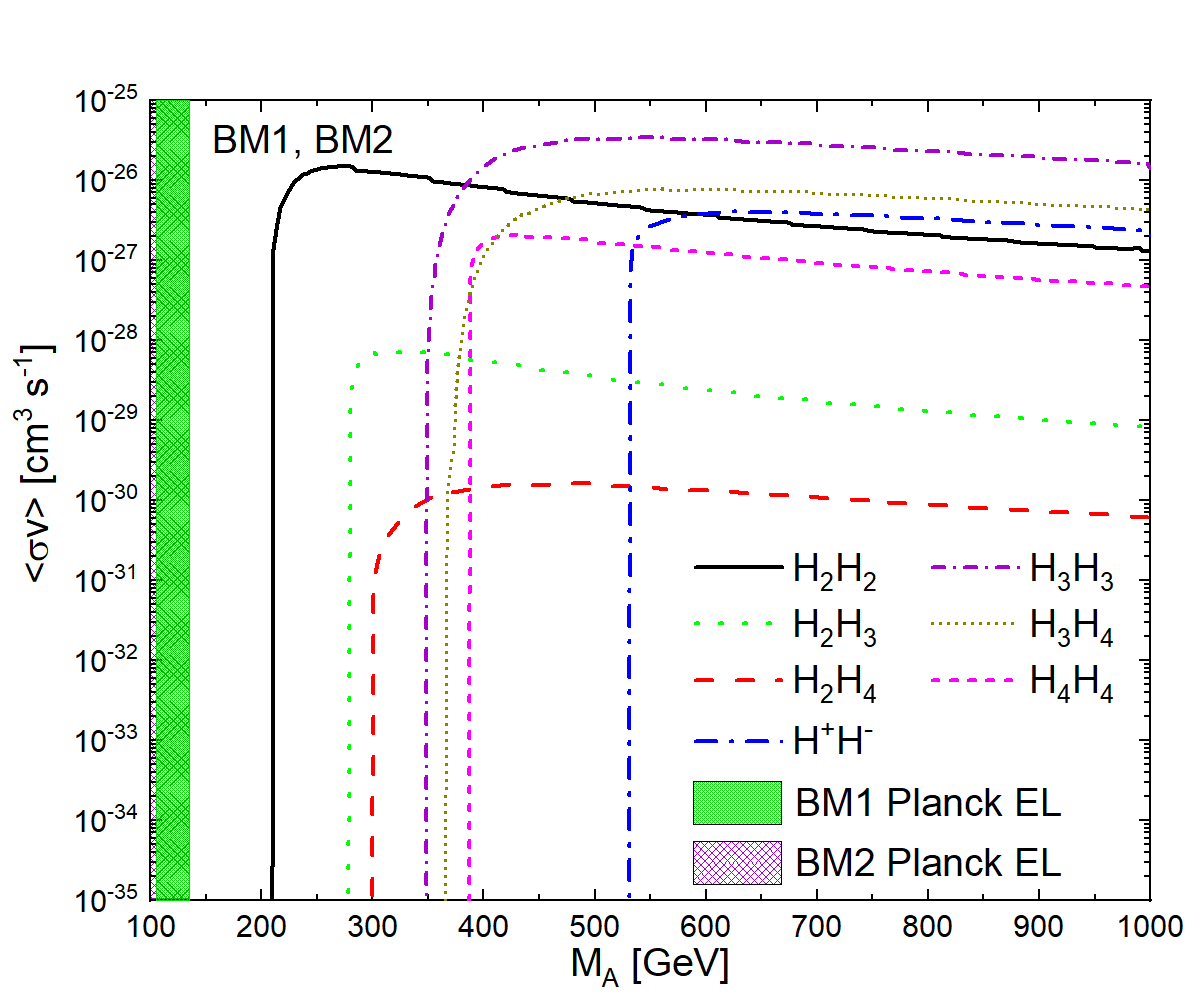}	
		\caption{}
	\end{subfigure}
	\quad
	\begin{subfigure}[t]{2.86in}
		\centering
\includegraphics[width=1\textwidth]{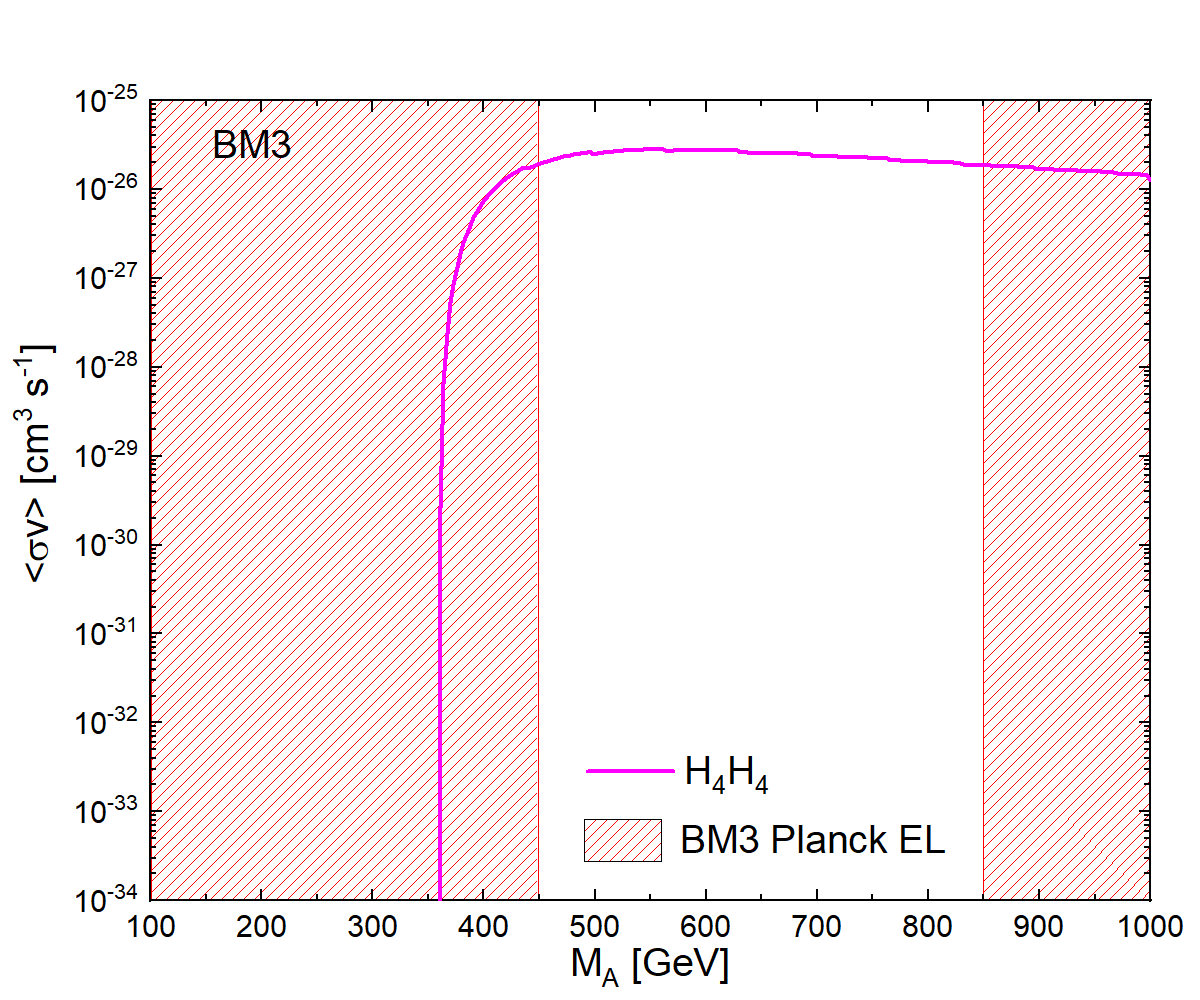}
		\caption{}\label{b}		
	\end{subfigure}
\caption{Channels of DM annihilation cross section into heavy Higgs bosons pairs for BM1, BM2 and BM3 are displayed in panels a, b respectively. The colourful hachuring denotes regions excluded by overabundance. In panel (a), the green and purple regions are mostly overlapping. The exact regions are green $\in$ [107 GeV, 135 GeV] and purple $\in$ [100 GeV, 135 GeV]. As it could be deduced from Table \ref{tab:hiA}, in the case of BM3 the remaining contributions are negligible.
}
\label{hihi}
\end{figure}

\begin{figure}[t]
\begin{center}
\includegraphics[width=0.9\textwidth]{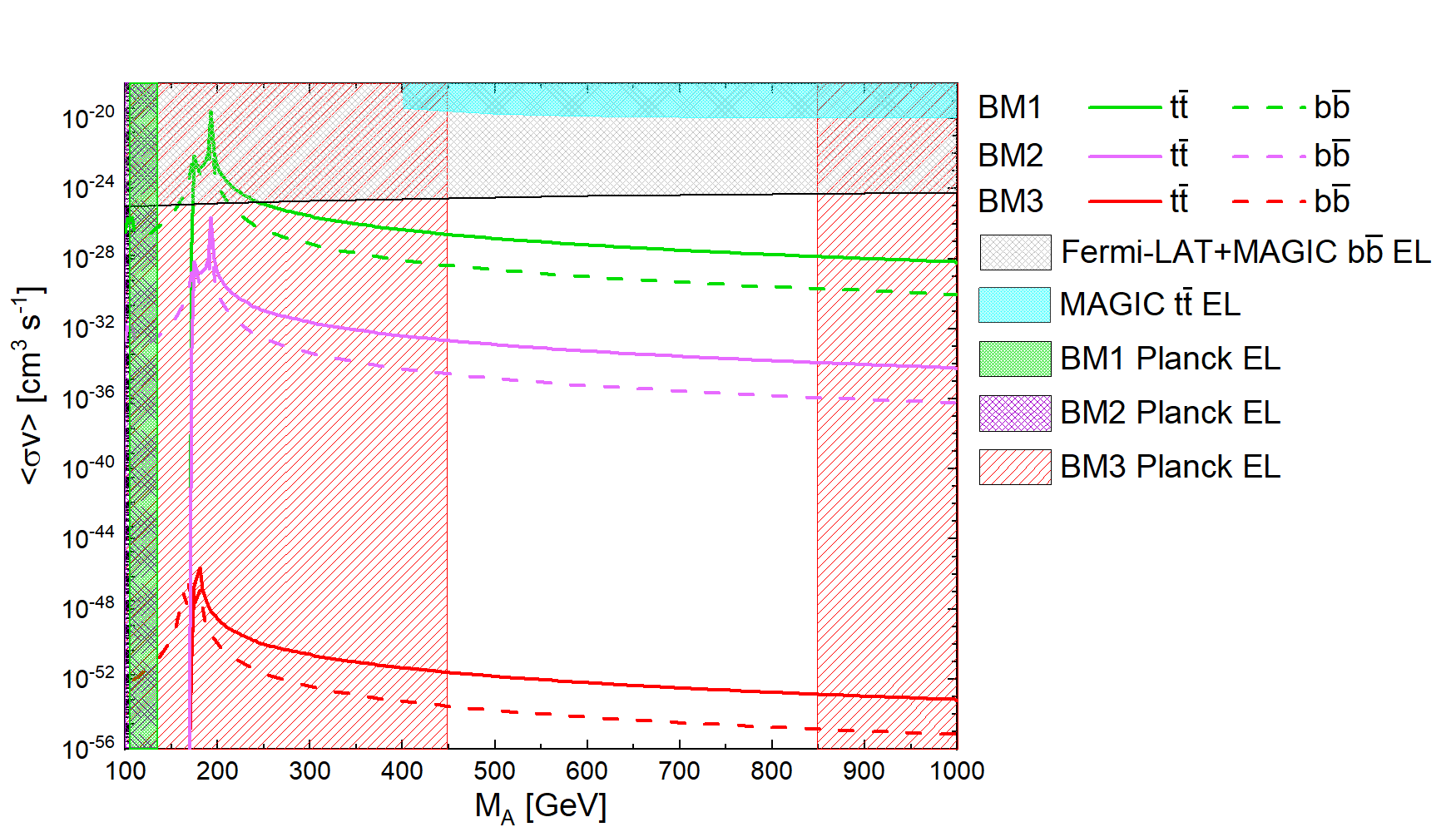}
\end{center}
\caption{The prompt photon annihilation cross section generated by DM annihilation into fermions are displayed for BM1, BM2 and BM3. The grey (limited from below by the solid black line) and cyan shaded regions stands for exclusion regions from $b\bar{b}$ by Fermi-LAT+MAGIC \cite{MAGIC:2016xys} and $t\bar{t}$ by MAGIC~\cite{Duangchan:2021vom}, respectively.}
\label{fermions}
\end{figure}

\begin{figure}[t]
\begin{center}
\includegraphics[width=0.9\textwidth]{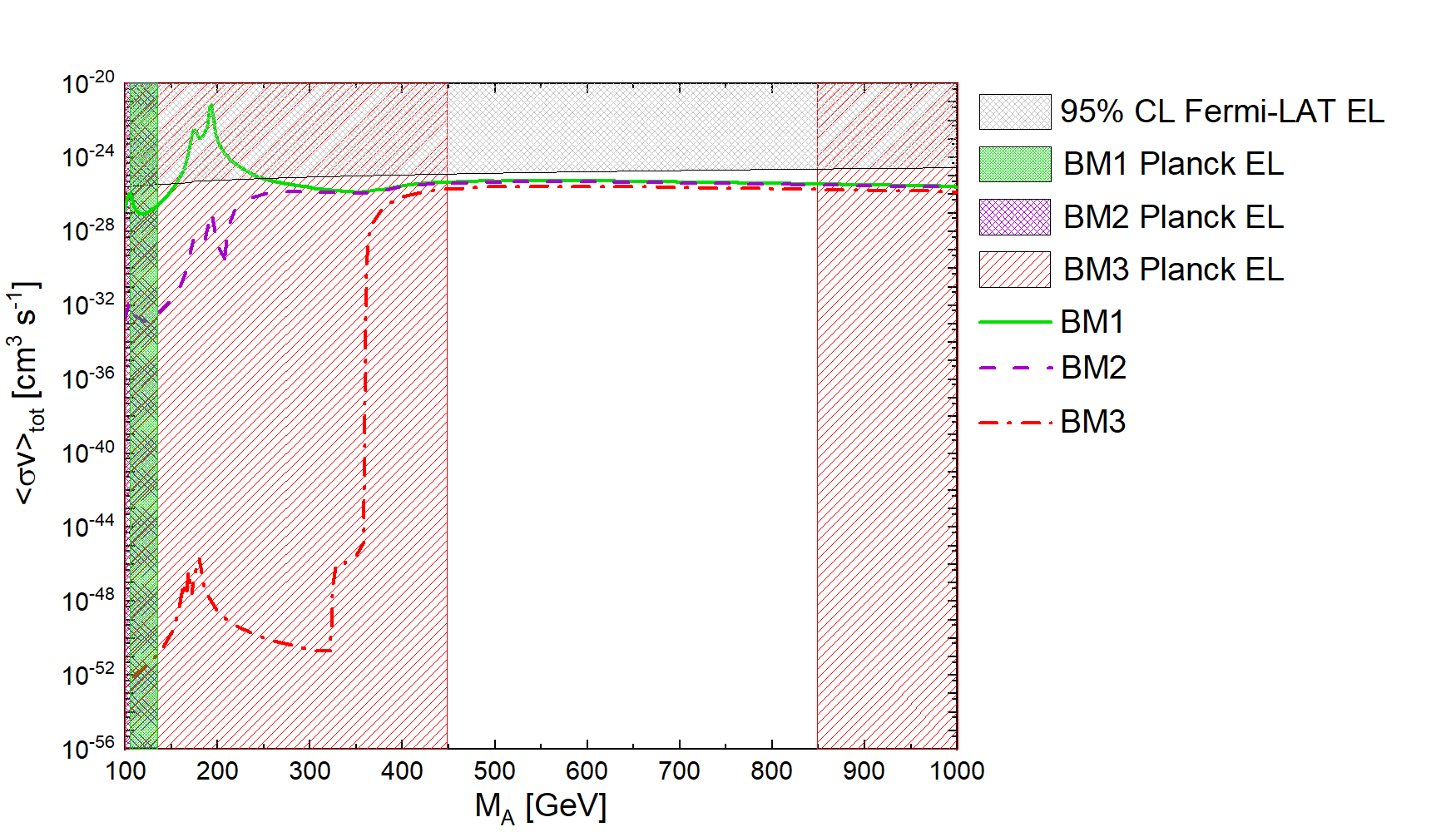}
\end{center}
\caption{Total annihilation cross sections for BM1, BM2 and BM3 are displayed. The shaded grey region stands for exclusion limit by the Fermi-LAT.}
\label{tot}
\end{figure}

In Figs.~\ref{hihi},~\ref{fermions} and~\ref{tot}, we display ${\langle \sigma v \rangle}-M_{A}$
 plane for prompt photon generated by pGDM annihilation into pair of fermions, Higgs bosons and the total cross-section ${\langle \sigma v \rangle}$ corresponding to all processes depicted in Fig.~\ref{relic-diag}. These plots include the combined Fermi-LAT exclusion limits on the ${\langle \sigma v \rangle-M_{A}}$ in the course of the combination of Segue I, Ursa Major II, Coma Berenices, Reticulum II, Ursa Minor and Draco~\cite{Arina:2020kko}. Also, the annihilation into pair of gauge bosons is omitted since these have infinitesimal contribution of order $10^{-70}\,{\rm cm}^3 {\rm s}^{-1}$. In these plots, peaks in the annihilation channels convey the enhancement due to the inter-mediating of heavier Higgs bosons. 	

As can be seen in Fig.~\ref{hihi}, the cross sections for annihilation of DM into $H_3H_3$ in the cases of BM1 and BM2 and $H_4H_4$ for BM3 have the highest values $\sim10^{-26}\,{\rm cm}^3 {\rm s}^{-1}$. In the case of BM3, this is 10 orders of magnitude larger than the cross section for annihilation into the other Higgs bosons. Expectedly, these contributions for BM1 and BM2 are identical. The high values of the cross sections for DM annihilation into $H_3H_3$ and $H_4 H_4$ in BM1 (similarly for BM2) and BM3 are due to the large couplings $g_{H_2 A A}\sim v_s \lambda_s \cos\alpha_5 \sin\alpha_6\sim v_s \lambda_s$ and $g_{H_4 A A}\sim v_s \lambda_s \cos\alpha_5 \cos\alpha_6\sim v_s \lambda_s$, respectively. The numerical values for these couplings are presented in Table~\ref{tab:hiA}. Moreover, the contributions of $H_1 H_1$ in BM1 (BM2) and BM3 are $\sim 10^{-61}\,{\rm cm}^3 {\rm s}^{-1}$ and $\sim10^{-60}\,{\rm cm}^3 {\rm s}^{-1}$ which in compare with the total cross section are negligible. This is a consequence of smallness of $g_{H_1 A A}$ and $g_{H_1 H_i H_j}$.

\begin{table}[]
\begin{center}
\begin{tabular}{|c|c|c|c|c|c|c|c|c|}
\hline 
 BM & $g_{H_1 AA}$ & $g_{H_2 AA}$ & $g_{H_3 AA}$ & $g_{H_4 AA}$ & $g_{H_1 H_1 AA}$ & $g_{H_2 H_2 AA}$ & $g_{H_3 H_3 AA}$ & $g_{H_4 H_4 AA}$ \\ \hline \hline 
 1,2 & 0.00 & 1.10 & 261.85& 94.71 & 0.00 & $-$0.15 & 0.56 & $-$0.09 \\
 3 & 0.00 & 0.00 & 0.00 & 263.85 & 0.00 & 0.00 & 0.00 & 0.53 \\
\hline 
\end{tabular}
\end{center}
\begin{center}
\begin{tabular}{|c|c|c|c|c|c|c|c|c|c|c|}
\hline 
 BM & $g_{H_1 H_1 H_1}$ & $g_{H_1 H_1 H_2}$ & $g_{H_1 H_1 H_3}$ & $g_{H_1 H_1 H_4}$ & $g_{H_1 H_2 H_2}$ & $g_{H_1 H_2 H_3}$ & $g_{H_1 H_2 H_4}$ \\ \hline \hline 
 1,2 & 191.14 & 0.00 & 0.00 & 0.00 & $-$1029.41& 7.65 & 1.88 \\
3 &191.14 & 0.00 & 0.00 & 0.00 & 81.41 & 0.00 & 0.00 \\
\hline 
\end{tabular}
\end{center}
\begin{center}
\begin{tabular}{|c|c|c|c|c|c|c|c|c|c|c|}
\hline 
 BM & $g_{H_1 H_3 H_3}$ & $g_{H_1 H_3 H_4}$ & $g_{H_1 H_4 H_4}$ & $g_{H_2 H_2 H_2}$ & $g_{H_2 H_2 H_3}$ & $g_{H_2 H_2 H_4}$ & $g_{H_2 H_3 H_3}$ \\ \hline \hline 
 1,2 & $-$32.18 & 78.63 & $-$162.65& 60.64 & $-$201.02 & 442.52 & 12.49 \\
3 &151.39 & 0.00 & 0.00 & $-$94.34 & 0.00 & 0.00 & $-$31.44 \\
\hline 
\end{tabular}
\end{center}
\begin{center}
\begin{tabular}{|c|c|c|c|c|c|c|c|c|c|c|}
\hline 
 BM & $g_{H_2 H_3 H_4}$ & $g_{H_2 H_4 H_4}$ & $g_{H_3 H_3 H_3}$ & $g_{H_3 H_3 H_4}$ & $g_{H_3 H_4 H_4}$ & $g_{H_4 H_4 H_4}$ \\ \hline \hline 
 1,2 & $-$7.42 & 16.52 & 675.46 & 372.14 & $-$351.97 & 1245.36 \\
3& 0.00 & 0.00 & 0.00& 0.00 & 0.00 & 791.57 \\
\hline 
\end{tabular}
 \end{center}
\caption{Cubic couplings in GeV units between Higgs bosons and DM for BM1, BM2 and BM3.}
\label{tab:hiA}
\end{table}

Concluding, in the cases of BM1 (BM2) and BM3 the largest $\langle \sigma v \rangle$ in the high-DM-mass regions $M_A>210,\, 360$ GeV, respectively, originate from DM annihilation into the heaviest Higgs boson pairs, while the highest shares for lighter DM are due to annihilation into $\bar{t}t$ and $\bar{b}b$. Note that for BM1, the low DM-mass region $\sim~155-250$ GeV is within the Fermi-LAT exclusion limit.
%%%%%%%%%%%%%%%%%%%%%%%%%%%%%%%
\begin{figure}[t]
\begin{center}
\includegraphics[width=0.76\textwidth]{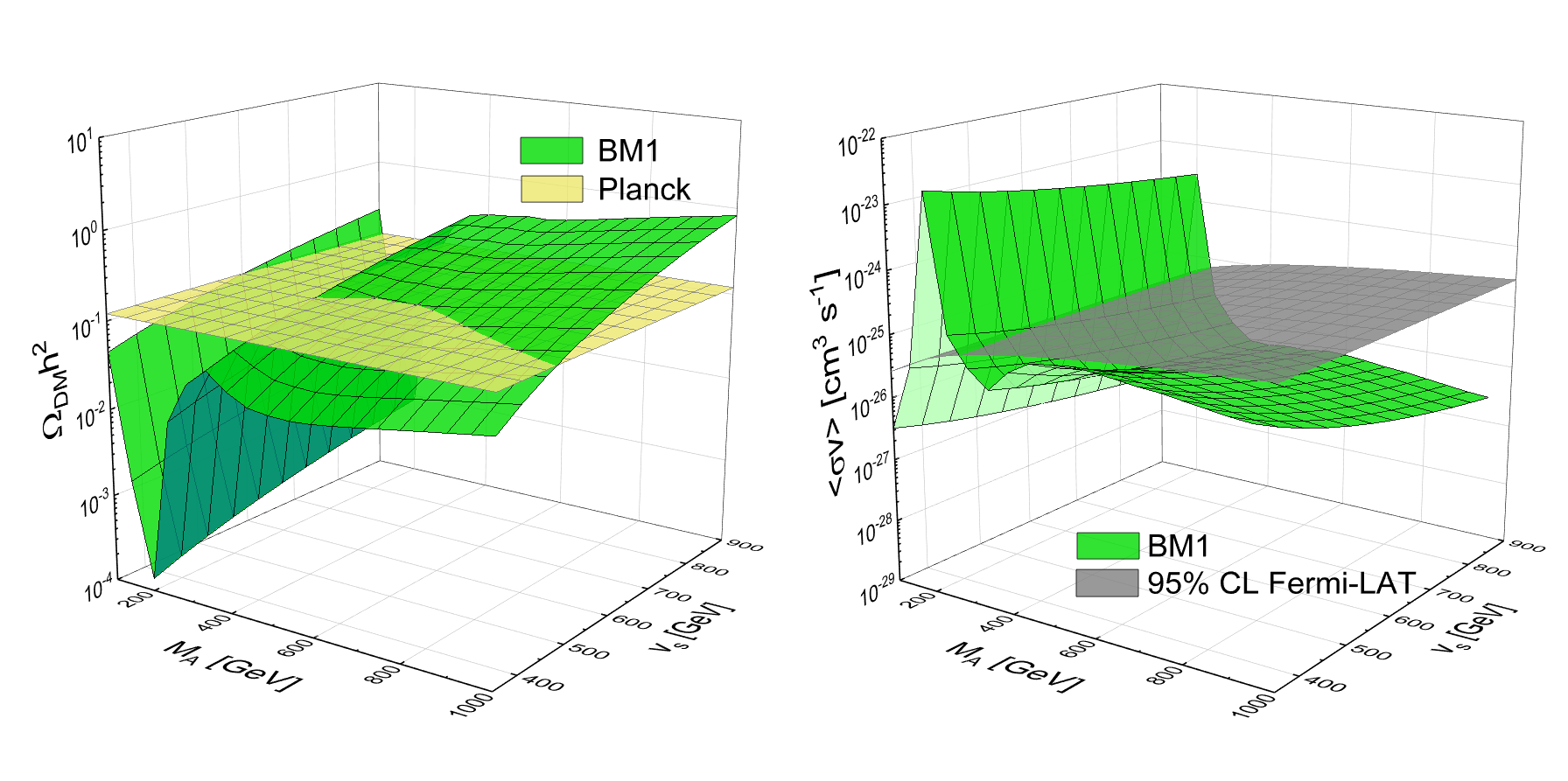}
\includegraphics[width=0.76\textwidth]{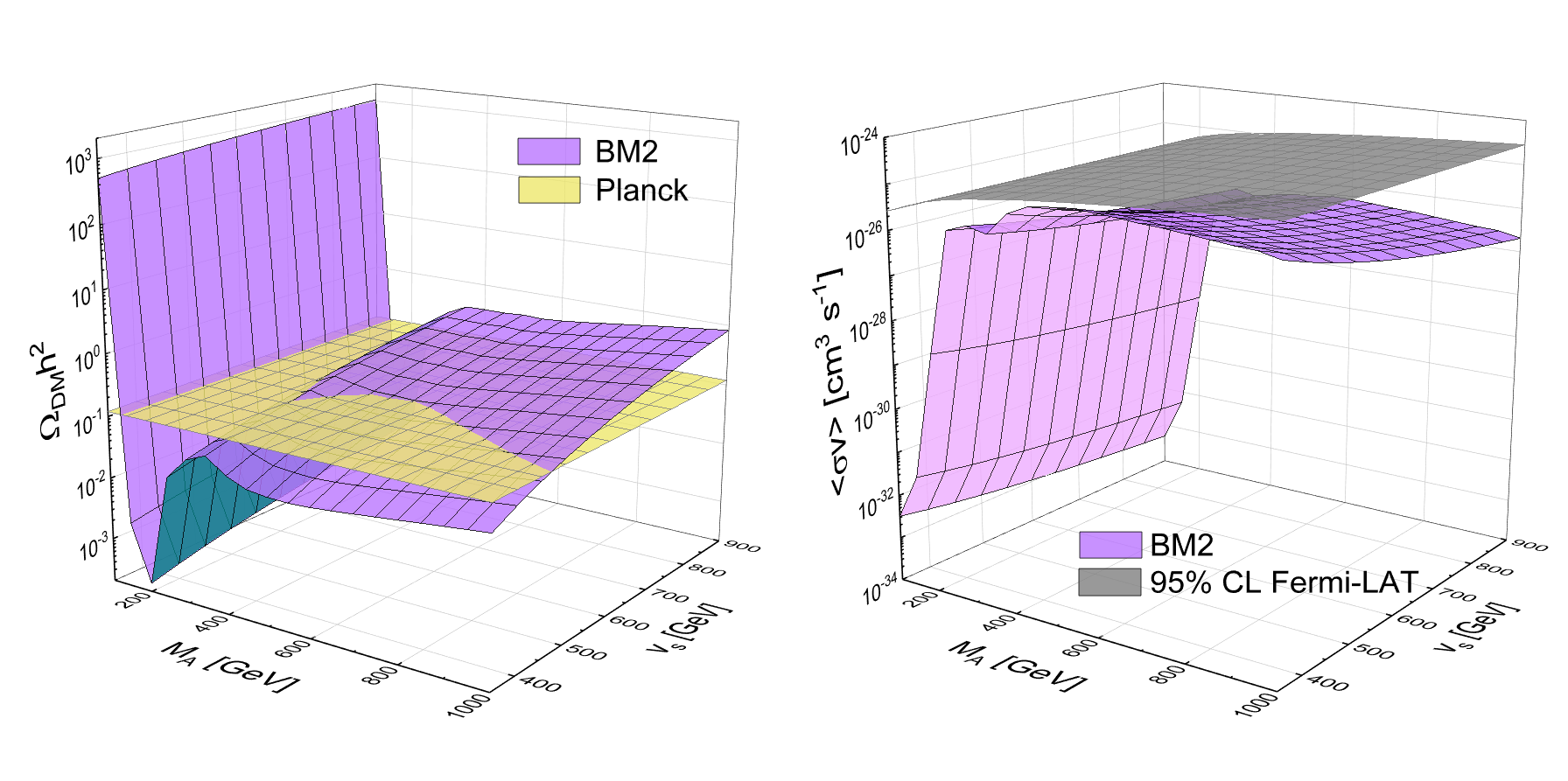}
\includegraphics[width=0.76\textwidth]{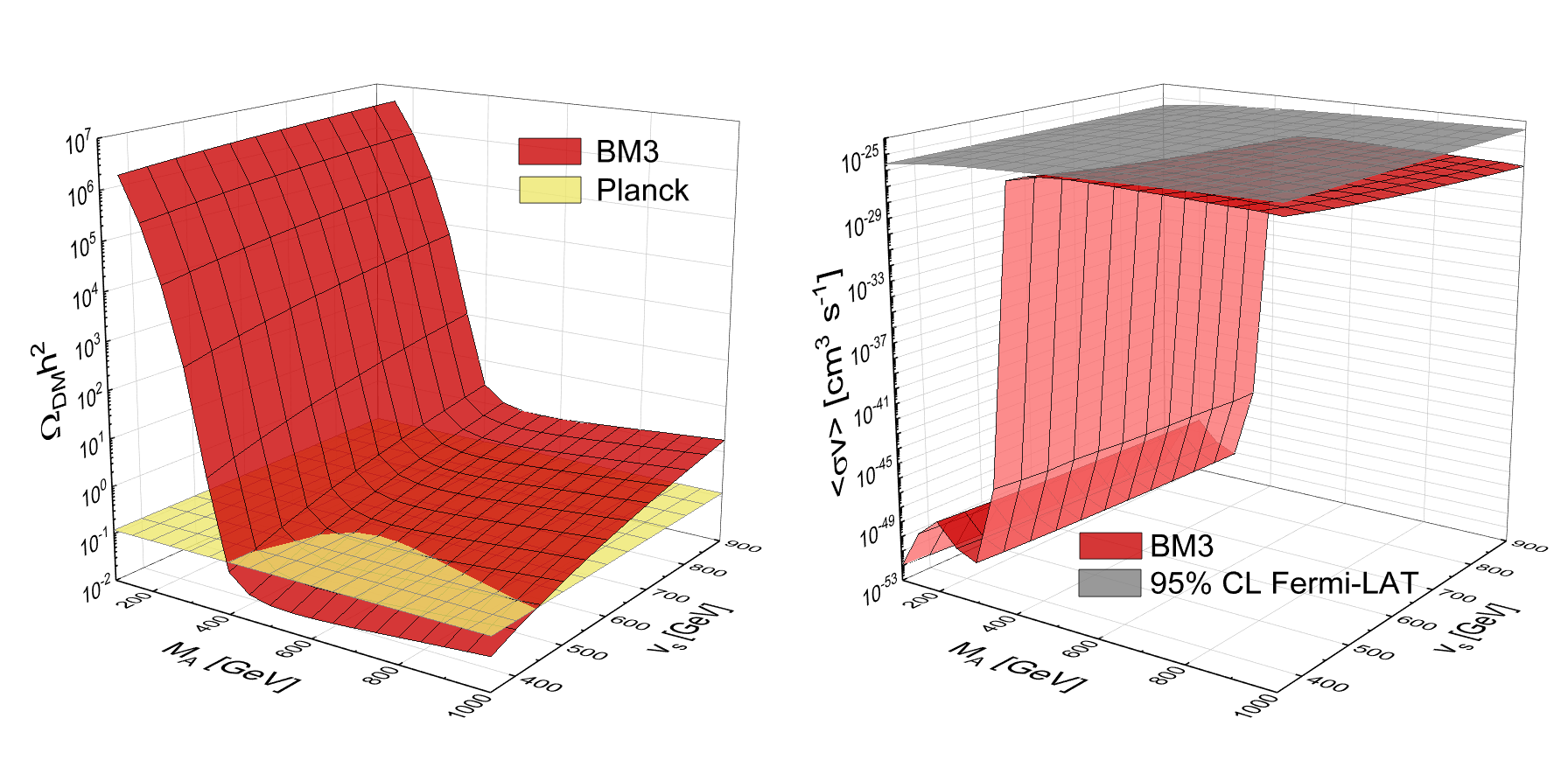}
\end{center}
\caption{The 3D plot for the profile of the pGDM relic density in terms of $v_s$ and $M_A$ (left panels) and for the total annihilation cross sections in terms of $v_s$ and $M_A$ (right panels) are displayed.}
\label{3D}
\end{figure}

Combining the results for BM1, BM2 and BM3 from Figs.~\ref{relic} and~\ref{tot}, we perceive that the DM masses which are fulfilling the Planck limit are also acceptable by Fermi-LAT upper bound. 
Note that DM annihilation into the heaviest Higgs bosons directly depends on $v_s$, which is fixed 
at relatively large value $\sim 450$~GeV. 
In Fig.~\ref{3D} we illustrate the results of a scan
with the input parameters \eqref{input} defined for BM1, BM2 and BM3 however allowing for a variation of $v_s$. The purpose of this final scan is to see whether annihilation cross sections are still below the Fermi-LAT upper limits even if $v_s$ is varying.
Since the leading contributions to annihilation into heavy Higgs bosons depend on $\lambda_s v_s \propto {1/v_s}$, larger $v_s$ implies abundance $\Omega_{DM}$ growing above the Planck limit at about $v_s \sim 400-500$~GeV.
As it is seen from the right panels at $v_s\gsim 400$~GeV predictions for the indirect detection decrease below the Fermi-LAT limits. Concluding, for $v_s\in[400, 500]$~GeV, DM masses satisfy the Planck data and stay below the Fermi-LAT limits.

%%%%%%%%%%%%%%%%%%%%%%%%%%%%%%%%%%%%%%%%%%%%%%%
\section{Conclusions}\label{conc}
We have studied an extension of the generic 2HDM with an extra complex singlet scalar filed (2HDMCS). The model allows both for CP-violation (CPV) in the scalar potential and accommodates extra sources of CPV in Yukawa couplings. The singlet $S$ is charged under a global $\rm U(1)$ symmetry which is softly broken by a mass term $\mu^2 S^2+\hc$ and consequently a pseudo-Goldston boson, a dark matter candidate (pGDM), arises.

It has been shown that even with the generic 2HDM Yukawa interactions the tree-level scattering amplitude of DM off nucleons at zero momentum transfer is vanishing within this model. In other words, the decoupling of DM from nucleais is present even in the generic 2HDM with violation of CP (CPV) located both in the potential and in Yukawa couplings. So, within this model, constraints imposed by negative results of direct detection experiments~\cite{Bertone:2004pz,Feng:2010gw,XENON:2018voc} are naturally satisfied.

The alignment limit (AL) in which the lightest Higgs boson has exactly Standard Model couplings to vector bosons has been formulated and discussed. It has been shown, in contrast to the $Z_2$-symmetric 2HDM, that even in the AL the three weak-basis CP invariants are non-zero, confirming the presence of CPV in the scalar potential.

The phenomenology of the model has been discussed. In particular, it has been shown that the electron EDM can either be dominated by 1-loop corrections (when bosonic and fermionic 2-loop contributions cancel each other), by standard Barr-Zee 2-loop contributions or by a mixture of them. It has been shown that the experimental upper limit on the electron EDM does not prohibit large values for CP invariants ${\rm Im}J_{1,2,3}$ leaving a substantial amount of CPV in the potential. 

Three benchmark points have been defined in the parameter space. For them, detailed analyses of relic DM abundance and indirect detection experiments have been performed in the AL. In particular, it has been shown that DM annihilation into fermions dominates in the low DM mass regions while annihilation into heavy Higgs bosons is maximal in high DM mass regions. Having validated our results with experimental constraints provided by Planck collaboration and Fermi-LAT experiment, we have presented regions of DM masses for each benchmark scenario that are in complete agreement with the Planck limit on relic density and indirect detection limits.

\section*{Acknowledgements}
\noindent
The work of BG is supported in part by the National Science Centre (Poland) as a research project, decisions no 2017/25/B/ST2/00191 and 2020/37/B/ST2/02746. 
The work of ND is supported by the Polish National Science Centre, decision no 2017/25/B/ST2/00191, by the National Natural Science Foundation of China (NSFC) under grants No. 12022514, No. 11875003 and No. 12047503, and CAS Project for Young Scientists in Basic Research YSBR-006, by the Development Program of China under Grant No. 2020YFC2201501 (2021/12/28) and by the CAS~President’s~International~Fellowship~Initiative~(PIFI)~grant.
\newpage
\appendix
%%%%%%%%%%%%%%%%%%%%%%%%%%%%%%%%%%%%%%%%%%%%
\section{Yukawa Couplings}\label{sec:YI}
The most general Yukawa Lagrangian in terms of the weak eigenstates reads (\cite{Haber:2006ue} and \cite{Grzadkowski:2018ohf})
\begin{align}
 - \mathcal{L}_Y\ &=\overline{q}_L \big(\Phi_1({\eta_{1,d}^{0}})^\dagger + \Phi_2 (\eta_{2,d}^{0})^\dagger \big) d_R\: +\: \overline{q}_L \big(\widetilde{\Phi}_1 \eta_{1,u}^{0} + \widetilde{\Phi}_2 \eta_{2,u}^{0} \big) u_R
 \nonumber \\
 & \:+\: {\overline{L}_L^{i}} \big(\Phi_1 (\eta_{1,l}^{0})^\dagger + \Phi_2 (\eta_{2,l}^{0})^\dagger \big) \,l_R\
 +\ \mathrm{H.c.\,},
\end{align}
where $q_L = \big( u_L\,, d_L\big)^{\sf T}$, $L_L = \big( \nu_L\,, l_L\big)^{\sf T}$ (with $i=1,2,3$), $\widetilde{\Phi}_{1,2} = i\sigma^2 \Phi^*_{1,2}$ is the hypercharge-conjugate of $\Phi_{1,2}$, and $\sigma^2$ is the second Pauli matrix. 
Now, splitting doublets into their charged and neutral components the Yukawa
couplings may be re-expressed in terms of mass eigenstates as
\begin{align}
- \mathcal{L}_Y\ &=\overline{d}_L \big(\Phi_1^0({\eta_{1,d}})^\dagger + \Phi_2^0 (\eta_{2,d})^\dagger \big) d_R\: -\: \overline{d}_L K \big(\Phi_1^-({\eta_{1,d}})^\dagger + \Phi_2^- (\eta_{2,d})^\dagger \big) u_R
 \nonumber \\
 &\: +\: \overline{u}_L \big((\Phi_1^0)^* \, {\eta_{1,u}} + \Phi_2^0 \,\eta_{2,u} \big) u_R\: +\: \overline{u}_L K^\dagger \big(\Phi_1^+\, {\eta_{1,u}} + \Phi_2^+ \, \eta_{2,u} \big) d_R
 \nonumber \\
 & \:+\: \overline{l}_L \big(\Phi_1^0({\eta_{1,l}})^\dagger + \Phi_2^0 (\eta_{2,l})^\dagger \big) l_R\: -\: \overline{l}_L K \big(\Phi_1^-({\eta_{1,l}})^\dagger + \Phi_2^- (\eta_{2,l})^\dagger \big) l_R +\ \mathrm{H.c.\,},
\end{align}
where $K$ is the CKM-matrix. 
It is convenient to decompose $\eta_{a,f}$ into a diagonal part $\kappa_f \equiv \sqrt{2}/v\cdot\text{diag}(m_{f_1},m_{f_2},m_{f_3})$ and the leftover $\rho^f$ following notation of \cite{Haber:2010bw}:
\begin{align}
\eta_{1,{f}}=\kappa_f {v_1\over v}-\rho^f {v_2\over v}, \label{eta1}
\\
\eta_{2,{f}}=\kappa_f {v_2\over v}+\rho^f {v_1\over v}\,, \label{eta2}
\end{align}
where $f=u,d,l$. Here, we choose $\rho_f$ diagonal to avoid irrelevant issue of FCNC. 

Consequently, the couplings to the up- and down-type quarks and leptons may be written as follows
{\allowdisplaybreaks
\begin{align}
\bar{u}_k u_k H_j \; &: \; - {m_{u_k} \over v^2} e_j - {1 \over 2\sqrt{2}v}
	\Big[ \big( \rho^u_{kk} \big)^* (1-\gamma_5) \mathcal{F}_j^* + \rho^u_{kk} (1+\gamma_5) \mathcal{F}_j \Big],
\non \\
\bar{d}_k d_k H_j \; &: \; - {m_{d_k} \over v^2} e_j - {1 \over 2\sqrt{2}v}
	\Big[ \big( \rho^d_{kk} \big)^* (1+\gamma_5) \mathcal{F}_j^* + \rho^d_{kk} (1-\gamma_5) \mathcal{F}_j \Big],
\non \\
	\bar{l}_k l_k H_j \; &: \; - {m_{l_k} \over v^2} e_j -{1 \over 2\sqrt{2}v}
	\Big[ \big( \rho^l_{kk} \big)^* (1+\gamma_5) \mathcal{F}_j^* + \rho^l_{kk} (1-\gamma_5) \mathcal{F}_j \Big],
\non \\
\bar{u}_m d_k H^+ \; &: \; {1 \over 2} K_{mk}
	\Big[ \big( \rho^u_{mm} \big)^* (1-\gamma_5) - \big( \rho^d_{kk} \big)^* (1+\gamma_5) \Big],
\label{f-Higgs} \\
\bar{d}_k u_m H^- \; &: \; {1 \over 2} K_{mk}^*
	\Big[ \rho^u_{mm} (1+\gamma_5) - \rho^d_{kk} (1-\gamma_5) \Big]
\non \\
\bar{l}_k \nu_k H^- \; &: \;-{1 \over 2} \rho^l_{kk} (1-\gamma_5),
\non \\
\bar{\nu}_k l_k H^+ \; &: \;- {1 \over 2} \big( \rho^l_{kk} \big)^* (1+\gamma_5),
\non 
\end{align}
}
where the parameter $\mathcal{F}_j$ is defined as
 \begin{align}
\mathcal{F}_j=& R_{j2}v_1-R_{j1}v_2-iR_{j3}v,
\label{fi}
 \end{align} 
with $j=1,2,3,4$.

%%%%%%%%%%%%%%%%%%%%%%%%%%%%%%%%%%%%%%%%%%
\section{Dark Matter and Higgs Bosons Trilinear Couplings}
\label{app:couplings}

The Lagrangian involving mixing of DM candidate $A$ and Higgs bosons in the limit $G^{\pm,0} \to 0$ is given as follows
\begin{align}
\mathcal{L}_{A^2} &= {1\over 2} \Big[ R_{i1} \kappa_1 v_1+ R_{i2} \kappa_2 v_2 + 
R_{i4} v_s \lambda_s \Big] A^2 H_i \nonumber \\
&+{1\over 2 v^2} \Big[ R_{i1}^2 v^2 \kappa_1+ R_{i2}^2 v^2 \kappa_2 +
 R_{i3}^2 (v_2^2 \kappa_1+ v_1^2 \kappa_2) + 
 R_{i4}^2 v^2 \lambda_s \Big] A^2 H_i^2
 \nonumber \\
&+{1\over 2 v^2} \Big[R_{i1} R_{j1} v^2 \kappa_1+ 
 R_{i2} R_{j2} v^2 \kappa_2 
 + R_{i3} R_{j3} (v_2^2 \kappa_1+ v_1^2 \kappa_2) + R_{i4} R_{j4} v^2 \lambda_s \Big] A^2 H_i H_j. 
 \label{lag:A2}
\end{align}
$\mathcal{L}_{A^2}$ can be re-written in the following form
\begin{align}
\mathcal{L}_{A^2} &= {1\over 2}\big(g_{H_iA^2} H_i+ g_{{H_i}^2A^2} H_i^2
+g_{H_i H_jA^2} H_i H_j + g_{A^2 H^+ H^-} H^+ H^-\big) A^2.
 \label{lag:A2}
\end{align}

In addition to the couplings identified in Eqs.~\eqref{e1},~\eqref{AL-hqq}, and~\eqref{ghh-al}, the other relevant couplings for the annihilation channels identified in Fig.\ref{relic-diag} are given in this appendix.

The Higgs couplings to the pair of gauge bosons in term of $e_1$ as defined in \eqref{e1} may be given by
\begin{align}
g_{H_iZ_\mu Z_\nu} &= {-g^2 \over 2 \cos^2{\theta_W}} e_i g_{\mu \nu},
\qquad
 g_{H_iW^{+}_\mu W^{-}_\nu} = {-g^2 \over 2} e_i g_{\mu \nu},
\end{align}
whilst the other couplings $e_2$, $e_3$ and $e_4$ vanish.
The coupling $ZH_iH_j$ takes on the following form
\begin{align}
g_{ZH_iH_j}=& {i g \over 2 v \cos{\theta_W}}(p_i-p_j)^\mu \big[(R_{i2}R_{j3}-R_{i3}R_{j2}){v_1}+
 (R_{i3}R_{j1}-R_{i1}R_{j3}){v_2}\big]
 \nonumber \\
=& {-i g \over 2 v^2 \cos{\theta_W}}(p_i-p_j)^\mu {\rm Im}(\mathcal{F}_i\mathcal{F}_j^*).
 \end{align}
Thus $\mathcal{F}_i\mathcal{F}_j^*$ may be expressed in the following form 
 \begin{align}
\mathcal{F}_i\mathcal{F}_j^*=&(\delta_{ij} - R_{i4}R_{j4})v^2-e_i e_j-iv \big[(R_{i2}R_{j3}-R_{i3}R_{j2}){v_1}+
 (R_{i3}R_{j1}-R_{i1}R_{j3}){v_2}\big].
 \end{align} 
 
Trilinear couplings of neutral scalars to the charged scalar $H^{\pm}$ may enter the $H_i \gamma \gamma$ coupling through the $H^{\pm}$ loop. These coupling may be obtained as
\begin{align}
g_{H_i H^+ H^-} ={1\over v^2} \Bigg[ &R_{i1} v_1 v_2^2 (\lambda_1 - \lambda_4-{\rm Re}\lambda_5 +{v_1^2\over v_2^2} \lambda_3+{v_2^2-2v_1^2\over v_1 v_2}{\rm Re}\lambda_6+{v_1\over v_2}{\rm Im} \lambda_7)
 \nonumber \\
+& R_{i2} v_1^2 v_2 (\lambda_2 - \lambda_4-{\rm Re}\lambda_5+{v_2^2\over v_1^2} \lambda_3+{v_1^2-2v_2^2\over v_1 v_2}{\rm Re}\lambda_7+{v_2\over v_1}{\rm Im} \lambda_6) \nonumber \\
+&
 R_{i3} v(v_1 v_2 {\rm Im} \lambda_5 + {v_2^2} {\rm Im} \lambda_6 + {v_1^2} {\rm Im} \lambda_7)+ R_{i4} v_s (v_2^2 \kappa_1 + v_1^2 \kappa_2) \Bigg],
\end{align}
with $i=1,2,3,4$.

Finally, the trilinear neutral Higgs couplings may be expressed in the following form
{\allowdisplaybreaks
\begin{align}
\small
\begin{autobreak}
g_{H_i H_j H_k}=
{1\over 2 v^3}\bigg[2 \lambda_1 v v_1 \Big(3 R_{i1} R_{j1} R_{k1} v^2 
+(R_{i3} R_{j3} R_{k1} 
+R_{i3} R_{j1} R_{k3} 
+R_{11} R_{j3} R_{k3}) v_2^2 \Big) 
+ 2 \lambda_2 v \Big(3 R_{i2} R_{j2} R_{k2} v^2 
+ (R_{i3} R_{j3} R_{k2} 
+ R_{i3} R_{j2} R_{k3} 
+ R_{i2} R_{j3} R_{k3}) v_1^2 \Big) v_2 
+ 2 \lambda_{345} v \Big((R_{i2} R_{j2} R_{k1} 
+ R_{i2} R_{j1} R_{k2} 
+ R_{i1} R_{j2} R_{k2}) v^2 v_1 + (R_{i3} R_{j3} R_{k1} 
+ R_{i3} R_{j1} R_{k3}
+ R_{i1} R_{j3} R_{k3}) v_1^3 
+ (R_{i2} R_{j1} R_{k1} 
+ R_{i1} R_{j2} R_{k1}
+ R_{i1} R_{j1} R_{k2}) v^2 v_2 
+ (R_{i3} R_{j3} R_{k2} 
+ R_{i3} R_{j2} R_{k3} 
+ R_{i2} R_{j3} R_{k3}) v_2^3\Big) 
- 4 {\rm Re \,}\lambda_{5} v^3 \Big(R_{j3} R_{k3} (R_{i1} v_1 
+ R_{i2} v_2) 
+R_{i3} (R_{j3} R_{k1} v_1 
+ R_{j1} R_{k3} v_1 
+ R_{j3} R_{k2} v_2
+ R_{j2} R_{k3} v_2)\Big) 
- 2 {\rm Im \,}\lambda_{5} \Big((R_{i3} R_{j2} R_{k1} 
+ R_{i2} R_{j3} R_{k1} 
+ R_{i3} R_{j1} R_{k2} 
+ R_{i1} R_{j3} R_{k2} 
+ R_{i2} R_{j1} R_{k3} 
+ R_{i1} R_{j2} R_{k3}) v^2 v_1^2 
+ (R_{i3} R_{j1} R_{k1}
+ R_{i1} R_{j3} R_{k1} 
+ R_{i3} R_{j2} R_{k2}
+ R_{i2} R_{j3} R_{k2}
+ R_{i1} R_{j1} R_{k3} 
+ R_{i2} R_{j2} R_{k3}) v^2 v_1 v_2 
- 3 R_{i3} R_{j3} R_{k3} v_1^3 v_2 
+ (R_{i3} R_{j2} R_{k1}
+ R_{i2} R_{j3} R_{k1} 
+ R_{i3} R_{j1} R_{k2} 
+ R_{i1} R_{j3} R_{k2}
+ R_{i2} R_{j1} R_{k3}
+ R_{i1} R_{j2} R_{k3}) v^2 v_2^2
- 3 R_{i3} R_{j3} R_{k3} v_1 v_2^3\Big) 
+ 6 \lambda_{s} R_{14} R_{24} R_{34} v^3 v_s 
+ 2 \kappa_1 v \Big((R_{14} R_{24} R_{k1} 
+ R_{14} R_{j1} R_{34} 
+ R_{i1} R_{24} R_{34}) v^2 v_1 
+ (R_{14} R_{j1} R_{k1} 
+ R_{i1} R_{24} R_{k1}
+R_{i1} R_{j1} R_{34}) v^2 v_s 
+ (R_{14} R_{j3} R_{k3} 
+ R_{i3} R_{24} R_{k3}
+ R_{i3} R_{j3} R_{34}) v_2^2 v_s\Big) 
+ 2 \kappa_2 v \Big((R_{14} R_{24} R_{k2} 
 + R_{14} R_{j2} R_{34}
+ R_{i2} R_{24} R_{34}) v^2 v_2 + (R_{14} R_{j2} R_{k2} 
+ R_{i2} R_{24} R_{k2}
+R_{i2} R_{j2} R_{34}) v^2 v_s 
+ (R_{14} R_{j3} R_{k3} 
+ R_{i3} R_{24} R_{k3}
+ R_{i3} R_{j3} R_{34}) v_1^2 v_s\Big) 
+ 2 {\rm Re\,}\lambda_{6} v \Big(3 (R_{i2} R_{j1} R_{k1}
+ R_{i1} R_{j2} R_{k1}
+ R_{i1} R_{j1} R_{k2}) v^2 v_1 
+ 3 R_{i1} R_{j1} R_{k1} v^2 v_2 - 2 (R_{i3} R_{j3} R_{k1} 
+ R_{i3} R_{j1} R_{k3} 
+ R_{i1} R_{j3} R_{k3}) v_1^2 v_2 
+ (R_{i3} R_{j3} R_{k2} 
+ R_{i3} R_{j2} R_{k3} 
+ R_{i2} R_{j3} R_{k3}) v_1 v_2^2 
+ (R_{i3} R_{j3} R_{k1} 
+ R_{i3} R_{j1} R_{k3} 
+ R_{i1} R_{j3} R_{k3}) v_2^3\Big) 
- 2 {\rm Im \,}\lambda_{6} \Big(3 (R_{i3} R_{j1} R_{k1} 
+ R_{i1} R_{j3} R_{k1} 
+ R_{i1} R_{j1} R_{k3}) v^2 v_1^2 + (R_{i3} R_{j2} R_{k1}
+ R_{i2} R_{j3} R_{k1} 
+ R_{i3} R_{j1} R_{k2} 
+ R_{i1} R_{j3} R_{k2} 
+ R_{i2} R_{j1} R_{k3} 
+ R_{i1} R_{j2} R_{k3}) v^2 v_1 v_2 
+ ((R_{i3} R_{j1} R_{k1} 
+ R_{i1} R_{j3} R_{k1}
+ R_{i1} R_{j1} R_{k3}) v^2 
+ 3 R_{i3} R_{j3} R_{k3} v_1^2) v_2^2 
+ 3 R_{i3} R_{j3} R_{k3} v_2^4\Big) 
+ 2 {\rm Re\,}\lambda_{7} v \Big(3 R_{i2} R_{j2} R_{k2} v^2 v_1 
+ (R_{i3} R_{j3} R_{k2}
+ R_{i3} R_{j2} R_{k3} 
+ R_{i2} R_{j3} R_{k3}) v_1^3 
+ 3 (R_{i2} R_{j2} R_{k1} 
+ R_{i2} R_{j1} R_{k2} 
+ R_{i1} R_{j2} R_{k2}) v^2 v_2 + (R_{i3} R_{j3} R_{k1}
+ R_{i3} R_{j1} R_{k3} 
+ R_{i1} R_{j3} R_{k3}) v_1^2 v_2 
- 2 (R_{i3} R_{j3} R_{k2} 
+ R_{i3} R_{j2} R_{k3} 
+ R_{i2} R_{j3} R_{k3}) v_1 v_2^2\Big) 
- 2 {\rm Im \,}\lambda_{7} \Big((R_{i3} R_{j2} R_{k2} 
+ R_{i2} R_{j3} R_{k2}
+ R_{i2} R_{j2} R_{k3}) v^2 (v_1^2 +3 v_2^2)
+ 3 R_{i3} R_{j3} R_{k3} v_1^2(v_1^2-v_2^2) 
+ (R_{i3} R_{j2} R_{k1} 
+ R_{i2} R_{j3} R_{k1}
+ R_{i3} R_{j1} R_{k2}
+ R_{i1} R_{j3} R_{k2} 
+ R_{i2} R_{j1} R_{k3} 
 + R_{i1} R_{j2} R_{k3}) v^2 v_1 v_2
\Big) \bigg].
\end{autobreak}
\end{align}
}

%%%%%%%%%%%%%%%%%%%%%%%%%%%%%%%%%%%%%%%%%%%%
\section{Details for CP Invariants \boldmath{$J_1$}, \boldmath{$J_2$} and \boldmath{$J_3$}}\label{app:Ji}
As discussed in Section \ref{CPI}, the CP-sensitive basis-invariant quantities can be obtained using Eqs.~\eqref{j1},~\eqref{j2} and~\eqref{j3}. The complete form of these relations are given below:
\begin{align}
{\rm Im\,} J_i = \sum^{p+q+r<3}_{p,q,r\geq0} a_{i}^{pqr} ({\rm Im\,} \lambda_5)^p ({\rm Im} \lambda_6\,)^q ({\rm Im\,} \lambda_7)^r.
\end{align}

The non zero coefficient for ${\rm Im\,} J_1$ are $a^{100}_{1}$, $a^{010}_{1}$ and $a^{001}_{1}$, as
{\allowdisplaybreaks
\begin{align}
\begin{autobreak}
a^{100}_{1} = 
\frac{v_{1} v_{2}}{v^4} \Big[
v_{1} v_{2} \big(\lambda_{2} - \lambda_{1} \big)
+2({\rm Re\,} \lambda_6+{\rm Re\,} \lambda_7 ) (v_{1}^2
-v_{2}^2 )\Big],
\end{autobreak}
\\
\begin{autobreak}
a^{010}_{1} = 
\frac{1}{v^4} \Big[v_{1} v_{2} \big(2 (\kappa_{1}-\kappa_{2}) v_{s}^2
+v_{1}^2 (\lambda_{1}
+\lambda_{2}-2\lambda_{345})
+2 v_{2}^2 (-\lambda_{2}+\lambda_{345})
+4( {\rm Re\,} \lambda_6-{\rm Re\,} \lambda_7) v_{1} v_{2}\big)
+2 {\rm Re\,} \lambda_7 \big(v_{1}^2
-v_{2}^2\big)^2 \Big],
\end{autobreak}
\\
\begin{autobreak}
a^{001}_{1}= 
\frac{1}{v^4} \Big[v_{1} v_{2} \big(2 (\kappa_{1}-\kappa_{2}) v_{s}^2
-v_{2}^2 (\lambda_{1}
+\lambda_{2}-2\lambda_{345})
+2 v_{1}^2 (\lambda_{1}-\lambda_{345})
-4( {\rm Re\,} \lambda_7-{\rm Re\,} \lambda_6) v_{1} v_{2}\big)
-2 {\rm Re\,} \lambda_6 \big(v_{1}^2
-v_{2}^2\big)^2 \Big],
\end{autobreak}
\end{align}
}
%%%%%%%%%%%%%%%%%%%%%%%%%%%%%%%%%%%%%%%%%%%%%%%%%%%%%%%%%%%%%%%%%%%%
The non-zero coefficients in ${\rm Im\,} J_2$ are given, as 
{\allowdisplaybreaks
\small
\begin{align}
\begin{autobreak}
a^{100}_{2}= 
\frac{v_{2}}{4 v^8} \bigg[ 4 \lambda_{3} v_{1} \Big(v_{2} (2 v^2 v_{1} v_{s}^2 (\kappa_{1}-\kappa_{2})
-2 \lambda_{2} v_{1} v_{2}^4
+2 \lambda_{4} v_{1} (v_{2}^4-v_{1}^4)
+2{\rm Re\,} \lambda_7 v_{2} (v_{1}^2-v_{2}^2)^2)
-2{\rm Re\,} \lambda_6 (v_{1}^3-v_{1} v_{2}^2)^2 \Big)
+4 v_{1} v_{s}^2 \Big(2 \lambda_{4} v_{1}^3 v_{2} (\kappa_{1}-\kappa_{2})
+2 \lambda_{4} v_{1} v_{2}^3 (\kappa_{1}-\kappa_{2})
+v_{1}^2 v_{2}^2 \big(2 {\rm Re\,} \lambda_7 (2 \kappa_{1}+\kappa_{2})
-2 {\rm Re\,} \lambda_6 (\kappa_{1}+2 \kappa_{2}) \big)
+2 {\rm Re\,} \lambda_6 v_{1}^4 (2 \kappa_{1}-\kappa_{2})
+2 {\rm Re\,} \lambda_7 v_{2}^4 \big(\kappa_{1}-2 \kappa_{2}\big) \Big)
+4 v_{1}^2 v_{2} v_{s}^4 (\kappa_{1}
-\kappa_{2}) (\kappa_{1}+\kappa_{2})
-4 \lambda_{1}^2 v_{1}^6 v_{2}
+8 \lambda_{1} v_{1}^4 \big(v_{1}^2 v_{2} (\lambda_{3}+\lambda_{4})
-{\rm Re\,} \lambda_5 v_{2}^3
+v_{1} v_{2}^2 (2{\rm Re\,} \lambda_7
-4 {\rm Re\,} \lambda_6)
+2{\rm Re\,} \lambda_6 v_{1}^3\big)
+2 v_{1}^2 v_{2}^5 \big( 
2 \lambda_{2}^2 
-4 \lambda_{2} \lambda_{4} 
+2 \lambda_{4}^2 
-2 |\lambda_{5}|^2 
-4 {\rm Re\,} \lambda_6 {\rm Re\,} \lambda_7 
-16 |\lambda_{7}|^2 
\big)
+ v_{1}^4 v_{2}^3 \big(
8 \lambda_{2} {\rm Re\,} \lambda_5 +60 |\lambda_{7}|^2 
\big)
- 4 v_{1}^3 v_{2}^4 \big(
4 \lambda_{2} ({\rm Re\,} \lambda_6-2{\rm Re\,} \lambda_7) 
+2 \lambda_{4} ({\rm Re\,} \lambda_6 +2 {\rm Re\,} \lambda_7)
+8{\rm Re\,} \lambda_5 {\rm Re\,} \lambda_6
\big)
- 4 v_{1} v_{2}^6 \big(
2 {\rm Re\,} \lambda_7 (2 \lambda_{2}-\lambda_{4}) 
\big)
+4 \lambda_{3}^2 v_{1}^2 v_{2} (v_{2}^4 -v_{1}^4)
- 2 v_{1}^6 v_{2} \big(
2 \lambda_{4}^2 
-2 |\lambda_{5}|^2
- 4{\rm Re\,} \lambda_6 {\rm Re\,} \lambda_7 
\big)
+ 4 v_{1}^5 v_{2}^2 \big(
2 \lambda_{4}(2 {\rm Re\,} \lambda_6 + {\rm Re\,} \lambda_7 )
+8 {\rm Re\,} \lambda_5 {\rm Re\,} \lambda_7
\big)
-8 \lambda_{4} {\rm Re\,} \lambda_6 v_{1}^7
+4 |\lambda_{7}|^2 v_{2}^7 
-4 |\lambda_{6}|^2 v_{2} \big({v_{1}^8\over v_2^2}-8 v_{1}^6 
+15 v_{1}^4 v_{2}^2\big) \bigg],
\end{autobreak}
\\
\begin{autobreak}
a^{010}_{2} = 
\frac{v_{1}v_{2}}{v^8} \bigg[
\lambda_{1} v_{1}^3 (2 v_{1} v_{s}^2 (3 \kappa_{1}-2 \kappa_{2})
-2 v_{1} v_{2}^2 ( 2 \lambda_{2} 
-3 \lambda_{3} 
-3 \lambda_{4})
- 2 v_{1}^3 ( \lambda_{3} + \lambda_{4})
+2v_{1}^2 v_{2} (5 {\rm Re\,} \lambda_6 -4 {\rm Re\,} \lambda_7 )
+6 {\rm Re\,} \lambda_7 v_{2}^3)
-2 v_{1} v_{s}^2 \Big(v_{1} v_{2}^2 \big( 2 \kappa_{1} \lambda_{2}
-4 \kappa_{1} (\lambda_{3} +\lambda_{4}) -\kappa_{2} \lambda_{2}
+3 \kappa_{2} (\lambda_{3} +\lambda_{4}) \big)
-4 v_{1}^2 v_{2} (\kappa_{1}-\kappa_{2}) (2 {\rm Re\,} \lambda_6-{\rm Re\,} \lambda_7)
+4 {\rm Re\,} \lambda_7 v_{2}^3 (\kappa_{2}
-\kappa_{1})+\kappa_{1} v_{1}^3 (\lambda_{3}+\lambda_{4}) \Big)
+4 \kappa_{1} v_{1}^2 v_{s}^4 (\kappa_{1}-\kappa_{2})
+2 \lambda_{1}^2 v_{1}^6
+2 v_{1}^2 v_{2}^4 \big( \lambda_{2} ( \lambda_{2} -3 \lambda_{3} -3 \lambda_{4})
+2 \lambda_{3} ( \lambda_{3} +2 \lambda_{4} )+2 \lambda_{4}^2\big)
+2 v_{1}^4 v_{2}^2 \big(
\lambda_{2} \lambda_{3} 
+\lambda_{2} \lambda_{4} 
-2\lambda_{3}^2 
-4 \lambda_{3} \lambda_{4}
-2 \lambda_{4}^2 \big)
+ v_{1}^5 v_{2} \big(
2{\rm Re\,} \lambda_6 ( \lambda_{2} -6 \lambda_{3} -6 \lambda_{4} )
+8 {\rm Re\,} \lambda_7 ( \lambda_{3} +\lambda_{4}) \big)
-v_{1}^3 v_{2}^3 \big(
16 {\rm Re\,} \lambda_6 ( \lambda_{2}
-\lambda_{3}
-\lambda_{4} )
- 2{\rm Re\,} \lambda_7 ( 7 \lambda_{2} - 10 \lambda_{3} - 10 \lambda_{4} )\big)
-8 v_{1} v_{2}^5 {\rm Re\,} \lambda_7 \big(
 \lambda_{2} 
- \lambda_{3}
- \lambda_{4} \big)
+2 |\lambda_{5}|^2 v_{1}^2 (v_{1}^2-v_{2}^2)^2
-4|\lambda_{6}|^2 v_{1}^6
+12 |\lambda_{6}|^2 v_{1}^4 v_{2}^2
+8 {\rm Re\,} \lambda_6 {\rm Re\,} \lambda_7 (v_{1}^6
-5 v_{1}^4 v_{2}^2
+2v_{1}^2 v_{2}^4)
+4 |\lambda_{7}|^2 (6 v_{1}^4 v_{2}^2-5 v_{1}^2 v_{2}^4+ v_{2}^6 )
+{\rm Re\,} \lambda_5 v_{1} 
 \Big(2 v_{1} v_{s}^2 \big(
\kappa_{1} (3 v_{2}^2-2 v_{1}^2 )
+\kappa_{2} (v_{1}^2-2 v_{2}^2)\big)
+4 \lambda_{1} v_{1}^3 (v_{2}^2-v_{1}^2)
+4 v_{1}^3 v_{2}^2 (\lambda_{2} - 2 \lambda_{4})
-2 v_{1} v_{2}^4 ( 2 \lambda_{2} - 3 \lambda_{4})
+2 \lambda_{3} \big( v_{1}^5
+2{\rm Re\,} \lambda_7 (7 v_{1}^4 v_{2}
-8 v_{1}^2 v_{2}^3
+3 v_{2}^5)\big)
+2{\rm Re\,} \lambda_6 (v_{1}^6
-8 v_{1}^4 v_{2}^2
+5 v_{1}^2 v_{2}^4) \Big)\bigg],
\end{autobreak}
\\
\begin{autobreak}
a^{001}_{2} = 
-\frac{v_{2}^2}{v^8} \bigg[ 
\lambda_{1} v_{1}^2 \Big(v_{2} \big(2 v_{1} v_{s}^2 (\kappa_{1}
-2 \kappa_{2})
-2 v_{1} v_{2}^2 ( 2\lambda_{2} 
-\lambda_{345}
- {\rm Re\,} \lambda_5
+8 {\rm Re\,} \lambda_7)
-2 v_{1}^3 ( 3 \lambda_{345} 
- {\rm Re\,} \lambda_5)
+2{\rm Re\,} \lambda_7 v_{2}^3\big)
+2{\rm Re\,} \lambda_6 \big(7 v_{1}^2 v_{2}^2
-4 v_{1}^4\big) \Big)
-2 v_{1} v_{s}^2 
\Big(\kappa_{1} v_{2}^3 (2 \lambda_{2}
-{\rm Re\,} \lambda_5)
+\kappa_{2} v_{2}^3 (
-3 \lambda_{2}
+\lambda_{345}
+ {\rm Re\,} \lambda_5) 
+v_{1}^2 v_{2} ( (\kappa_{1}- \kappa_{2}) (\lambda_{3}+\lambda_{4})+(2 \kappa_{1} 
-3 \kappa_{2}) \lambda_{345}
-2 v_{1} v_{2}^2 (\kappa_{1}
-2\kappa_{2}) ({\rm Re\,} \lambda_6
-2 {\rm Re\,} \lambda_7)
+4 {\rm Re\,} \lambda_6 v_{1}^3 \big(\kappa_{1}
-\kappa_{2}\big) \Big)
+4 \kappa_{2} v_{1} v_{2} v_{s}^4 (\kappa_{2}
-\kappa_{1})
+v_{1}^3 v_{2}^3 \big(
6 \lambda_{2} (\lambda_{3} +\lambda_{4} )
-4 \lambda_{3} ( \lambda_{3} + 2\lambda_{4} )
-4 |\lambda_{5}|^2 
+12 |\lambda_{7}|^2 
-4 \lambda_{4}^2 
-8 \lambda_{4} {\rm Re\,} \lambda_5 
-40 {\rm Re\,} \lambda_6 {\rm Re\,} \lambda_7
-8 \lambda_{3} {\rm Re\,} \lambda_5 
+4 \lambda_{2} {\rm Re\,} \lambda_5 \big)
+v_{1} v_{2}^5 \big(
8 {\rm Re\,} \lambda_6 {\rm Re\,} \lambda_7 
+2\lambda_{4} {\rm Re\,} \lambda_5 
-4|\lambda_{7}|^2 
-2 \lambda_{2} (\lambda_{345} -2 {\rm Re\,} \lambda_5 )
+2 |\lambda_{5}|^2 
+2 \lambda_{2}^2 
+2 \lambda_{3} {\rm Re\,} \lambda_5 \big)
+v_{1}^4 v_{2}^2 \big( 
2{\rm Re\,} \lambda_6 ( 3 \lambda_{2} 
-10 \lambda_{345} +2 {\rm Re\,} \lambda_5 )
+ 2{\rm Re\,} \lambda_7 ( 
8 \lambda_{345} -3{\rm Re\,} \lambda_5 \big)
-v_{1}^2 v_{2}^4 \big(
2 {\rm Re\,} \lambda_6 ( 4 \lambda_{2} 
-4 \lambda_{345} -3 {\rm Re\,} \lambda_5 )
-2 {\rm Re\,} \lambda_7 ( 5 \lambda_{2} 
-6 \lambda_{345} -2 {\rm Re\,} \lambda_5)\big)
+2 v_{1}^5 v_{2} \big(
\lambda_{1}^2 
+2 \lambda_{345}^2 
- {\rm Re\,} \lambda_5 ( \lambda_{345} 
+{\rm Re\,} \lambda_5 )
+ |\lambda_{5}|^2 
+8 {\rm Re\,} \lambda_6 {\rm Re\,} \lambda_7 \big)
+2 {\rm Re\,} \lambda_6 v_{1}^6 \big(
4 \lambda_{345} 
- {\rm Re\,} \lambda_5 \big)
+2 {\rm Re\,} \lambda_5 {\rm Re\,} \lambda_7 v_{2}^6
+4 |\lambda_{6}|^2 v_{2}({v_{1}^7\over v_2^2}
-5 v_{1}^5 
+6 v_{1}^3 v_{2}^2) \bigg],
\end{autobreak}
\\
\begin{autobreak}
a^{210}_{2}= 
-\frac{2}{v^8} v_{2} v_{1}^3
\big((v_{1}^2-v_{2}^2)^2-v^2 (v_{1}^2-3 v_{2}^2) \big),
\end{autobreak}
\\
\begin{autobreak}
a^{120}_{2}= 
\frac{2}{v^8} v_{1}^4 \big(v_{1}^4
-8 v_{1}^2 v_{2}^2
+3 v_{2}^4
\big),
\end{autobreak}
\\
\begin{autobreak}
a^{201}_{2} = 
\frac{2}{v^8} v_{2}^3 v_{1} \big(v^2 (3 v_{1}^2
-v_{2}^2)+(v_{1}^2-v_{2}^2)^2\big), 
\end{autobreak}
\\
\begin{autobreak}
a^{102}_{2} = 
-\frac{2}{v^8} v_{2}^4 \big(v_{2}^4-8 v_{1}^2 v_{2}^2+3 v_{1}^4\big),
\end{autobreak}
\\
\begin{autobreak}
a^{021}_{2} = 
\frac{8}{v^8} v_{1} v_{2} (v_{1}^2 - v_{2}^2)
\big(
-2 v_{1}^2 v_{2}^2 
+ v_{1}^4
\big),
\end{autobreak}
\\
\begin{autobreak}
a^{012}_{2}= 
\frac{8}{v^8} v_{1} v_{2} (v_{1}^2 - v_{2}^2)
\big(
-2 v_{1}^2 v_{2}^2 + v_{2}^4
\big),
\end{autobreak}
\\
\begin{autobreak}
a^{111}_{2} = 
\frac{10}{v^8} v_{1}^2 v_{2}^2 (v_{1}^4
-v_{2}^4).
\end{autobreak}
\end{align}
}

The non-zero coefficients in ${\rm Im\,} J_3$ are given as
{\allowdisplaybreaks
\small
\begin{align}
\begin{autobreak}
a^{100}_3 = \frac{1}{v^4} 
\Big[
 4 v_{1} v_{2} ({\rm Re} \; \lambda_6+{\rm Re} \; \lambda_7) \big(
 \lambda_{1} v_{2}^2-\lambda_{2} v_{1}^2+2 \lambda_{4} (v_{2}^2-v_{1}^2)
 \big)
 +v_{1}^2 v_{2}^2 (\lambda_{1}-\lambda_{2}) (\lambda_{1}+\lambda_{2}+4 \lambda_{4})
 +4 (v_{2}^4-v_{1}^4) (|\lambda_6|^2+2 {\rm Re} \; \lambda_6 {\rm Re} \; \lambda_7+|\lambda_7|^2)
\Big],
\end{autobreak}
\\
\begin{autobreak}
a^{010}_3 = \frac{4}{v^4} 
\bigg[
 -v_{1} v_{2}^3 
 \big(
 \lambda_{1} {\rm Re} \; \lambda_5
 - \lambda_{2}^2
 + 2 \lambda_{2} \lambda_{3}
 + 2 \lambda_{4} (2 \lambda_{345}-{\rm Re} \; \lambda_5)
 \big)
 + v_{1}^3 v_{2} \Big(
 {\rm Re} \; \lambda_5 (\lambda_{2}+2 \lambda_{4})
 -(\lambda_{1}+2 \lambda_{4}) \big( \lambda_{2}-2 (\lambda_{3}+\lambda_{4}) \big)
 \Big)
 -2 v_{1}^2 v_{2}^2 ({\rm Re} \; \lambda_6-2 {\rm Re} \; \lambda_7) (\lambda_{1}+\lambda_{2}+4 \lambda_{4})
 +2 v_{1}^4 \big( {\rm Re} \; \lambda_5 ({\rm Re} \; \lambda_6 + {\rm Re} \; \lambda_7)-{\rm Re} \; \lambda_7 (\lambda_{1}+2 \lambda_{4}) \big)
 -2 v_{2}^4 \big( {\rm Re} \; \lambda_7 (\lambda_{2}+2 \lambda_{4}+{\rm Re} \; \lambda_5)+{\rm Re} \; \lambda_5 {\rm Re} \; \lambda_6 \big)
 -4 v^2 v_{1} v_{2} \big( |\lambda_6|^2-|\lambda_7|^2 \big)
\bigg],
\end{autobreak}
\\
\begin{autobreak}
a^{001}_3 = \frac{4}{v^4} 
\bigg[
 v_{1}^3 v_{2} \Big(
 -\lambda_{1}^2
 +2 \lambda_{1} \lambda_{3}
 +\lambda_{2} {\rm Re} \; \lambda_5
 +2 \lambda_{4} \big(
 2 (\lambda_{3}+\lambda_{4})
 +{\rm Re} \; \lambda_5 \big)
 \Big)
 +v_{1} v_{2}^3 \big((\lambda_{2}
 +2 \lambda_{4}) (\lambda_{1}
 -2 (\lambda_{3}+\lambda_{4}))
 -{\rm Re} \; \lambda_5 (\lambda_{1}+2 \lambda_{4})\big)
 -2 v_{1}^2 v_{2}^2 (2 {\rm Re} \; \lambda_6-{\rm Re} \; \lambda_7) (\lambda_{1}+\lambda_{2}+4 \lambda_{4})
 +2 v_{1}^4 \big( {\rm Re} \; \lambda_6 (\lambda_{1}
 +2 \lambda_{4}+{\rm Re} \; \lambda_5)
 +{\rm Re} \; \lambda_5 {\rm Re} \; \lambda_7 \big)
 +2 v_{2}^4 \big( \lambda_{2} {\rm Re} \; \lambda_6
 +2 \lambda_{4} {\rm Re} \; \lambda_6
 -{\rm Re} \; \lambda_5 ({\rm Re} \; \lambda_6
 +{\rm Re} \; \lambda_7) \big)
 -4 v^2 v_{1} v_{2} \big( |\lambda_6|^2-|\lambda_7|^2 \big)
\bigg],
\end{autobreak}
\end{align}
}
and 
\begin{align}
a^{120}_{3}=a^{102}_{3} =a^{111}_{3} =\frac{8}{v^4} (v_{1}^4-v_2^4).
\end{align}
Obviously, the above relations vanish if simultaneously the parameters ${\rm Im\,}\lambda_{5,6,7}$ become zero.

%%%%%%%%%%%%%%%%%%%%%%%%%%%%%%%%%%%%%%%%%%
\section{The Electron EDM}
\label{app:eEDM}
The 2-loop contributions to the electron EDM shown in Fig.~\ref{diag-EDM} are summarized below.

The diagram (a)~\cite{Pilaftsis:1999qt,Abe:2013qla,Kanemura:2020ibp}:
 {\small \begin{align}
{{(d_e})^{\gamma H_i}_f/ e}&={N_C e^2\over 32 \pi^4} \sum_{f=\tau,t,b} {Q_f^2 \over m_f} \sum_{i=1}^4 \big[f({m_f^2\over M_{H_i}^2})c_{if} \tilde{c}_{ie}+g({m_f^2\over M_{H_i}^2})c_{ie} \tilde{c}_{if}\big],
\\
{{(d_e})^{ZH_i}_f/ e}&={N_C g_{Zee} \over 32 \pi^4} \sum_{f=\tau,t,b} {Q_f g_{Zff}\over m_f} \sum_{i=1}^4\big[F({m_f^2\over M_{H_i}^2},{m_f^2\over M_{Z}^2})c_{if} \tilde{c}_{ie}+G({m_f^2\over M_{H_i}^2},{m_f^2\over M_{Z}^2})c_{ie} \tilde{c}_{if}\big],
 \end{align}}
 where $g_{Zff}= 2 e (T^3_f-2Q_f \sin^2\theta_W)/ \sin 2\theta_W$. 
 
The diagram (b)~\cite{Bowser-Chao:1997kjp,Nakai:2016atk,Oredsson:2019mni}:
 {\small \begin{align}
{{(d_e})^{W H^\pm}_{ff'}/ e}&={N_C e^2\over 512 \pi^4 \sin^2\theta_W (M_{H^\pm}^2-M_W^2)} \sum_{ff'} |K_{ff'}|^2 \int_0^1 dx \big((Q_{f}-Q_{f'}) x^2 +Q_{f'}x\big) 
\nonumber \\
&\times \big[T({m_{f}^2\over M_{H^\pm}^2},{m_{f'}^2\over M_{H^\pm}^2})-T({m_{f}^2\over M_{W}^2},{m_{f'}^2\over M_{W}^2})\big] 
\bigg\{(c_\pm\tilde{c}_{q}+\tilde{e}_{q}c_q) (1+x) \, m_{f}
\nonumber \\
&+ (c_\pm\tilde{c}_{q}-\tilde{e}_{\pm}c_q) (1-x) \, m_{f'}\bigg\},
 \end{align}}
where $c_{q}$, $\tilde{c}_{q}$, $c_{\pm}$ and $\tilde{c}_{\pm}$ are defined in Eq.~\eqref{ce}.

The diagram (c):
 {\small\begin{align}
{{(d_e})^{\gamma H_i}_W/ e}=-{e^2 \over 128 \pi^4 v^2} \sum_{i=1}^4 \big[ (6+ {M_{H_i}^2\over M_W^2}) f({M_W^2\over M_{H_i}^2})+ (10- {M_{H_i}^2\over M_W^2}) g({M_W^2\over M_{H_i}^2})\big] e_i \tilde{c}_{ie},
 \end{align}
 \begin{align}
{ {(d_e})^{Z H_i}_W/ e}&={e g_{Zee} g_{ZWW}\over 128 \pi^4 \cos^4\theta_W v^2}\times
\nonumber \\
&\sum_{i=1}^4 
\Big[ \big(-1+6 \cos^ 2\theta_W- {M_{H_i}^2\over 2 M_W^2}(1-2 \cos^ 2\theta_W \big) F({M_W^2\over M_{H_i}^2},\cos^ 2\theta_W)
\nonumber \\
&- \big(3-10 \cos^ 2\theta_W+ {M_{H_i}^2\over 2 M_W^2}(1-2 \cos^ 2\theta_W)\big) G({M_W^2\over M_{H_i}^2},\cos^ 2\theta_W)\Big] e_i \tilde{c}_{ie},
 \end{align}}
 where $g_{ZWW}= e \cot \theta_W$.
 
The diagram (d):
 {\small \begin{align}
{{(d_e})^{\gamma H_i}_{H^\pm}/ e}&=-{e^2 \over 128 \pi^4 M_{H^\pm}^2} \sum_{i=1}^4 \big[ f({M_{H^\pm}^2 \over M_{H_i}^2})- g({M_{H^\pm}^2\over M_{H_i}^2})\big] g_{H^\pm H_i}\tilde{c}_{ie},
\\
{{(d_e})^{Z H_i}_{H^\pm}/ e}&=-{g_{Zee} \,g_{ZH^\pm} \over 128 \pi^4 M_{H^\pm}^2} \sum_{i=1}^4 \big[ F({M_{H^\pm}^2 \over M_{H_i}^2},{M_{H^\pm}^2 \over M_{Z}^2})- G({M_{H^\pm}^2 \over M_{H_i}^2},{M_{H^\pm}^2 \over M_{Z}^2})\big] g_{H^\pm H_i}\tilde{c}_{ie},
 \end{align}}
with $g_{ZH^\pm}=e \cot 2\theta_W$.
 
The diagrams (e) and (f):
 {\small \begin{align}
{(d_e)^{H_i}_{W,H^\pm}/ e}&={1 \over 512 \pi^4 v^2} \sum_{i=1}^4 \Big[{g^2}H({M_{H_i}^2,M_{H^\pm}^2})e_i -2 G({M_{H_i}^2,M_{H^\pm}^2})g_{H^\pm H_i}\Big]\tilde{c}_{ie}.
 \end{align}}
The integral functions adopted above are given in Appendix \ref{app:int}.

Thus, adding together the contributions listed above, the total electron EDM is given by
 \begin{align}
 d_e=\delta &\Big[(d_e)^{\gamma H_i}_f+(d_e)^{Z H_i}_f+ {(d_e})^{W H^\pm}_{ff'}+(d_e)^{\gamma H_i}_W+ (d_e)^{Z H_i}_W
\nonumber \\
 &+(d_e)^{\gamma H_i}_{H^\pm}+(d_e)^{Z H_i}_{H^\pm}+(d_e)^{H_i}_{W,H^\pm}\Big],
 \end{align}
with $\delta=1.96\times 10^{-13}$~[cm$\cdot$GeV], giving total $d_e$ the dimension~[e$\cdot$cm].

%%%%%%%%%%%%%%%%%%%%%%%%%%%%%%%%%%%%%%%%%%%%%%%%%%%%%%%%%%%%%%%%%%%%%%%%%%%%%
\section{Loop Integrals for EDMs}\label{app:int}

 The integral functions used in Section~\ref{sec:EDM} for the calculation of the electron EDM are \cite{Barr:1990vd,Abe:2013qla}
 {\small \begin{align}
 D_2(z,z)&=\int_0^1 dx{x^2\over 1-x+zx^2},
 \\
 f(z)&=z/2\int_0^1 dx {1-2x(1-x) \over x(1-x)-z} \log({x(1-x)\over z}),
 \\
 g(z)&=z/2\int_0^1 dx {1\over x(1-x)-z} \log({x(1-x)\over z}),
 \\
 F(x,y)&={y f(x) \over y-x}-{x f(x) \over y-x},
 \\
G(x,y)&={y g(x) \over y-x}-{x g(x) \over y-x},
\\
T(a_1,a_2)&={\log(a_1x+a_2(1-x))-\log(x(1-x)) \over x(1-x)-a_1x-a_2(1-x)},
\\
H(m_1^2,m_2^2)&={M_{W}^2 \over M_{H^\pm}^2-M_{W}^2}\big[ A(M_{W}^2, m_1^2)- A(M_{H^\pm}^2, m_1^2) \big],
\\
G(m_1^2,m_2^2)&={M_{W}^2 \over M_{H^\pm}^2-M_{W}^2}\big[ B(M_{W}^2, m_1^2)- B(M_{H^\pm}^2, m_1^2) \big],
 \end{align}}
 with
 {\small
 \begin{align}
A(m_1^2,m_2^2)&=\int_0^1 dz (1-z)^2 (z-4+z {M_{H^\pm}^2-m_2^2 \over M_{W}^2}
\times
\nonumber \\
& {m_1^2 \over M_{W}^2(1-z)+m_2^2 z -m_1^2 z (1-z)}\log({M_{W}^2 (1-z)+m_2^2 z \over m_1^2 z(1-z)}),
\\
B(m_1^2,m_2^2)&=
 {m_1^2 z (1-z)^2 \over M_{W}^2(1-z)+m_2^2 z -m_1^2 z (1-z)}\log({M_{W}^2 (1-z)+m_2^2 z \over m_1^2 z(1-z)}).
 \end{align}}

\bibliography{biblio}

\end{document}